\documentclass[reprint,aip,cha,onecolumn,groupedaddress,nobibnotes,nofootinbib]{revtex4-1}
\usepackage{amsmath, amsfonts, amssymb, amsthm, mathtools} 
\usepackage[caption=false]{subfig}
\usepackage{algorithm}
\usepackage{color}
\usepackage{mathtools}
\usepackage[
  unicode=true,
  bookmarks=false,
  pdfpagelabels=false,
  hyperfootnotes=false,
  hyperindex=false,
  pageanchor=false,
  colorlinks=false,
]{hyperref}
\usepackage[table]{xcolor}
\usepackage{pbox}
\usepackage{cleveref}

\DeclareMathOperator{\vol}{vol}
\DeclareMathOperator{\cut}{cut}
\DeclareMathOperator{\NCut}{NCut}

\begin{document}
\title{A Critical Comparison of Lagrangian Methods for Coherent Structure Detection}

\author{Alireza Hadjighasem}
\email{alirezah@mit.edu}
\altaffiliation{Department of Mechanical Engineering, MIT, 77 Massachusetts
Ave., Cambridge, MA 02139, USA}

\author{Mohammad Farazmand}
\email{mfaraz@mit.edu}
\altaffiliation{Department of Mechanical Engineering, MIT, 77 Massachusetts
Ave., Cambridge, MA 02139, USA}

\author{Daniel Blazevski}
\email{daniel.blazevski@gmail.com}
\altaffiliation{Insight Data Science, 45 W 25th St, New York, NY 10010, USA}

\author{Gary Froyland}
\email{g.froyland@unsw.edu.au}
\altaffiliation{School of Mathematics and Statistics, University of New South Wales, Sydney NSW 2052, Australia}

\author{George Haller}
\email{georgehaller@ethz.ch (Email address for correspondence)}
\altaffiliation{Institute of Mechanical Systems, Department of Mechanical and Process Engineering, ETH Z\"{u}rich, Leonhardstrasse 21, 8092 Z\"{u}rich, Switzerland}

\date{\today}


\begin{abstract}
We review and test twelve different approaches to the detection of
finite-time coherent material structures in two-dimensional, temporally
aperiodic flows. We consider both mathematical methods and diagnostic
scalar fields, comparing their performance on three benchmark examples:
the quasiperiodically forced Bickley jet, a two-dimensional turbulence
simulation, and an observational wind velocity field from Jupiter's
atmosphere. A close inspection of the results reveals that the various
methods often produce very different predictions for coherent structures,
once they are evaluated beyond heuristic visual assessment. As we
find by passive advection of the coherent set candidates, false positives
and negatives can be produced even by some of the mathematically justified
methods due to the ineffectiveness of their underlying coherence principles
in certain flow configurations. We summarize the inferred strengths
and weaknesses of each method, and make general recommendations for
minimal self-consistency requirements that any Lagrangian coherence
detection technique should satisfy. 
\end{abstract}
\maketitle
\begin{quotation}
\textbf{Coherent Lagrangian (material) structures are ubiquitous in
unsteady fluid flows, often observable indirectly from tracer patterns
they create, for example, in the atmosphere and the ocean. Despite
these observations, a direct identification of these structures from
the flow velocity field (without reliance on seeding passive tracers)
has remained a challenge. Several heuristic and mathematical detection
methods have been developed over the years, each promising to extract
materially coherent domains from arbitrary unsteady velocity fields
over a finite time interval of interest. Here we review a number of
these methods and compare their performance systematically on three
benchmark velocity data sets. Based on this comparison, we discuss
the strengths and weaknesses of each method, and recommend minimal
self-consistency requirements that Lagrangian coherence detection
tools should satisfy.} 
\end{quotation}

\section{Introduction}

Coherent structures, such as eddies, jet streams and fronts, are ubiquitous
in fluid dynamics. They tend to enhance or inhibit material transport
between distinct flow regions. Their Lagrangian (trajectory-based)
analysis has improved our understanding of a number of fluid mechanics
problems, including ocean mixing~\cite{Beron_Vera08b,Haller13,Harrison13,Dong14},
the swimming of marine animals~\cite{Dabiri05,Peng08,Huhn15} and
fluid-structure interactions~\cite{Lipinski08,Green10,Le13}.

A number of different approaches to Lagrangian structure detection
have been proposed over the past two decades (see Refs.~\onlinecite{Peacock10,Peacock13,Peacock15,Shadden12,Haller15,Allshouse15}
for reviews). The volume and variety of these methods have made it
difficult for the practitioner to choose the appropriate tool that
fits their needs best. In addition, purely heuristic tools with unclear
assumptions and mathematical methods supported by theorems have rarely
been contrasted, creating a general feeling that all Lagrangian methods
give pretty much the same results. All this creates a need for taking
stock in the area of material structure detection by comparing the
methods on challenging benchmark problems. The purpose of this paper
is to address this need by surveying a large number of Lagrangian
coherent structure detection methods. We aim to provide a comparative
guide to practitioners who wish to use these techniques in specific
flow problems.

\begin{figure}
\subfloat[\label{fig:Bickley_AR}]{\includegraphics[width=0.4\textwidth]{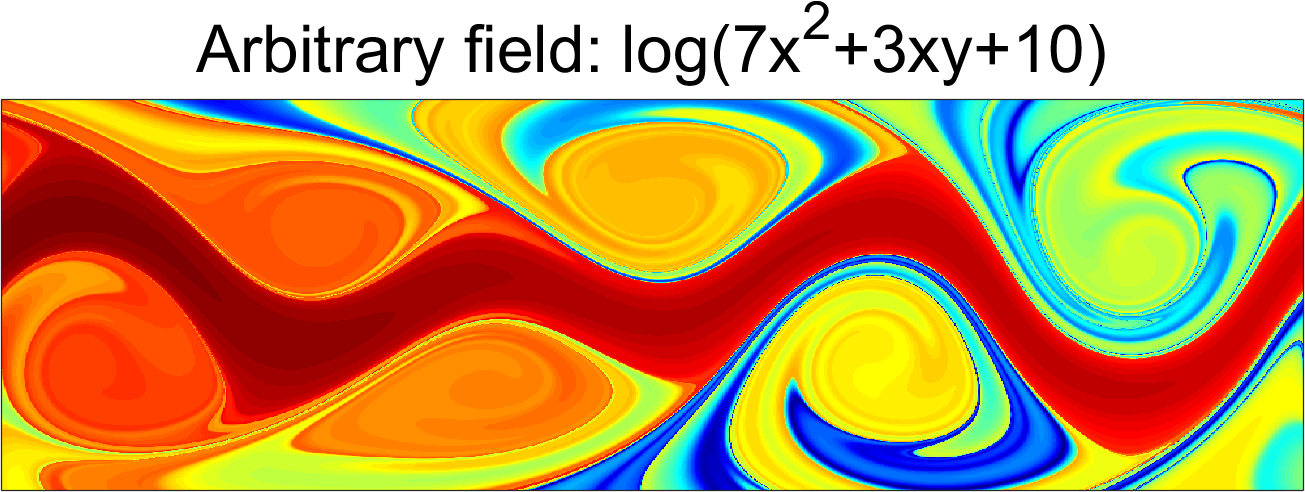}}\\
\; \subfloat[\label{fig:2Dturb_AR}]{\includegraphics[width=0.25\textwidth]{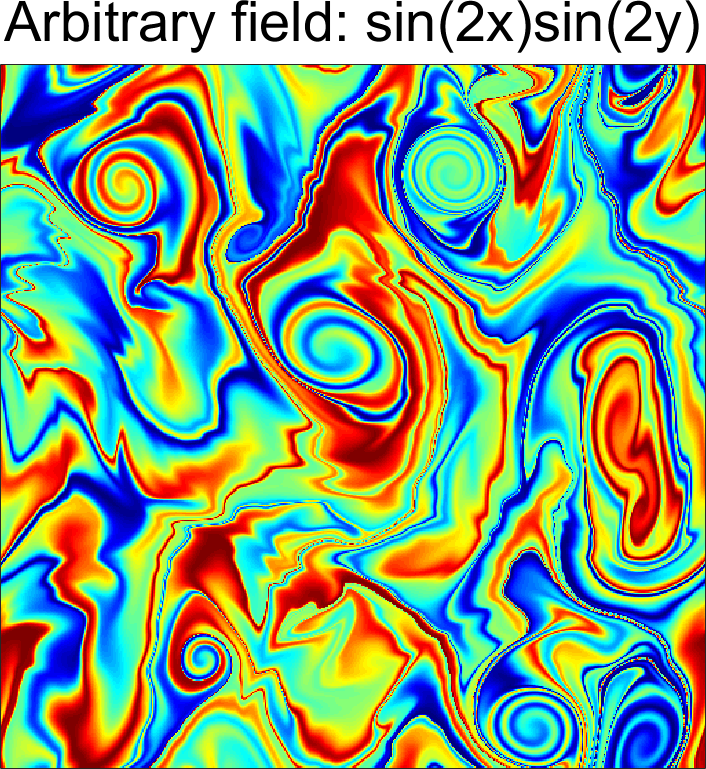}}
\; \subfloat[\label{fig:Jupiter_Fake_field}]{\includegraphics[width=0.3\textwidth]{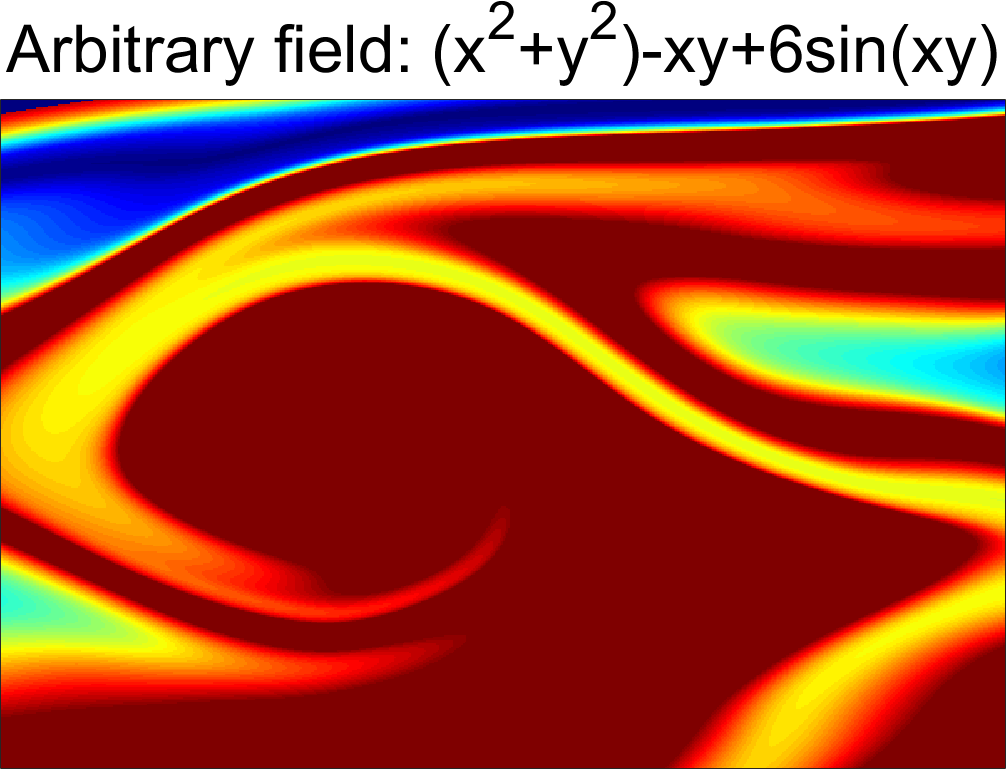}}
\caption{Three arbitrary advected scalar fields evaluated on advected particle
positions $(x,y)$ at the end time, then plotted over the initial
positions $(x_{0},y_{0})$ of the particles. (a) quasiperiodic Bickley
jet (b) two-dimensional turbulence (c) observed wind field of Jupiter.
\label{fig:Three-arbitrary-advected}}
\end{figure}

In this comparison, we consider twelve coherent structure detection
methods. After a brief introduction to each method, we compare their
outputs on three examples, then summarize our findings in a list of
strengths and weaknesses for each method. We classify the twelve methods
into two broad categories: 
\begin{enumerate}
\item Diagnostic methods: They propose a scalar field, derived from physical
intuition, whose features are expected to highlight coherent structures.
These methods are reviewed in Section~\ref{sec:Lagrangian-coherence-diagnostics}. 
\item Analytic methods: They define the coherent structures as the solutions
of mathematically formulated coherence problems. These methods are
reviewed in Section~\ref{sec:Mathematical-approaches-to}. 
\end{enumerate}
Being diagnostic or analytic in nature is not an a priori positive
or negative feature for a method. As we point out in Section~\ref{sec:Mathematical-approaches-to},
a heuristic but insightful diagnostic method might outperform a rigorous
mathematical coherence principle that has been formulated with disregard
to the underlying physics. On the computational side, a consistently
performing diagnostic may also be preferred as a tool for quick exploration
over a rigorous mathematical approach with heavy computational cost.
On the other hand, diagnostic tools offering purely visual inference
of structures must meet a minimum expectation: they must consistently
outperform visual inference from randomly chosen scalar fields, such
as those shown for our three examples in Fig. \ref{fig:Three-arbitrary-advected}.

These three examples include the following: 
\begin{enumerate}
\item \emph{The Bickley jet:} an analytically defined velocity field with
quasi-periodic time dependence. 
\item \emph{Two-dimensional turbulence:} a high-resolution data set obtained
from the direct numerical simulation of the Navier\textendash Stokes
equations in two dimensions. 
\item \emph{Jupiter's wind field:} an observational data set of Jupiter's
atmospheric velocities, reconstructed from video footage taken by
the Cassini spacecraft.. 
\end{enumerate}
These examples are ordered in an increasing level of difficulty, given
how much information is available about the flow in each of them.
The Bickley jet velocity field is temporally aperiodic but recurrent,
and known analytically at all locations and times. The two-dimensional
turbulence data set is slightly more challenging, as the velocity field
is fully aperiodic, known only at discrete points in space and time.
One could, however, still increase the resolution of the data by solving
the Navier\textendash Stokes equations over finer grids (or, equivalently,
by including more Fourier modes). Furthermore, the temporal duration
of the data set can also be increased at will. The third example involving
the Jupiter's atmospheric velocities poses the greatest challenge,
as the spatial and temporal length and resolution of this fully aperiodic
data set is limited by the available video footage recorded by the
Cassini mission.

Comparisons of a limited number of methods on specific structures
in individual examples have already appeared \cite{Beron_Vera13,Haller16,Ma15}.
Here the objective is to perform a systematic comparison on a variety
of challenging flow fields in which a ground truth can nevertheless
be reasonably established. Our scope is also broader in that we cover
all known types of Lagrangian structures in two-dimensions: elliptic
(vortex-type), hyperbolic (repelling or attracting) and parabolic
(jet-core-type) material structures.

The rest of this paper is organized as follows. In \Cref{sec:Setup,sec:Lagrangian-coherence-diagnostics,sec:Mathematical-approaches-to},
we introduce the twelve diagnostic and analytic methods considered
in this comparison. Despite our efforts to keep the method descriptions
to a minimum, the introduction of analytic methods necessarily takes
up more space due to the need to explain the mathematical principles underlying them.
In Section~\ref{sec:compare}, the methods are applied to the three
examples, with different aspects of their performance compared. Our
overall assessment of the strengths and weaknesses of each method
appears in Section \ref{sec:Assessment}, and a proposed set of minimal
requirements for Lagrangian coherence detection methods is given in
Section~\ref{sec:conclude}.

\section{General setup}

\label{sec:Setup}

We consider here flows defined by two-dimensional unsteady velocity
fields $v(x,t)$ known over a finite time interval $[t_{0},t_{1}]$.
The fluid particle motions satisfy the differential equation 
\begin{equation}
\dot{x}=v(x,t),\quad x\in U\subset\mathbb{R}^{2},\quad t\in[t_{0},t_{1}],\label{eq:ch6:dynsys}
\end{equation}
whose trajectories are denoted by $x(t;t_{0},x_{0})$, with $x_{0}$
referring to their initial position at time $t_{0}$. Our focus here
is Lagrangian, concerned with coherent behavior exhibited by sets
of trajectories of \eqref{eq:ch6:dynsys}. This is in contrast to
the classic Eulerian approach taken in fluid mechanics which focuses
on coherent features of $v(x,t)$.

Central to all Lagrangian approaches is the flow map 
\begin{equation}
F_{t_{0}}^{t}(x_{0}):x_{0}\mapsto x(t;t_{0},x_{0}),\label{eq:ch6:flow_map}
\end{equation}
mapping initial positions $x_{0}$ to their current positions $x$
at time $t$. Several Lagrangian coherence-detection methods also
rely on the flow gradient $\nabla F(x_{0})$ (or deformation gradient),
the derivative of the flow map with respect to the initial condition
$x_{0}$. The stretching induced by the flow gradient is captured
by the right Cauchy\textendash Green strain tensor $C_{t_{0}}^{t}$ of the deformation field, defined as \cite{Truesdell04} 
\begin{equation}
C_{t_{0}}^{t}(x_{0})=\left[\nabla F_{t_{0}}^{t}\right]^{\intercal}\nabla F_{t_{0}}^{t},\label{eq:ch6:CG}
\end{equation}
with the symbol $\intercal$ indicating matrix transposition. In our
present two-dimensional setting, the symmetric and positive definite
tensor $C_{t_{0}}^{t}$ has two positive eigenvalues $0<\lambda_{1}\leq\lambda_{2}$
and an orthonormal eigenbasis $\left\{ \xi_{1},\xi_{2}\right\} $
satisfying 
\begin{align}
 & C_{t_{0}}^{t}(x_{0})\xi_{i}(x_{0})=\lambda_{i}(x_{0})\mathbf{\xi}_{i}(x_{0}),\quad\left|\xi_{i}(x_{0})\right|=1,\quad i=1,2,\nonumber \\
 & \xi_{2}(x_{0})=\Omega\xi_{1}(x_{0}),\quad\Omega=\left(\begin{array}{cc}
0 & -1\\
1 & 0
\end{array}\right).\label{eq:ch6:eig_def}
\end{align}

\section{Diagnostics for Lagrangian coherence}\label{sec:Lagrangian-coherence-diagnostics}

We first briefly review five Lagrangian diagnostic scalar fields that
have been proposed for material coherence detection in the literature.
They are classified as Lagrangian because their pointwise value at
a point $x_{0}$ of the flow domain depends solely on the trajectory
segment running from the location $x_{0}$ at time $t_{0}$ up to
the location $F_{t_{0}}^{t_{1}}(x_{0})$ at time $t_{1}$. Based on
simple geometric or physical arguments, these diagnostics are expected
to highlight coherence or lack thereof in the flow. Most of them,
however, offer neither a strict definition of the coherent flow structures
they seek, nor a precise mathematical connection between their geometric
features and those flow structures.

A basic expectation for such diagnostic scalar fields is that they
should at least outperform generic passively advected scalar fields
in their diagnostic abilities. By definition, Lagrangian coherent
structures (LCSs) create coherent trajectory patterns \cite{Haller15},
and hence the footprint of LCSs should invariably appear in \emph{any}
generic tracer distribution advected by trajectories. To this end,
in our comparisons performed on given examples, we have also included
ad hoc passive scalar fields as baselines for the efficacy of diagnostic
and mathematical approaches (see \Cref{fig:Three-arbitrary-advected}).

Another expectation for Lagrangian diagnostics stems from the fact
that LCSs are composed of the same\emph{ material} trajectories, irrespective
of what coordinate system we use to study them. Therefore, the assessment
of whether or not a trajectory is part of an LCS is inherently independent
of the frame of the observer\cite{Peacock15}. Any self-consistent
two-dimensional LCS method should, therefore, identify the same set
of trajectories as LCSs under all Euclidean observer changes of the
form $x=Q(t)y+b(t),$ where $y\in\mathbb{R}^{2}$ is the coordinate
in the new frame, $Q(t)\in SO(2)$ represents time-dependent rotation,
and $b(t)\in\mathbb{R}^{2}$ represents time-dependent translation.
A similar requirement holds, with appropriate modifications, for three-dimensional
LCS-detection methods.

Frame-invariance is particularly important in truly unsteady flows,
which have no distinguished frame of reference \cite{lugt79}. Within
this class, geophysical fluid flows represent an additional challenge,
because they are defined in a rotating frame. The detection or omission
of a feature by a diagnostic in such flows, therefore, should clearly
not be an artifact of the co-rotation of the frame with the Earth.
For each surveyed diagnostic below, we will discuss its objectivity
or lack thereof.

\subsection{Finite-time Lyapunov exponent (FTLE)}

Haller \cite{Haller00,Haller02} proposed that the time $t_{0}$ positions
of the strongest repelling LCSs over the time interval $[t_{0},t_{1}]$
should form ridges of the finite-time Lyapunov exponent (FTLE) field
\begin{equation}
\begin{split}\text{FTLE}_{t_{0}}^{t_{1}}(x_{0}) & =\frac{1}{\left|t_{1}-t_{0}\right|}\log\left\Vert \nabla F_{t_{0}}^{t_{1}}(x_{0})\right\Vert \\
 & =\frac{1}{\left|t_{1}-t_{0}\right|}\log\sqrt{\lambda_{2}(x_{0})}.
\end{split}
\label{FTLE}
\end{equation}
Similarly, time $t_{1}$ positions of the strongest attracting LCSs
over $[t_{0},t_{1}]$ are expected to be marked by ridges of the backward-time
FTLE field $\text{FTLE}_{t_{1}}^{t_{0}}$. Repelling and attracting
LCSs are usually referred to as hyperbolic LCSs, as they generalize
the notion of hyperbolic invariant manifolds to finite-time dynamics.
The FTLE field is objective by the objectivity of the invariants of the
Cauchy--Green strain tensor~\cite{Gurtin82}.

The FTLE field \eqref{FTLE} measures the largest finite-time growth
exponent experienced by infinitesimal perturbations to the initial
condition $x_{0}$ over the time interval $[t_{0},t_{1}]$. It is
therefore a priori unclear if a given FTLE ridge indeed marks a repelling
material surface, or just a surface of high shear (cf. Ref.~\onlinecite{Haller15}
for an example). Nevertheless, time-evolving FTLE ridges computed
over sliding intervals $[t_{0}+T,t_{1}+T]$ with varying $T$ are
often informally identified with LCSs. There are, however, both conceptual
and mathematical issues with such an identification, and the evolving
ridges so obtained may be far from Lagrangian \cite{Haller15}.

Motivated by the fact that material stretching is minimal along jet steams, FTLE trenches have been proposed for detection of unsteady jet cores (or parabolic LCSs)\cite{BeronVera10,Beron-Ver12}. While, in many examples, the jet cores are closely approximated by the FTLE trenches, there exist counterexamples where an FTLE trench does not coincide with the jet \cite{Farazmand14}.

The FTLE diagnostic is not geared towards detecting elliptic (vortex-type)
LCSs in finite-time flow data. While the FTLE values are expected
to be low near elliptic LCS, a sharp boundary for vortex-type structures
does not generally emerge from this diagnostic, as seen in our examples
below.

\subsection{Finite-Size Lyapunov Exponent (FSLE)}

An alternative assessment of perturbation growth in the flow is provided
by the Finite-Size Lyapunov exponent (FSLE). To define this quantity,
we first select an initial separation $\delta_{0}>0$ and a separation
factor $r>1$ of interest. The separation time $\tau(x_{0};\delta_{0},r)$
is then defined as the minimal time in which the distance between
a trajectory starting from $x_{0}$ and some neighboring trajectory
starting $\delta_{0}$-close to $x_{0}$ first reaches $r\delta_{0}$.
The FSLE associated with the location $x_{0}$ is then defined as
\cite{Artale97,Aurell97,Joseph02}

\begin{equation}
\text{FSLE}(x_{0};\delta_{0},r)=\frac{\log r}{\tau(x_{0};\delta_{0},r)}.
\end{equation}

In contrast to the FTLE field, the FSLE field focuses on separation
scales exceeding the threshold $r$, and hence can be used for selective
structure detection. A further conceptual advantage of the FSLE field
is that its computation requires no a priori choice of an end time
$t_{1}$.

By analogy with FTLE ridges, FSLE ridges have also been proposed as
indicators of hyperbolic LCSs (see Refs.~\onlinecite{Joseph02,Dovidio14,Bettencourt13}).
This analogy is mathematically justified for sharp enough FSLE ridges
of nearly constant height \cite{Karrasch13}. A general correspondence
between FSLE and FTLE ridges, however, does not exist. This is because
$\text{FSLE}(x_{0};\delta_{0},r)$ lumps trajectory separation events
occurring over different time intervals into the same scalar field,
and hence has no general relationship to the single finite-time flow
map $F_{t_{0}}^{t}(x_{0})$.

The FSLE field has generic jump discontinuities and a related sensitivity
to the computational time step (see Ref.~\onlinecite{Karrasch13}
for details). The FSLE, however, is still an objective field, given
that it is purely a function of particle separation.

\subsection{Mesochronic analysis}\label{Mesochronic_method}

Mezi{\'{c}} et al. \cite{Mezic10} proposed the eigenvalue configuration
of the deformation gradient $\nabla F_{t_{0}}^{t_1}(x_{0})$ as a diagnostic
for qualitatively different regions of material mixing. Specifically,
their \emph{mesochronic} classification considers regions where $\nabla F_{t_{0}}^{t_1}(x_{0})$
have real eigenvalues as \emph{mesohyperbolic}, and regions where
$\nabla F_{t_{0}}^{t_1}(x_{0})$ has complex eigenvalues as \emph{mesoelliptic}.
Mesohyperbolic regions are further divided into two categories as
follows. Since $\nabla F_{t_{0}}^{t_1}(x_{0})$ is an orientation preserving
diffeomorphism, we necessarily have $\det\nabla F_{t_{0}}^{t_1}(x_{0})>0$,
which implies that real eigenvalues of $\nabla F_{t_{0}}^{t_1}(x_{0})$
are either both negative or both positive. Mezic et al. \cite{Mezic10}
refer to the case where the real eigenvalues are both positive as
\emph{mesohyperbolic without (a $180^{\circ}$) rotation}. If the
eigenvalues are real and negative, the trajectory is called \emph{mesohyperbolic
with (a $180^{\circ}$) rotation}.

Data collected in the aftermath of Deepwater Horizon Spill \cite{Mezic10}
shows that mixing zones in the ocean are predominantly mesohyperbolic
when the integration time is selected to be about 4 days. Longer studies
of ocean data suggest that oceanic flows are predominantly mesoelliptic
over time scales beyond four days \cite{Beron_Vera13}. This is in
line with the expectation that accumulated material rotation along
general trajectories unavoidably creates nonzero imaginary parts for
the eigenvalues of $\nabla F_{t_{0}}^{t_1}(x_{0})$, even if the underlying
trajectory starting from $x_{0}$ is of saddle-type.

From a mathematical point of view, the linear mapping $\nabla F_{t_{0}}^{t_1}(x_{0})$
is a two-point-tensor between the tangent spaces $T_{x_{0}}\mathbb{R}^{2}$
and $T_{F_{t_{0}}^{t_1}(x_{0})}\mathbb{R}^{2}$ of $\mathbb{R}^{2}$.
Posing an eigenvalue problem for $\nabla F_{t_{0}}^{t_1}(x_{0})$ is,
therefore, only meaningful when these tangent spaces coincide, i.e.,
$x_{0}=F_{t_{0}}^{t_1}(x_{0})$ lies on a trajectory that returns \emph{exactly}
to its starting point at time $t$. For this reason, it is difficult
to attach a mathematical meaning to the mesochronic partition in general
unsteady flows in which such returning trajectories are nonexistent.

The mesochronic partition of the flow domain is not objective due
to the frame-dependence of the deformation gradient $\nabla F_{t_{0}}^{t_1}(x_{0})$
(see e.g., Ref.~\onlinecite{Liu03}). As a consequence, the elliptic-hyperbolic
classification of trajectories obtained from this method will change
under changes of the observer.

The mesochronic notions of hyperbolicity and ellipticity differ from
classic hyperbolicity and ellipticity concepts for Lagrangian trajectories.
Even regions of concentric closed orbits in a steady flow (a classic
case of elliptic particle motion) are marked by nested sequences of
alternating mesoelliptic and mesohyperbolic annuli (see Ref.~\onlinecite{Mezic10}
for an example). No published account of a coherent vortex definition
from these plots is yet available, but outermost boundaries of smooth
and nested elliptic, hyperbolic-with-rotation annulus sequences have
recently been suggested as coherent structure boundaries \cite{Mezic14}
for general unsteady flow. We will adopt this definition (i.e., at
least three nested annuli of different mesochronic types, containing
no saddle-type critical points of $\det\nabla F_{t_{0}}^{t_1}(x_{0})$)
for a mesoelliptic coherent structure in our comparison study.

\subsection{Trajectory length}\label{Mfunction_method} Mancho et al. \cite{Mancho13} propose that
abrupt variations (i.e., curves of high gradients) in the arc-length
function 
\[
M_{t_{0}}^{t_{1}}(x_{0})=\int_{t_{0}}^{t_{1}}\left|v(x(s;t_{0},x_{0}),s)\right|ds
\]
of the trajectory $x(s;t_{0},x_{0})$ indicate the time $t_{0}$ positions
of boundaries of qualitatively different dynamics. The $M_{t_{0}}^{t_{1}}(x_{0})$
function is arguably the quickest to compute of all Lagrangian diagnostics
considered here. It also naturally lends itself to applications to
float data, given that the arclength of a trajectory can be computed
without any reliance on a velocity field or on neighboring trajectories.

As any scalar field computed along trajectories, $M_{t_{0}}^{t_{1}}(x_{0})$
is generally expected to show an imprint of Lagrangian coherent structures,
as indeed found by Mancho et al. \cite{Mancho13}. There is, however,
no established mathematical connection between material coherent structures
and features of $M_{t_{0}}^{t_{1}}(x_{0})$. Indeed, several counter-examples
to coherent structure detection based on trajectory length are available.\cite{Ruiz-Herrera15,Ruiz-Herrera16}.

The function $M_{t_{0}}^{t_{1}}(x_{0})$ is not objective or even
Galilean invariant. For instance, in a frame co-moving with \emph{any}
selected trajectory $x(s;t_{0},x_{0})$, the trajectory itself has
zero arclength, and hence its initial condition $x_{0}$ will generically
be a global minimum. The level curve structure of $M_{t_{0}}^{t_{1}}(x_{0})$
is not objective either, because the integrand of its gradient field
$\nabla M_{t_{0}}^{t_{1}}(x_{0})$ consists of elements that are frame-dependent.

\subsection{Trajectory complexity}

Rypina et al. \cite{Rypina11} propose a partitioning of the flow
domain into regions where trajectories exhibit different levels of
complexity. They quantify individual trajectory complexity over a finite time interval $[t_{0},t_{1}]$ using the \textit{ergodicity defect} (cf. Ref.~\onlinecite{Scott09})
\begin{equation}
d(s;x_{0},t_{0})=\sum\limits _{j=1}^{K}\,\,\left[\frac{N_{j}(s)}{N}-s^{2}\right]^{2},f\label{eq:ergod_defect}
\end{equation}
where $N$ is the total number of trajectories, and $N_{j}(s)$ is
the the number of trajectory points that lie inside the $j^{th}$
element of a square grid of side-length $s$. The integer $K=1/s^{2}$
denotes the total number of boxes forming the grid. The total area
of the full flow domain is normalized to unity. Mathematically, formula
\eqref{eq:ergod_defect} is just the $L^{2}$ deviation of a histogram
based on the trajectory points from a constant histogram. 

The ``most non-ergodic trajectory'' is a fixed point, for which
we obtain $d=1$. In contrast, for an ``ergodic trajectory'' (uniformly
distributed trajectory), one should obtain $\lim_{s\to0}d(s;x_{0},t_{0})=0$.\footnote{The terms \emph{ergodic} and \emph{non-ergodic} used by Rypina et
al. \cite{Rypina11} are to be understood at an informal level here,
given that \emph{all} infinite trajectories (including fixed or periodic
points of a map) support ergodic invariant measures.} Rypina et al. \cite{Rypina11} define the average ergodicity defect
over different scales of $s$ as 
\begin{equation}
\bar{d}(s;x_{0},t_{0})=\mathrm{mean}_{s}(d(s;x_{0},t_{0})),
\end{equation}
where the mean is taken over a broad range of spatial scales $s$
of interest. 

While no mathematical connection is known between the ergodicity defect
and finite-time coherent structures, locations of abrupt changes (large
gradients) in the topology of $\bar{d}(s;x_{0},t_{0})$ as a function
of $x_{0}$ are expected to mark boundaries between qualitatively
different flow regions. The quantity $\bar{d}(s;x_{0},t_{0})$ is
objective, because presence in, or absence from, a grid cell is invariant
under rotations and translations, as long as the same rotations and
translations are applied to both the trajectories and the grid cells.
The approach is simple to implement and has proven itself effective
on low-resolution data \cite{Rypina11}.

\subsection{Shape coherence}

Ma and Bollt \cite{Ma14} seek coherent set boundaries as closed material
lines at time $t_{0}$ that are nearly congruent\footnote{Two geometric objects are called \emph{congruent} if one can be transformed
into the other by a combination of rigid-body motions.} with their advected images at time $t_{1}$. Such near-congruence
is ensured by classic results if the curvature distributions along
the original and advected curve are close enough.

Motivated by examples of steady linear flows, Ma and Bollt \cite{Ma14}
propose finding shape-coherent curves as minimizers of the angle between
the dominant eigenvectors of the forward-time and the backward-time
Cauchy\textendash Green strain tensors. Stated in our present context,
the position of the boundary of a shape-coherent set at time $\hat{t}=(t_{0}+t_{1})/2$
is a closed curve along which the splitting angle function 
\begin{equation}
\theta(\hat{x}_{0})=\arcsin\left(|\xi_{2}^{fw}(\hat{x}_{0})\times\xi_{2}^{bw}(\hat{x}_{0})|\right),\qquad\hat{x}_{0}=F_{t_{0}}^{\hat{t}}(x_{0}),\label{eq:ch6:shape_coherent}
\end{equation}
vanishes. Here we used the definitions 
\[
C_{\hat{t}}^{t_{1}}(\hat{x}_{0})\xi_{2}^{fw}(\hat{x}_{0})=\lambda_{2}^{fw}(\hat{x}_{0})\xi_{2}^{fw}(\hat{x}_{0}),\qquad C_{\hat{t}}^{t_{0}}(\hat{x}_{0})\xi_{2}^{bw}(\hat{x}_{0})=\lambda_{2}^{bw}(\hat{x}_{0})\xi_{2}^{bw}(\hat{x}_{0}),\qquad\left|\xi_{2}^{fw}\right|=\left|\xi_{2}^{bw}\right|=1.
\]

Ma and Bollt \cite{Ma14,Ma15} argue that level curves of eq. \eqref{eq:ch6:shape_coherent}
with $\left|\theta(\hat{x}_{0})\right|\ll1$ should show significant
shape coherence over a finite time interval. They support this expectation
with examples of steady, linear velocity fields.

For unsteady flows with general time dependence, the smallness of
$\left|\theta(\hat{x}_{0})\right|\ll1$ along closed structure boundaries
remains a heuristic assertion that we will test here on temporally
aperiodic examples. Locating closed level curves of $\theta(\hat{x}_{0})$ reliably is a highly challenging numerical problem to which Refs.~\onlinecite{Ma14,Ma15} offer no immediate solution. For a direct comparison with other methods, we
will simply identify the set $\left|\theta(\hat{x}_{0})\right|\ll1$ for
initial conditions $\hat{x}_{0}$ seeded at time $\hat{t}$, then
advect these initial conditions under the flow map $F_{\hat{t}}^{t_{0}}$
to time $t_{0}$. The resulting open set must then necessarily contain the structure boundary curves envisioned by Refs.~\onlinecite{Ma14,Ma15}. The splitting angle diagnostic \eqref{eq:ch6:shape_coherent}
is objective, given that it only depends on the angle between appropriate
Cauchy\textendash Green eigenvectors.

\section{Mathematical approaches to Lagrangian coherence}\label{sec:Mathematical-approaches-to}

Here we recall approaches that locate coherent structures by providing
precise solutions to mathematically formulated coherence principles.
These approaches, however, are only precise relative to their starting
coherence principle. One still needs to test whether those coherence
principles capture observed coherent trajectory patterns consistently
and effectively in various finite-time data sets. Indeed, a heuristic
but well-motivated diagnostic tool may consistently outperform a rigorous
mathematical approach that is based on an ineffective coherence principle.

As in the case of diagnostics, we consider frame-indifference (or
objectivity) to be a fundamental requirement for the self-consistency
of mathematical approaches to Lagrangian coherence. All mathematical
approaches considered below satisfy this requirement.

\subsection{Transfer operator method}

Transfer operator approaches provide a global view of density evolution
in the phase space, identifying maximally coherent or minimally dispersive
regions over a finite time interval $[t_{0},t_{1}]$. Such regions
are known as \textit{almost-invariant sets} for autonomous systems
\cite{Dellnitz99,Froyland05,Froyland09} or \textit{coherent sets}
for non-autonomous systems \cite{FLS10,Froyland10,Froyland13} and
minimally mix with the surrounding phase space.

\subsubsection{Probabilistic transfer operator method}

Following the approach from Ref.~\onlinecite{Froyland13}, we let $M\subset\mathbb{R}^{d}$
be a compact domain and let $\mu$ denote a reference probability
measure on $M$ representing the distribution or concentration of
a quantity of interest. In many cases, one would select $\mu$ to
be the normalized volume on $M$; this would treat all parts of the phase
space equally. In other cases, one might select $\mu$ to be the distribution
of a chemical in a fluid or the distribution of a (compressible) air
mass in the atmosphere. 

We now imagine advection-diffusion dynamics; this could arise from
purely advective dynamics with some additional small amplitude $\epsilon$-diffusion,
as in the examples considered in this comparative study, or this could
be genuine advection-diffusion dynamics. Specializing to the former
case, we have a flow map $F_{t_{0}}^{t_{1}}:M\to F_{t_{0}}^{t_{1}}(M)$
from some initial time $t_{0}$ to some final time $t_{1}$. Roughly
speaking, we wish to identify subsets $A_{t_{0}}\subset M$, $A_{t_{1}}\subset F_{t_{0}}^{t_{1}}(M)$
that maximize the quantity 
\[
\mu(A_{t_{0}}\cap(F_{t_{0}}^{t_{1}})^{-1}A_{t_{1}})/\mu(A_{t_{0}}),
\]
subject $\mu(A_{t_{0}})\le1/2$ ($A_{t_{0}}$ comprises not more than
half of $M$). The numerator represents the $\mu$-proportion of $A_{t_{0}}$
that is mapped into $A_{t_{1}}$, and the entire expression is therefore
the fraction of $\mu$-mass that is mapped from $A_{t_{0}}$ to $A_{t_{1}}$. 

The determination of these sets is achieved by computing the second
singular value of a normalized transfer operator $\cal{L}_\epsilon$ and
extracting the sets $A_{t_{0}}$ and $A_{t_{1}}$ from level sets
of the corresponding left and right singular vectors, respectively;
see Ref.~\onlinecite{Froyland13} or the survey Ref.~\onlinecite{Froyland14} for details. 

One can characterise the amount of mixing that has occurred during the interval $[t_0,t_1]$ as
\begin{equation} \label{eq:var}
\rho:=\max_{A_{t_0},A_{t_1}\subset M}\left\{\frac{\langle \mathcal{L}_\epsilon \mathbf{1}_{A_{t_0}},\mathbf{1}_{A_{t_1}}\rangle}{\mu(A_{t_0})}+\frac{\langle \mathcal{L}_\epsilon \mathbf{1}_{A_{t_0}^c},\mathbf{1}_{A_{t_1^c}}\rangle}{\mu(A_{t_0^c})}\right\}.
\end{equation}
The quantity $\rho$ probabilistically quantifies the degree to which one can find agreement between pairs of sets $F^{t_1}_{t_0}A_{t_0}$ and $A_{t_1}$ (and between their complements).
Larger $\rho$ means sets can be found with greater agreement and that less mixing has occurred.
One has the theoretical upper bound $\rho\le 1+\sigma_2$, where $\sigma_2$ is the second singular value of $\mathcal{L}_\epsilon$ (Theorem 2 [48]).
One can represent \eqref{eq:var} as an $L^2$ maximisation problem, the solutions of which are left and right singular vectors of $\mathcal{L}_\epsilon$;  see [48].
The objective of this maximisation problem is an $L^2$ relaxation of \eqref{eq:var} and using a standard approach, one recovers feasible solutions of \eqref{eq:var} as optimal level sets (optimal according to the objective \eqref{eq:var}) of the solutions of the relaxation; in this case, level sets of the left and right singular vectors. Further singular vectors can be used to find
multiple coherent sets by either (i) thresholding individual singular
vectors as in the numerics section, or (ii) clustering several vectors
embedded in Euclidean space as in Ref.~\onlinecite{Froyland05}.

In practice, a common way to numerically compute $\mathcal{L}$ is
to use Ulam's method. One (i) partitions $M$ and $F_{t_{0}}^{t_{1}}(M)$
into a fine grid of sets, (ii) samples several initial points in each
grid set, (iii) numerically integrates these initial points, and (iv)
computes grid set to grid set transition probabilities by counting
how many initial points from each grid set $A$ enter another grid
set $B$. If there are $m$ grid sets in $M$ and $n$ grid sets in
$F_{t_{0}}^{t_{1}}(M)$, one obtains a sparse $m\times n$ stochastic
transition matrix $P$, which may be identified as a Markov chain
transition matrix with each grid set considered a state. One now normalizes
this matrix $P$ to produce a matrix $L$ approximating $\mathcal{L}$
and computes singular vectors (see Refs.~\onlinecite{Froyland10,Froyland13} for
details). The small additional $\epsilon$-diffusion need not be explicitly
simulated because numerical diffusion already arises from the discretization
of $M$ and $F_{t_{0}}^{t_{1}}(M)$ into grid sets. 

Alternative, non-Ulam numerical implementations of variations of the
transfer operator method from Ref.~\onlinecite{Froyland13} include Ref.~\onlinecite{Williams15}
which uses approximate Galerkin projection onto a basis of thin-plate
splines; Ref.~\onlinecite{Denner2015} which uses spectral collocation, and Ref.~\onlinecite{BanischKoltai2016},
which uses diffusion map constructions.

\subsubsection{Dynamic Laplace operator method}

Considering the $\epsilon\to0$ (i.e., zero diffusion amplitude) limit
in the previous section leads to a geometric theory of finite-time
coherent sets, which targets the \emph{boundaries} of coherent families
of sets. For simplicity of presentation, assume that the flow map
$F_{t_{0}}^{t_{1}}:M\to F_{t_{0}}^{t_{1}}(M)$ from the previous section
is volume-preserving. The goal of the dynamic Laplacian approach \cite{F15}
is to seek surfaces $\Gamma\subset M$ that disconnect a bounded phase
space $M$ in such a way that the advected disconnecting surface $F_{t_{0}}^{t_{1}}(\Gamma)$
remains as short as possible relative to the volume of the disconnected
parts for $t\in[t_{0},t_{1}]$. Thus, the region enclosed by $\Gamma$
(or by $\Gamma$ and by the boundary of the phase space) is coherent because
filamentation of the boundary is minimized under nonlinear evolution
of the dynamics. Specifically, for a finite subset $\mathcal{T}$
of $[t_{0},t_{1}]$ containing $t_{0}$ and $t_{1}$, the quantity $(\frac{1}{|\mathcal{T}|}\sum_{t\in\mathcal{T}}\ell_{d-1}(F_{t_{0}}^{t}(\Gamma)))/\min\lbrace\ell(M_{1}),\ell(M_{2})\rbrace$
is minimized over smooth disconnecting $\Gamma$, where $\ell$ is
the volume measure on the phase space, $\ell_{d-1}$ is the induced
volume measure for hypersurfaces, and $M_{1},M_{2}$ partition phase
space with shared smooth boundary $\Gamma$. 

To solve this problem, one considers the \emph{dynamic Laplace operator}
\[
\triangle^{D}:=\frac{1}{|\mathcal{T}|}\sum_{t\in\mathcal{T}}F_{t_{0}}^{t}\circ\triangle\circ(F_{t_{0}}^{t})^{-1}
\]
on $M$. The standard Laplace-Beltrami operator $\triangle$ is extensively
used in manifold learning or nonlinear dimensionality reduction via
Laplace eigenmaps and spectral clustering \cite{Belkin2003}.
The second and lower eigenvectors of $\triangle^{D}$ reveal further
geometric information in analogy to the eigenvectors of the standard
(static) Laplace operator \cite{Belkin2003} and multiple coherent
sets can be extracted using the methods described in the previous
section for transfer operators. In practice, one approximates the
above operator with a numerical method appropriate for elliptic self-adjoint
operators (e.g.\ finite difference \cite{F15}, radial basis function
collocation \cite{Froyland15_2}, or others). 

Because this method arises as a zero-diffusion limit \cite{F15} of
the probabilistic transfer operator method discussed in the previous
section, the numerical results obtained from the dynamic Laplace operator
approach are very similar and will not be discussed separately in
our comparison. Both the probabilistic transfer operator and dynamic
Laplace operator methods are objective by construction. An advantage
of the dynamic Laplace operator approach is the flexibility in the
method of approximation of the operators. Higher-order schemes may
be employed when the dynamics is smooth in order to exploit the smoothness
and reduce the input and computational requirements \cite{Froyland15_2}.
The theory and constructions for general non-volume-preserving $F_{t_{0}}^{t_{1}}$
and general reference measure $\mu$ are developed in Ref.~\onlinecite{FK16}.
 Ref.~\onlinecite{KK16} describes a related theory based on a single Riemannian
metric.

\subsection{Hierarchical coherent pairs}

The transfer operator method described in Ref.~\onlinecite{Froyland10} focused primarily on identifying two sets, $A_{0}$ and its complement $\widetilde{A}_{0}$,
that partition a given region of interest into two coherent sets.
Ma and Bollt \cite{Ma13} propose an extension of this idea that enables
the identification of multiple coherent pairs in a given domain. The
extension is based on an iterative and hierarchical refinement of
coherent pairs using a reference measure of probability $\mu$. Specifically,
Ma and Bollt \cite{Ma13} refine the coherent pairs $A_{0}$ and $\widetilde{A}_{0}$
identified earlier over several steps by applying the probabilistic
transfer operator method restricted to these sets. This iterative
refinement of coherent pairs can be stopped once $\mu$ shows no appreciable
improvement compared to the earlier iterations. We refer to this method
as \emph{hierarchical transfer operator method} throughout our comparison.
This ``repeated bisection'' approach is an alternative to extracting
multiple coherent sets using multiple singular vectors of $\mathcal{L}$
as described in the previous section. 

\subsection{Fuzzy cluster analysis of trajectories}

Recently, Froyland and Padberg-Gehle \cite{Froyland15} proposed a
method based on traditional fuzzy C-means clustering \cite{Bezdek81,Dunn73}
to identify finite-time coherent regions from incomplete and sparse
trajectory data set. Their method locates coherent sets as clusters
of trajectories according to the dynamic distance $D(x,y)=\int_{t_{0}}^{t_{1}}\|x(t)-y(t)\|^{2}\ dt$,
where $x(t),y(t)$ are a pair of trajectories over a finite time interval
$[t_{0},t_{1}]$.

To identify such coherent sets, Ref.~\onlinecite{Froyland15} first constructs
a trajectory array $X\in\mathbb{R}^{n\times dm}$ whose rows are vectors
$(X_{i})_{i=1,\ldots,n}$ containing concatenated positions of $n$
Lagrangian particles over $m$ discrete time intervals in $d$-dimensional
space; that is, $x_{i}=(x_{i,t_{0}},\ldots,x_{i,t_{1}})$. Second, Ref.~\onlinecite{Froyland15} applies the fuzzy C-means (FCM) algorithm to the
trajectory array $X$, which seeks to split the trajectories into
$K$ clusters based on the distance between a given trajectory point
$X_{i}$ and initial cluster centers $(C_{j})_{j=1,\ldots,K}$ predefined
in $\mathbb{R}^{dm}$, using the following objective function: 
\begin{equation}
\min\sum_{i=1}^{n}\sum_{j=1}^{c}u_{ij}^{m}\left\Vert X_{i}-C_{j}\right\Vert ^{2}=\min u_{ij}^{m}\sum_{t=t_{0}}^{t_{1}}\|x_{i,t}-c_{j,t}\|^{2},\label{eq:fcm_objective}
\end{equation}
where $u_{i,j}$ is the membership value defined as 
\begin{equation}
u_{i,j}=\left[\sum\limits _{j=1}^{K}\left(\frac{\Vert X_{i}-C_{k}\Vert}{\Vert X_{i}-C_{j}\Vert}\right)^{\frac{2}{m-1}}\right]^{-1},\qquad0\leq u_{i,j}\leq1,\qquad1\leq m<\infty.\label{eq:fcm_membership}
\end{equation}
The membership value $u_{i,j}$ describes the likelihood that a trajectory
point $X_{i}$ belongs to a cluster associated with the cluster center
$C_{j}$, for a fixed parameter $m$ specified in advance.

The parameter $m$ determines the fuzziness of cluster boundaries,
that is how much clusters are allowed to overlap. A large $m$ results
in less extreme membership values $u_{k,j}$, and consequently fuzzier
clusters. In the limit $m=1$, the memberships converge to $0$ or
$1$, and hence the FCM results in non-overlapping clusters in a fashion
similar to the K-means algorithm \cite{Lloyd82}. The cluster center
is the mean of all trajectory points, weighted by the degree of belonging
to each of the $K$ clusters. Specifically, the $j^{th}$ cluster
center is defined as

\begin{equation}
C_{j}=\frac{\sum\limits _{i=1}^{n}\left(u_{i,j}\right)^{m}X_{i}}{\sum\limits _{i=1}^{n}\left(u_{i,j}\right)^{m}}.\label{eq:fcm_centers}
\end{equation}

To optimize \eqref{eq:fcm_objective}, the FCM algorithm iteratively
computes membership values \eqref{eq:fcm_membership} and relocates
the cluster centers using \eqref{eq:fcm_centers}, until the objective
function \eqref{eq:fcm_objective} shows no substantial improvement.
Finally, given the membership values $u_{i,j}$ and cluster centers
$C_{j}$, each trajectory is assigned to only one cluster based on
the maximum membership value it carries.

Those trajectories carrying low membership values for all clusters,
with respect to a given threshold (selected as $0.9$ in all our examples
below), are occasionally considered to be non-coherent \cite{Froyland15}. The incomplete data case (e.g., some or all trajectories
have missing ``gaps'') is also described in Ref.~\onlinecite{Froyland15}. We finally
note that the fuzzy cluster analysis of trajectories is an objective
approach, as the label of trajectories remains invariant under any
affine coordinate transformation \cite{Froyland15}.

\subsection{Spectral clustering of trajectories}

Hadjighasem et al. \cite{Hadjighasem16b} propose spectral clustering
to identify coherent structures by grouping Lagrangian particles into
coherent and incoherent clusters. Specifically, they define a \emph{coherent
structure} as a distinguished set of Lagrangian trajectories that
maintain short distances among themselves relative to their distances
to trajectories outside the structure.

The spectral clustering approach starts with $n$ trajectories whose
positions are available at $m$ discrete times $t_{0}<t_{1}<\ldots<t_{k}<\ldots<t_{m-1}=t_{f}$
in a two-dimensional spatial domain. This information is stored in
an $n\times m\times2$-dimensional numerical array with elements $\mathbf{x}_{k}^{i}\coloneqq\mathbf{x}^{i}(t_{k})\in\mathbb{R}^{2}$.
The \emph{dynamical distance} $r_{ij}$ between Lagrangian particles
$\mathbf{x}^{i}$ and $\mathbf{x}^{j}$ is then defined as 
\begin{equation}
\begin{split}r_{ij} & \coloneqq\frac{1}{t_{f}-t_{0}}\int_{t_{0}}^{t_{f}}\left|\mathbf{x}^{i}(t)-\mathbf{x}^{j}(t)\right|\mathrm{d}t\\
 & \approx\frac{1}{t_{f}-t_{0}}\sum_{k=0}^{m-2}\frac{t_{k+1}-t_{k}}{2}\left(\left|\mathbf{x}_{k+1}^{i}-\mathbf{x}_{k+1}^{j}\right|+\left|\mathbf{x}_{k}^{i}-\mathbf{x}_{k}^{j}\right|\right),
\end{split}
\label{eq:ch6:clustering_metric}
\end{equation}
where $\left|\cdot\right|$ denotes the spatial Euclidean norm. Note
that the dynamic distance \eqref{eq:ch6:clustering_metric} is an
objective metric, as it only depends on the distance of trajectory points.

Next, Ref.~\onlinecite{Hadjighasem16b} constructs \emph{a similarity
graph} $G=(V,E,W)$, which is specified by the set of its nodes $V={v_{1},\ldots,v_{N}}$,
the set of edges $E\subseteq V\times V$ between nodes, and a symmetric
\emph{similarity matrix} $W\in\mathbb{R}^{n\times n}$ which assigns
weights $w_{ij}$ to the edges $e_{ij}$. The similarity matrix entries
(or \emph{weights)} $w_{ij}\geq0$ give the probability of nodes $v_{i}$
and $v_{j}$ to be in the same cluster. In the context of coherent
structure detection, the graph nodes $V$ are Lagrangian particles
themselves, with the associated similarity weights defined as 
\begin{equation}
w_{ij}=1/r_{ij}\quad\text{for}\;i\neq j.\label{eq:ch6:similarity_conversion}
\end{equation}

With the similarity weights at hand, the \emph{degree} of a node $v_{i}\in V$
is defined as 
\[
\text{deg}(v_{i})\coloneqq\sum_{j=1}^{n}w_{ij}.
\]
The subsequent \emph{degree matrix} $D$ is then constructed as a
diagonal matrix with the degrees $\deg(v_{i})$ in the diagonal. Given
a subset of nodes $A\in V$, the size of $A$ is measured by 
\[
\vol(A)\coloneqq\sum_{i\in A}\deg(v_{i}),
\]
with summation over the weights of all edges attached to nodes in
$A$.

With the notation developed so far, the problem of coherent structure
detection can now be posed in terms of a \emph{normalized graph cut}
problem: Given a similarity graph $G=(V,E,W)$, partition the graph
nodes $V$ into $k$ sets $A_{1},A_{2},\ldots,A_{k}$ such that the
following conditions hold: 
\begin{description}
\item [{Within-cluster similarity}] Nodes in the same cluster are similar
to each other, i.e., particles in a coherent structure have mutually short dynamical distances. 
\item [{Between-cluster dissimilarity}] Nodes in a cluster are dissimilar
to those located in the complementary cluster. In other words, particles
in a coherent structure have long dynamical distances from the rest
of the particles, particularity from those located in the mixing region
(i.e., \emph{noise cluster}) that fills the space outside the coherent
structures. 
\end{description}
The normalized cut that implements the above (dis)similarity conditions
can be formulated mathematically as 
\begin{equation}
\NCut(A_{i},...,A_{k})=\frac{1}{2}\sum_{i=1}^{k}\frac{\cut(A_{i},\overline{A}_{i})}{\vol(A_{i})},\qquad\cut(A_{1},...,A_{k})=\frac{1}{2}\sum_{i=1}^{k}W(A_{i},\overline{A}_{i}),\label{eq:ch6:Ncut}
\end{equation}
where $\overline{A}$ denotes the complement of set $A$ in $V$.
The minimization of the normalized cut exactly is an \emph{NP-complete
problem}. The solution of Ncut problem, however, can be approximated
by solutions of a generalized eigenproblem associated with the graph
Laplacian $L=D-W$, defined as \cite{Shi00}
\begin{equation}
Lu=\lambda Du.\label{eq:ch6:gen_eigenproblem}
\end{equation}

In particular, the first $k$ eigenvectors $u_{1},\ldots,u_{k}$, whose
corresponding eigenvalues are close to zero, minimize approximately
the Ncut objective \eqref{eq:ch6:Ncut}. The value of $k$, in this
case, is equal to the number of eigenvalues preceding the largest
gap in the eigenvalue sequence \cite{Bhatia97}. The first $k$ generalized
eigenvectors $u$ then offer an alternative representation of the
weighted graph data such that each leading eigenvector highlights
a single coherent structure in the computational domain. Finally,
these $k$ coherent structures beside the complementary incoherent
region can be extracted from the eigenvectors $u_{1},\ldots,u_{k}$
using a simple K-means algorithm \cite{Lloyd82} or more sophisticated
approaches, such as PNCZ \cite{Yu03}.

A related variational level-set formulation of the spectral clustering
approach is now available for two-dimensional flows \cite{Hadjighasem16c}.

\subsection{Stretching-based coherence: Geodesic theory of LCS}\label{section:geodesic}

The geodesic theory of LCSs is a collection of global variational
principles for material surfaces that form the centerpieces of coherent,
time-evolving tracer patterns \cite{Haller15}. Out of these material
surfaces, hyperbolic LCSs act as generalized stable and unstable manifolds,
repelling or attracting neighboring material elements with locally
the highest rate over a finite-time interval. Parabolic LCSs minimize
Lagrangian shear and hence serve as generalized jet cores. Finally,
elliptic LCSs extend the notion of Kolmogorov\textendash Arnold\textendash Moser
(KAM) tori and serve as generalized coherent vortex boundaries in
finite-time unsteady flows. Geodesic LCS theory is objective, as it
builds on material notions of strain and shear that are expressible
through the invariants of the right Cauchy\textendash Green strain
tensor.

Below we summarize the main results for two-dimensional flows from
Farazmand et al. \cite{Farazmand14} for hyperbolic and parabolic
LCSs, and from Haller and Beron\textendash Vera \cite{Haller13} for
elliptic LCSs. A general review with further mathematical LCS results,
as well as extensions to three-dimensional flows, can be found in
Ref.~\onlinecite{Haller15}.

\subsubsection{Stationary curves of the average shear: Hyperbolic and parabolic
LCSs}

A shearless LCS is a material curve whose average Lagrangian shear
shows no leading-order variation when compared to nearby $C^{1}$-close
material lines. Specifically, the time $t_{0}$ position of a shearless
LCS is a stationary curve for the material-line-averaged tangential
shear functional. Farazmand et al. \cite{Farazmand14} show that such
LCSs coincide with null-geodesics of the metric tensor

\begin{equation}
D_{t_{0}}^{t_{1}}(x_{0})=\frac{1}{2}\left[C_{t_{0}}^{t_{1}}(x_{0})\Omega-\Omega C_{t_{0}}^{t_{1}}(x_{0})\right],
\end{equation}
with the rotation matrix $\Omega$ given in \eqref{eq:ch6:eig_def}.
The tensor $D_{t_{0}}^{t_{1}}(x_{0})$ is Lorentzian (i.e., indefinite)
wherever $\lambda_{1}(x_{0})\neq\lambda_{2}(x_{0}).$ All null-geodesics
of $D_{t_{0}}^{t_{1}}(x_{0})$ are found to be trajectories of one
of the two line fields 
\begin{equation}
x_{0}^{\prime}=\xi_{j}(x_{0}),\qquad j=1,2.\label{eq:ch6:shrinkshear}
\end{equation}

We refer to trajectories of \eqref{eq:ch6:shrinkshear} with $j=1$
as \emph{shrink lines}, as they strictly shrink in arc-length under
the action of the flow map $F_{t_{0}}^{t_{1}}$. Similarly, we call
trajectories of \eqref{eq:ch6:shrinkshear} with $j=2$ \emph{stretch
lines}, as they strictly stretch under $F_{t_{0}}^{t_{1}}$. For lack
of a well-defined orientation for eigenvectors, equation \eqref{eq:ch6:shrinkshear}
defines a line field~\cite{Farazmand12_2}, not an ordinary differential
equation. Nevertheless, the trajectories of \eqref{eq:ch6:shrinkshear}
(i.e., curves tangent to the eigenvector field $\xi_{j}$) are well-defined
at all points where $\lambda_{1}(x_{0})\neq\lambda_{2}(x_{0})$.

\emph{Repelling LCS}s are defined as special shrink lines that start
from local maxima of $\lambda_{2}(x_{0})$; \emph{attracting LCSs},
by contrast, are special stretch lines that start from local minima
of $\lambda_{1}(x_{0})$. As a consequence of their definitions, repelling
and attracting LCSs (or \emph{hyperbolic LCS}s, for short) have a
role similar to that of stable and unstable manifolds of strong saddle
points in a classical dynamical system. Between any two of their points,
hyperbolic LCSs are solutions of the stationary shear variational
problem under fixed endpoint boundary conditions.

\emph{Parabolic LCS}s, in contrast, are composed of structurally stable
chains of alternating shrink\textendash stretch line segments that
connect tensorline singularities (i.e., points where $\lambda_{1}(x_{0})=\lambda_{2}(x_{0})$).
Out of all such possible chains, one builds parabolic LCSs (generalized
jet cores) by identifying tensorlines that are closest to being neutrally
stable (cf. Ref.~\onlinecite{Farazmand14} for further details).
Parabolic LCSs are more robust under perturbations than hyperbolic
LCSs, because they are solutions of the original stationary shear
variational principle under variable endpoint boundary conditions.

\subsubsection{Stationary curves of the average strain: Elliptic LCSs}

An \emph{elliptic LCS} is a closed material line across which the
material-line-averaged Lagrangian stretching shows no leading-order
variation when compared to closed, $C^{1}$-close material lines.
Specifically, the time $t_{0}$ position of an elliptic LCS is a stationary
curve for the material-line-averaged tangential strain functional.
As shown by Haller and Beron\textendash Vera \cite{Haller13}, such
stationary curves coincide with closed null-geodesics of the one-parameter
family of Lorentzian metric tensors 
\[
E_{\lambda}(x_{0})=\frac{1}{2}\left[C_{t_{0}}^{t}(x_{0})-\lambda I\right],
\]
where the real number $\lambda>0$ parametrizes the family. These
closed null-geodesics turn out to be closed trajectories (limit cycles)
of the two, one-parameter families of line fields

\begin{equation}
x_{0}^{\prime}=\eta_{\lambda}^{\pm}(x_{0})=\sqrt{\frac{\lambda_{2}(x_{0})-\lambda^{2}}{\lambda_{2}(x_{0})-\lambda_{1}(x_{0})}}\xi_{1}(x_{0})\pm\sqrt{\frac{\lambda^{2}-\lambda_{1}(x_{0})}{\lambda_{2}(x_{0})-\lambda_{1}(x_{0})}}\xi_{2}(x_{0}).\label{eta_vector}
\end{equation}

A simple calculation shows that all limit cycles of \eqref{eta_vector}
are infinitesimally uniformly stretching. Specifically, any subset
of such a limit cycle is stretched exactly by a factor of $\lambda$
over the time interval $[t_{0},t_{1}]$ under the flow map $F_{t_{0}}^{t_{1}}$.
As a result, elliptic LCSs exhibit no filamentation when advected
under the flow map $F_{t_{0}}^{t_{1}}$. Elliptic LCSs occur in nested
families due to their structural stability with respect to changes
in $\lambda$. The outermost member of such a nested limit cycle family
serves as a \emph{Lagrangian vortex boundary}.

For computing geodesic LCSs in the forthcoming examples, we use
the automated algorithm developed in Haller and Beron-Vera \cite{Haller13}
and Karrasch et al. \cite{Karrasch15}. A MATLAB implementation of
this method is provided in \url{https://github.com/LCSETH}. A simplified algorithm for computing geodesic LCSs without the use of the direction field is now available \cite{Serra16}, but will not be used in this paper. There
is no general extension of geodesic LCS theory to three dimensional
flows, but related local variational principles for hyperbolic and
elliptic LCSs are now available in three dimensions as well \cite{Blazevski14,Oettinger16b}.

\subsection{Rotational coherence from the Lagrangian-Averaged Vorticity Deviation
(LAVD)}

Farazmand \& Haller \cite{Farazmand15} introduce the notion of \emph{rotationally
coherent LCSs} as tubular material surfaces whose elements exhibit
identical mean material rotation over a finite time interval $[t_{0},t_{1}]$.
They use the classic polar decomposition to compute the polar rotation
angle (PRA) from the flow gradient $\nabla F_{t_{0}}^{t_{1}}$ for
this purpose. Outermost closed and convex level curves of the PRA
then define initial positions of rotationally coherent vortex boundaries.
The rotational LCSs obtained in this fashion are objective in two-dimensional
flows.

Polar rotations, however, are not additive: the total PRA computed
over a time interval $[t_{0},t_{1}]$ does not equal the sum of PRAs
computed over smaller sub-intervals \cite{Haller16}. As a consequence,
PRA does not match the experimentally observed mean material rotation
of finite-tracers in a fluid flow.

To resolve this dynamical inconsistency of the PRA, Haller \cite{Haller16b}
has recently developed a dynamic polar decomposition (DPD) as an alternative
to the classic polar decomposition. The DPD of the deformation gradient
is a unique factorization of the form 
\begin{equation}
\nabla F_{t_{0}}^{t_{1}}=O_{t_{0}}^{t_{1}}M_{t_{0}}^{t_{1}}=N_{t_{0}}^{t_{1}}O_{t_{0}}^{t_{1}},
\end{equation}
where $O_{t_{0}}^{t}$ is the \emph{dynamic rotation tensor} and $M_{t_{0}}^{t}$
and $N_{t_{0}}^{t}$ are the \emph{left dynamic stretch tensor} and
\emph{right dynamic stretch tensor}, respectively. Compared to the
classic polar decomposition, where the rotational and stretching components
are obtained from matrix manipulations, the dynamic rotation and stretch
tensors are obtained as solutions of linear differential equations.
Specifically, the dynamic rotation tensor $O_{t_{0}}^{t}=\nabla_{a_{0}}a(t)$
is the deformation gradient of a purely rotational flow $a(t)$ satisfying
\begin{equation}
\dot{a}=W\left(x(t;x_{0}),t\right)a,
\end{equation}
where the \emph{spin tensor} $W(x,t)$ is defined as $W(x,t)=\frac{1}{2}\left(\nabla v(x,t)-(\nabla v(x,t))^{T}\right)$.
The dynamic rotation tensor $O_{t_{0}}^{t}$ can further be factorized
into two deformation gradients: 
\begin{equation}
O_{t_{0}}^{t}=\Phi_{t_{0}}^{t}\Theta_{t_{0}}^{t}.
\end{equation}
Here the \emph{mean rotation tensor} $\Theta_{t_{0}}^{t}$ describes a
uniform rigid-body-type rotation, and the \emph{relative rotation
tensor} $\Phi_{t_{0}}^{t}$ represents the deviation from this uniform
rotation. The relative rotation tensor $\Phi_{t_{0}}^{t}=\nabla_{\alpha_{0}}\alpha(t)$
turns out to be the deformation gradient of the relative rotation
flow $\alpha(t)$ satisfying 
\begin{equation}
\dot{\alpha}=\left[W\left(x(t;x_{0}),t\right)-\bar{W}\left(t\right)\right]\alpha,
\end{equation}
where $\bar{W}(t)$ is the spatial average of the spin tensor. On
the other hand, the mean rotation tensor $\Theta_{t_{0}}^{t}=\nabla_{\beta_{0}}\beta(t)$
is the deformation gradient of the mean-rotation flow 
\begin{equation}
\dot{\beta}=\Phi_{t}^{t_{0}}\bar{W}\left(t\right)\Phi_{t_{0}}^{t}\beta.
\end{equation}
As the fundamental matrix solution of a classic linear system of ODEs,
the mean rotation tensor $\Theta_{t_{0}}^{t}$ is dynamically consistent,
implying that the \emph{intrinsic angle} $\psi_{t_{0}}^{t}(x_{0})$,
swept by $\Phi_{t_{0}}^{t}$ about its time-varying axis of rotation
over the time interval $[t_{0},t_{1}]$, is always the sum of $\psi_{t_{0}}^{t}(x_{0})$
and $\psi_{t}^{t_{1}}\left(F_{t_{0}}^{t}\left(x_{0}\right)\right)$
for any choice of $t\in[t_{0},t_{1}]$. The intrinsic rotation angle
$\psi_{t_{0}}^{t}(x_{0})$ is, therefore, a dynamically consistent
and objective extension of the PRA in both two- and three-dimensional
flows (see Ref.~\onlinecite{Haller16b} for more detail).

Using these results, Haller et al. \cite{Haller16} use the \emph{Lagrangian-Averaged
Vorticity Deviation} (LAVD), i.e., twice the value of the intrinsic
rotation angle $\psi_{t_{0}}^{t_{1}}(x_{0})$, to identify rotationally
coherent LCSs. The LAVD is defined as the trajectory-averaged, normed
deviation of the vorticity from its spatial mean, i.e., as 
\begin{equation}
\mathrm{LAVD}_{t_{0}}^{t_{1}}(x_{0})=\int_{t_{0}}^{t_{1}}\left|\omega(x(s;x_{0}),s)-\bar{\omega}(s)\right|\,\,ds,\label{eq:ch6:LAVD}
\end{equation}
where $\bar{\omega}$ is the spatial mean of the vorticity $\omega$.
As in the case of the PRA, initial positions of rotational LCSs are
defined as tubular level surfaces of the LAVD field along a singular
maximal level surface. By a tubular level surface, we mean here a
toroidal surface whose size exceeds a minimal length scale threshold
$l_{min}$ and whose convexity deficiency (i.e., whose distance from
its convex hull) stays below a maximal value $d_{min}$. LAVD-based
coherent Lagrangian vortex boundaries are then defined as outermost
members of nested families of tubular LAVD level surfaces. These boundaries
are objective by the objectivity of the LAVD field (cf. Ref.~\onlinecite{Haller16}).

By construction, the LAVD-based coherent vortex boundaries may display
tangential filamentation, but any developing filament necessarily
rotates at the same average rate with the vortex body, without a global
transverse breakaway \cite{Haller16}. As a notable implication for
experimental observations, centers of LAVD-based vortices (defined
by local maxima of the LAVD field) are proven to be the observed centers
of attraction or repulsion for inertial particles in the limit of
vanishing Rossby numbers (cf. Ref.~\onlinecite{Haller16}). To compute
the LAVD vortices, we use here a MATLAB implementation of the LAVD
method provided in \url{https://github.com/LCSETH}.

\section{Method comparisons on three examples}\label{sec:compare}
We now compare the performance of diagnostics and mathematical methods reviewed in Sections \Cref{sec:Lagrangian-coherence-diagnostics,sec:Mathematical-approaches-to}
on three specific examples. Our first example, the Bickley jet, is
an analytically defined velocity field with quasiperiodic time dependence
\cite{BeronVera10}. With its infinite time interval of definition
and recurrent time dependence, this example falls in the realm of
a classical dynamical systems problem with uniquely defined, infinite-time
invariant manifolds. The parameter setting we choose, however, is
not near-integrable, and hence the survival of the stable and unstable
manifolds and KAM tori of the unperturbed steady limit is a priori
unknown. In addition, the time dependence is recurrent but not periodic,
and hence the classic Poincaré map approach is not applicable to visualize
coherence in the flow.

Our second example is a finite-time velocity sample obtained from
a direct numerical simulation of two-dimensional turbulence \cite{Farazmand13}.
This flow captures most major aspects of a real-life coherence identification
problem: the velocity field is a data set; several coherent regions
exist, move around and even merge; and the time dependence of the
vector field is aperiodic and non-recurrent.

Our third example is a velocity field reconstructed from an enhanced
video footage of Jupiter, capturing Jupiter's Great Red Spot (GRS)
\cite{Hadjighasem16}. This last example has only a single vortical
structure, but the data set is short relative to rotation period of
the GRS. This shortness relative to characteristic time scales in
the data set is an additional challenge relative to our second example.

\Cref{table:ComputationalEffort} compares the computational effort
required by each method in terms of the number of particles advected.
We select the constants $n_{x}$, $n_{y}$ and $N_{s}$ in a way that
the total number of trajectories used in each method is the same for
each example. Beyond comparing the results in a single composite plot
for all methods in all three examples, we also illustrate different
aspects of select approaches on each example.

\Cref{table:userinput} compares the degree of autonomy for each
method in terms of the number of parameters it requires from the user.
Here, we only list major parameters, and ignore minor parameters such
as the integration time, grid resolution and ODE solver tolerance
conditions which are invariably required by all the methods. Moreover,
we specify some parameters as optional since they are not strictly
required for the implementation. Importantly, the number of parameters
required by each method should be viewed according to the functionality
of the method. For instance, the majority of diagnostic tools do not
offer any procedure for extracting coherent structures, while other
methods such as the geodesic, transfer operator/dynamic Laplacian,
LAVD, fuzzy clustering, and spectral clustering provide detailed coherence
structure boundaries in an automated fashion. Automated procedures
naturally require numerical control parameters, as opposed to simple
diagnostic tools, which are only evaluated visually and hence do not
deliver specific structure boundaries. 
\begin{table}[!htbp]
\centering
\setlength{\arrayrulewidth}{0.5mm}
\renewcommand{\arraystretch}{2.5}
{\rowcolors{2}{black!5!black!5}{black!15!black!15}
\begin{tabular}{|c|c|} 
\hline 
\centering
\textbf{Method}  & \textbf{\# particles}\tabularnewline
\hline 
\hline 
\pbox{20cm}{Trajectory length, Trajectory complexity, LAVD,\\ Fuzzy C-means clustering, Spectral clustering} & $n_{x}\times n_{y}$ \\
FTLE, Mesochronic, Shape coherence, Dynamic Laplacian, Geodesic  & $4\times n_{x}\times n_{y}$ \\
FSLE  & $(4+1)\times n_{x}\times n_{y}$ \\
Probabilistic transfer operator, Hierarchical coherent pairs & $N_{s}\times n_{x}\times n_{y}$ \\
\hline
\end{tabular}
}
\caption{Comparison of the minimum number of particles required by each method
to construct a Lagrangian field with the resolution $n_{x}\times n_{y}$.
The number $N_{s}$ is the number of sample points placed in each grid
box for the transfer operator method.}
\label{table:ComputationalEffort}
\end{table}

\begin{table}[!htbp]
\setlength{\arrayrulewidth}{0.5mm}
\renewcommand{\arraystretch}{2}
\centering
{\rowcolors{2}{black!5!black!5}{black!15!black!15}
\begin{tabular}{|c|c|c|}
\hline 
\textbf{Method}  & \textbf{\# parameters} & \textbf{Description}\tabularnewline
\hline 
\hline 
  FTLE & 0-1 &
  \begin{minipage}{3.8in}
    \vskip 2pt
    \begin{itemize}
    \setlength\itemsep{0.1em}
   \item (optional) auxiliary grid space to increase the accuracy of finite differencing \cite{Farazmand12_2}
   \end{itemize}
   \vskip 2pt
 \end{minipage}
 \\
   FSLE & 2 &
  \begin{minipage}{3.8in}
    \vskip 2pt
    \begin{itemize}
    \setlength\itemsep{0.1em}
   \item initial separation distance $\delta_{0}$
   \item separation factor $r$
   \end{itemize}
   \vskip 2pt
 \end{minipage}
 \\
   Mesochronic & 0-1 &
  \begin{minipage}{3.8in}
    \vskip 2pt
    \begin{itemize}
    \setlength\itemsep{0.1em}
   \item (optional) auxiliary grid space
   \end{itemize}
   \vskip 2pt
 \end{minipage}
 \\
   Trajectory length & 0-1 &
  \begin{minipage}{3.8in}
   \vskip 2pt
       \begin{itemize}
       \setlength\itemsep{0.1em}
   \item (optional) number $N_{t}$ of sampled points along each trajectory 
   \end{itemize}
	\vskip 2pt
 \end{minipage}
 \\
   Trajectory complexity & 2 &
  \begin{minipage}{3.8in}
    \vskip 2pt
    \begin{itemize}
    \setlength\itemsep{0.1em}
   \item number $N_{t}$ of sampled points along each trajectory
   \item vector specifying a range of spatial scales $s$
   \end{itemize}
   \vskip 2pt
 \end{minipage}
 \\
   Shape coherent & 0-1 &
  \begin{minipage}{3.8in}
    \vskip 2pt
    \begin{itemize}
    \setlength\itemsep{0.1em}
   \item (optional) auxiliary grid space
   \end{itemize}
   \vskip 2pt
 \end{minipage}
 \\
  \pbox{20cm}{Probabilistic transfer operator/ \\Dynamic Laplacian} & 1 &
  \begin{minipage}{3.8in}
    \vskip 2pt
    \begin{itemize}
    \setlength\itemsep{0.1em}
   \item number of sample points $N_{s}$ for initial boxes $B_{i}$
   \end{itemize}
   \vskip 2pt
 \end{minipage}
 \\
   Hierarchical coherent pairs & 2 &
  \begin{minipage}{3.8in}
    \vskip 2pt
    \begin{itemize}
    \setlength\itemsep{0.1em}
   \item number of sample points $N_{s}$ for initial boxes $B_{i}$
   \item threshold on a relative improvement of reference measure of probability $\mu$
   \end{itemize}
   \vskip 2pt
 \end{minipage}
 \\
   Fuzzy C-means clustering & 4 &
  \begin{minipage}{3.8in}
    \vskip 2pt
    \begin{itemize}
    \setlength\itemsep{0.1em}
   \item number $N_{t}$ of sampled points along each trajectory 
   \item number $K$ of clusters needs to be extracted 
   \item fuzzifier parameter $m$  
   \item minimum threshold on the maximum membership value a trajectory carrying in order to be considered coherent
   \end{itemize}
   \vskip 2pt
 \end{minipage}
 \\
   Spectral clustering & 1-2 &
  \begin{minipage}{3.8in}
    \vskip 2pt
    \begin{itemize}
    \setlength\itemsep{0.1em}
   \item (optional) number $N_{t}$ of sampled points along each trajectory 
   \item  graph sparsification radius $\epsilon$
   \end{itemize}
   \vskip 2pt
 \end{minipage}
 \\
   Geodesic & 6-7 &
  \begin{minipage}{3.8in}
    \vskip 2pt
    \begin{itemize}
	\setlength\itemsep{0.1em}
   \item (optional) auxiliary grid space
   \item minimum distance threshold between admissible singularities \cite{Karrasch15}
   \item radius of circular neighborhood around each singularity to determine its type \cite{Karrasch15}
   \item minimum distance threshold between a wedge pair \cite{Karrasch15}
   \item length for the Poincar\'{e} section
      \item number of initial conditions on each Poincar\'{e} section for which $\eta_{\lambda}^{\pm}(x_{0})$ will be computed
   \item range of stretching parameters $\lambda$ needs to be searched for identifying closed orbits
   \end{itemize}
   \vskip 2pt
 \end{minipage}
 \\
   LAVD & 2-3 &
  \begin{minipage}{3.8in}
    \vskip 2pt
    \begin{itemize}
    \setlength\itemsep{0.1em}
    \item (optional) auxiliary grid space for computing vorticity along trajectories, assuming the direct measure of vorticity is not available
   \item arclength threshold $l_{min}$ for discarding small-sized vortex boundaries
   \item convexity deficiency threshold $d_{min}$ for relaxing the strict convexity requirement
   \end{itemize}
   \vskip 2pt
 \end{minipage}
 \\
 \hline 
 \end{tabular}
}
\caption{Comparison of the minimum number of parameters required by each method
to construct a Lagrangian field with the resolution $m\times n$ over
the time interval $[t_{0},t_{1}]$. Here, we ignore trivial parameters
such as the ODE solver tolerance conditions, which are required by all
methods for advecting particles. Moreover, some parameters are specified
as optional since they are not strictly required for implementing
a method.}
\label{table:userinput} 
\end{table}

To carry out the computations, one inevitably must make a choice for the parameters listed in \Cref{table:userinput}. Given the large number of methods we consider, including the choice of the free parameters in the comparisons will be a cumbersome task. We therefore rely on our expertise and experience to choose a reasonable set of parameters for each method with the intention that (i) The choice of parameter(s) results in the most favorable outcome for the corresponding method and (ii) The outcome is robust, i.e., small variations in the parameters do not lead to drastic changes in the outcome.

Finally, a few words on how we will assess the efficacy of the methods in our comparison. If advection of various predictions in a given flow region confirms sustained material coherence for these predicted material structures, then we consider the very presence of a structure in that region as the established ground truth. (The geometric details of the predicted structure may vary from one method to the other.) 

Any method that fails to predict a structure in that same flow domain will then be deemed to yield a false negative in that domain. Likewise, if a method predicts a structure in a given region and our advection studies disprove the predicted coherence of this material domain under advection, then we consider a case of a false positive established for that method. Different methods seek to capture different aspects of coherence, but we only deem their efforts successful if they produce structures that remain arguably coherent under observations. Observed material coherence requires a lack of extensive folding and/or filamentation for the material structure.

\subsection{Quasi-periodically perturbed Bickley jet}

An idealized model for an eastward zonal jet in geophysical fluid
dynamics is the Bickley jet \cite{Delcastillo93,BeronVera10}, comprising
a steady background flow and a time-dependent perturbation. The time-dependent
Hamiltonian (stream function) for this model is given by 
\begin{equation}
\psi(x,y,t)=\psi_{0}(y)+\psi_{1}(x,y,t),
\end{equation}
where 
\begin{equation}
\psi_{0}(y)=-UL\tanh\left(\frac{y}{L}\right)
\end{equation}
is the steady background flow and 
\begin{equation}
\psi_{1}(x,y,t)=UL\mathrm{sech}^{2}\left(\frac{y}{L}\right)\mbox{Re}\left[\sum_{n=1}^{3}f_{n}(t)\exp(\mbox{i}k_{n}x)\right]\label{eq:ch6:psi_1}
\end{equation}
is the perturbation. The constants $U$ and $L$ are characteristic
velocity and length scales, with values adopted from Ref.~\onlinecite{BeronVera10}
as 
\begin{equation}
U=62.66\;\mbox{ms}^{-1},\ \ L=1770\;\mbox{km},\ \ k_{n}=2n/r_{0}.
\end{equation}
Here $r_{0}=6371\;\mbox{km}$ is the mean radius of the earth. For
$f_{n}(t)=\epsilon_{n}\exp(-\mbox{i}k_{n}c_{n}t)$, the time-dependent
part of the Hamiltonian consists of three Rossby waves with wave-numbers
$k_{n}$ travelling at speeds $c_{n}$. The amplitude of each Rossby
wave is determined by the parameters $\epsilon_{n}$. In line with
Ref.~\onlinecite{BeronVera10}, we take $f_{n}(t)=\epsilon_{n}\exp(-\mbox{i}k_{n}c_{n}t)$,
with constant amplitudes $\epsilon_{1}=0.075$, $\epsilon_{2}=0.4$, $\epsilon_{3}=0.3$ and speeds $c_{3}=0.461 U$, $c_{2}=0.205 U$, $c_{1}=c_{3}+((\sqrt(5)-1)/2)(k_{2}/k_{1})(c_{2}-c_{3})$. The time interval of interest is $t\in[0,11]$
day.

\begin{turnpage}
\begin{figure}
\label{fig:scalar_fields_bickley} \subfloat[\label{fig:BickleyFTLE}]{\includegraphics[width=0.42\textwidth]{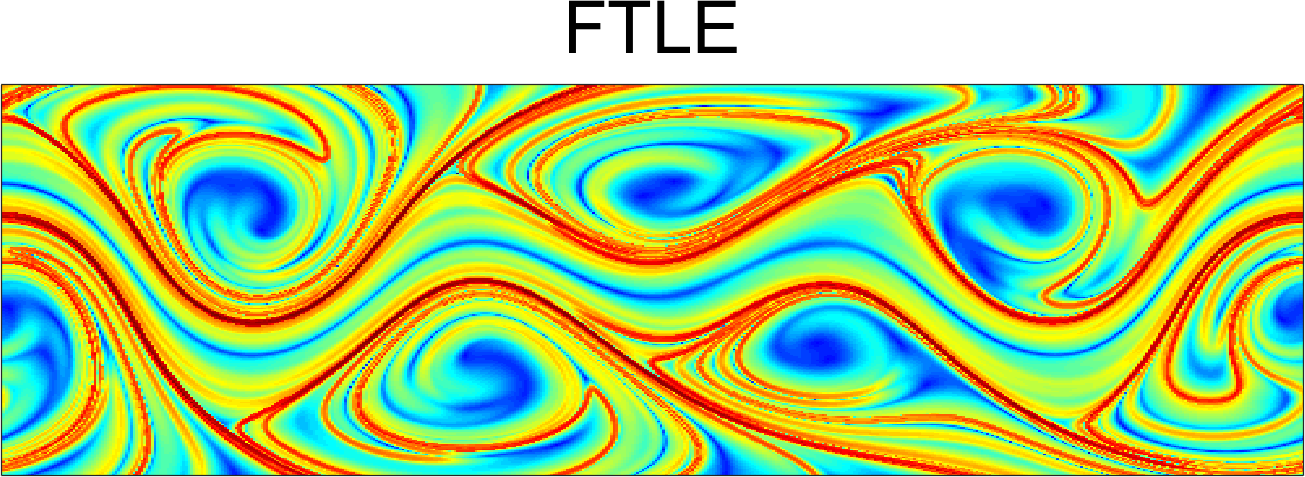}}\; \subfloat[\label{fig:BickleyFSLE}]{\includegraphics[width=0.42\textwidth]{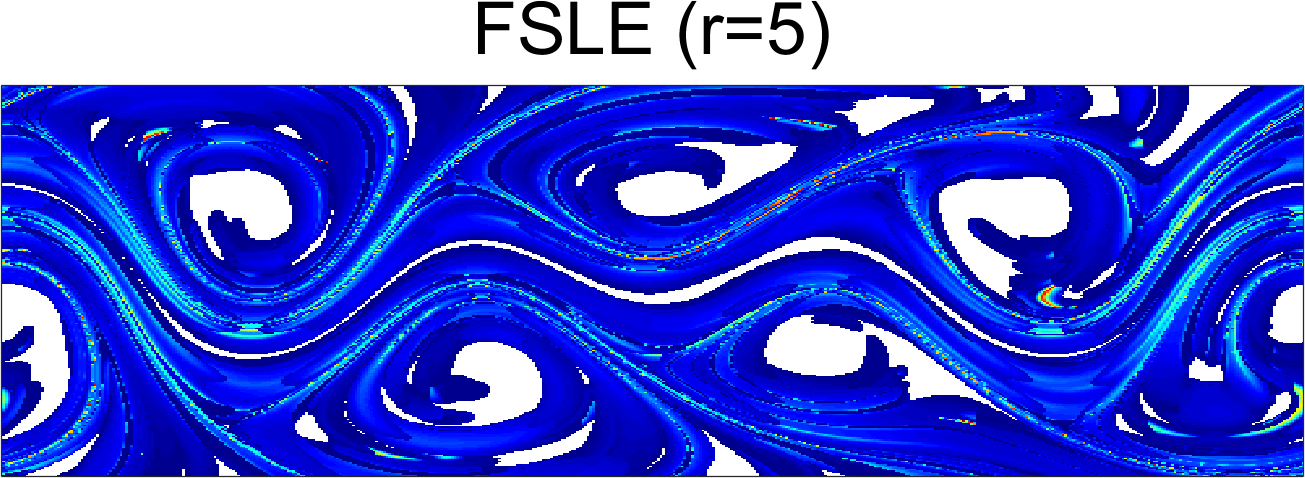}}\; 
\subfloat[\label{fig:BickleyMesochronic}]{\includegraphics[width=0.42\textwidth]{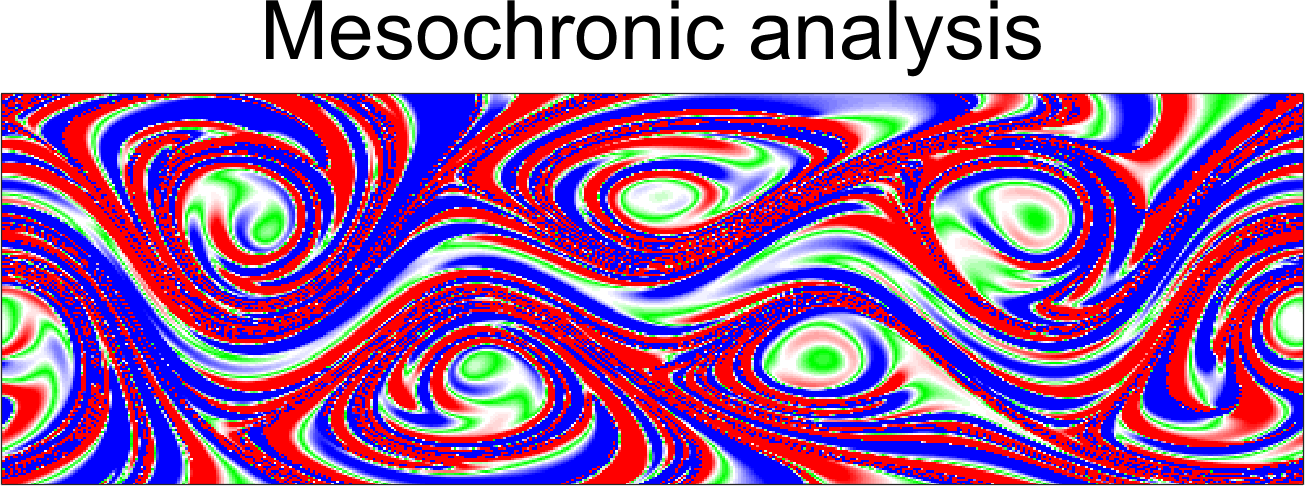}}\\ 
\subfloat[\label{fig:BickleyMfunction}]{\includegraphics[width=0.42\textwidth]{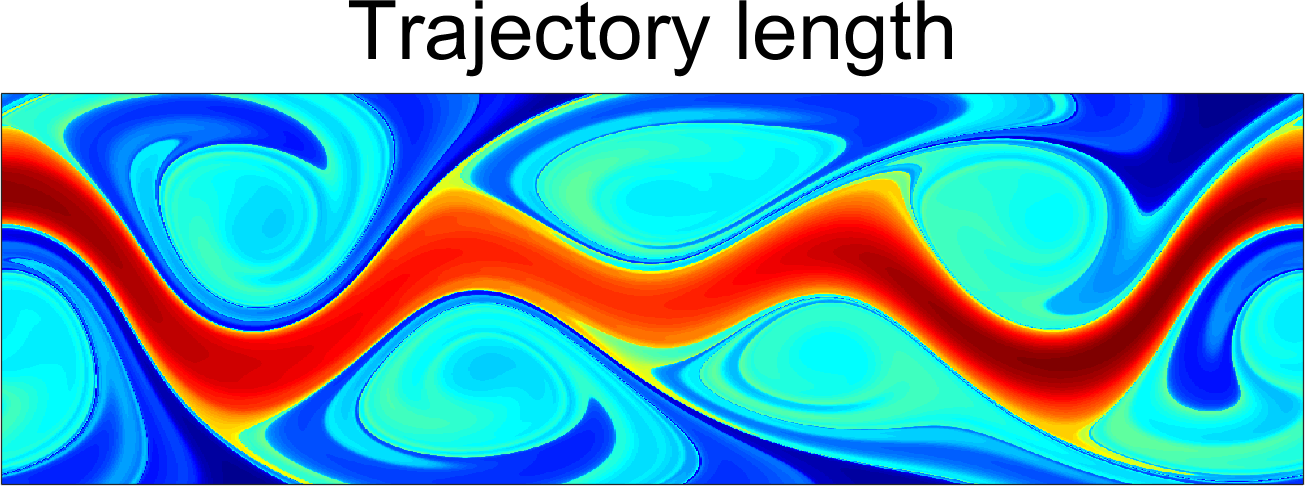}}\; \subfloat[\label{fig:BickleyCM}]{\includegraphics[width=0.42\textwidth]{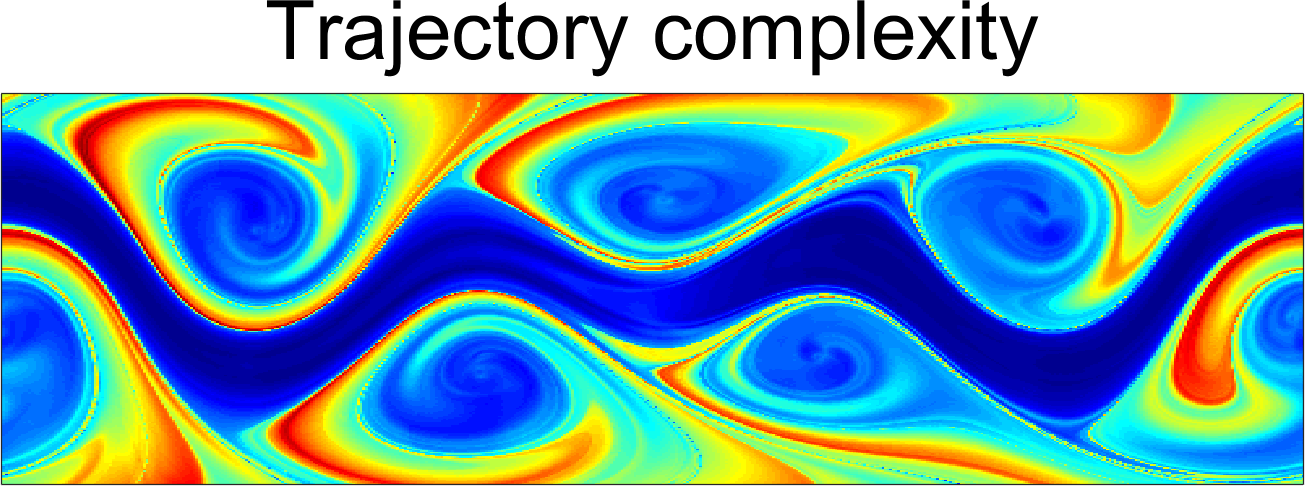}}\;
\subfloat[\label{fig:BickleyShapeCoherent}]{\includegraphics[width=0.42\textwidth]{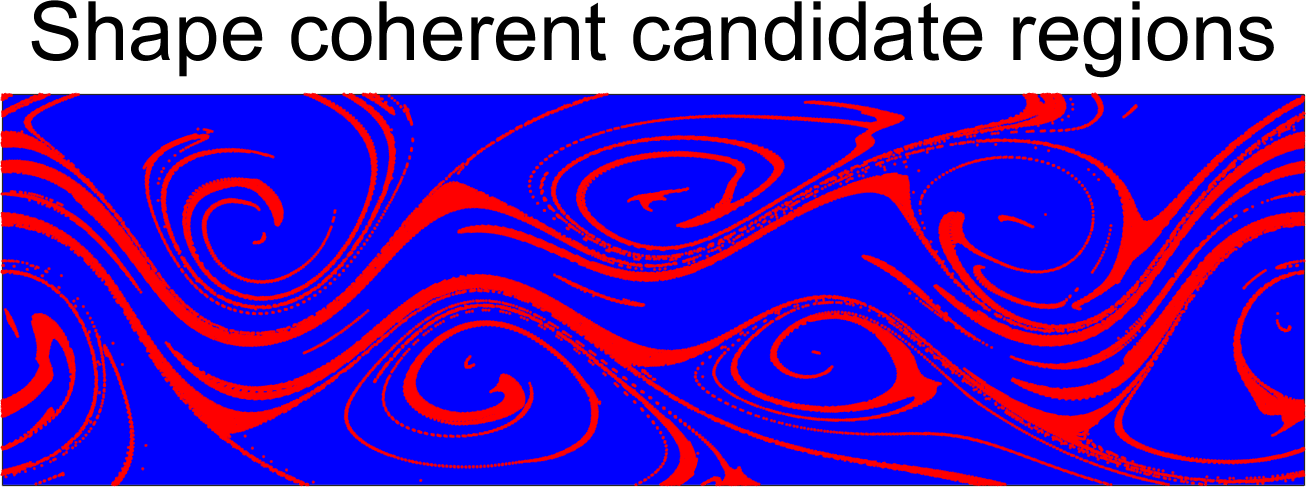}}\\
\subfloat[\label{fig:BickleyTransferOperator}]{\includegraphics[width=0.42\textwidth]{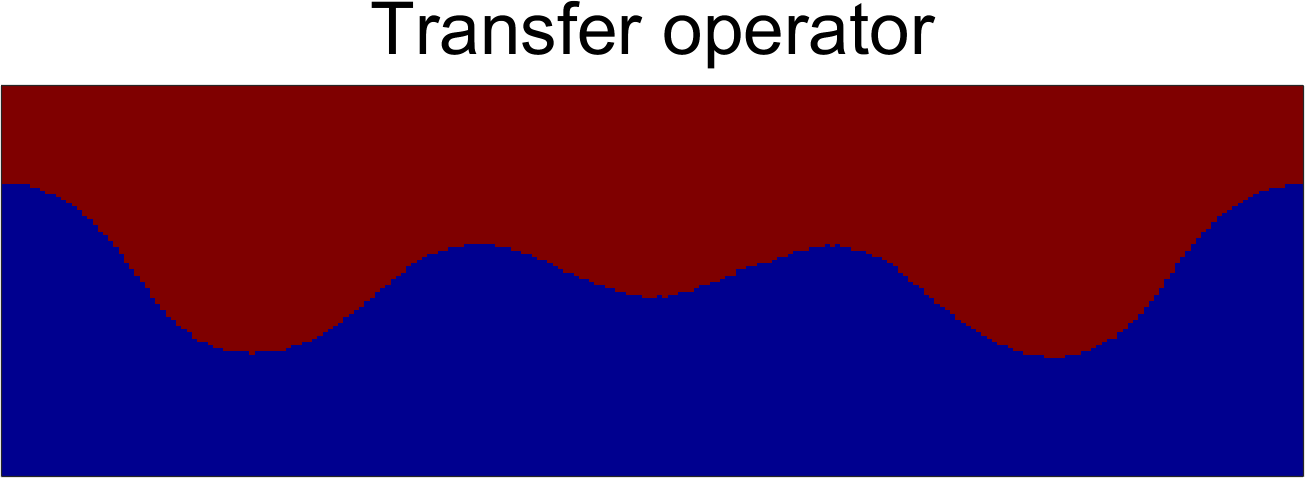}}\; \subfloat[\label{fig:BickleyHierarchy}]{\includegraphics[width=0.42\textwidth]{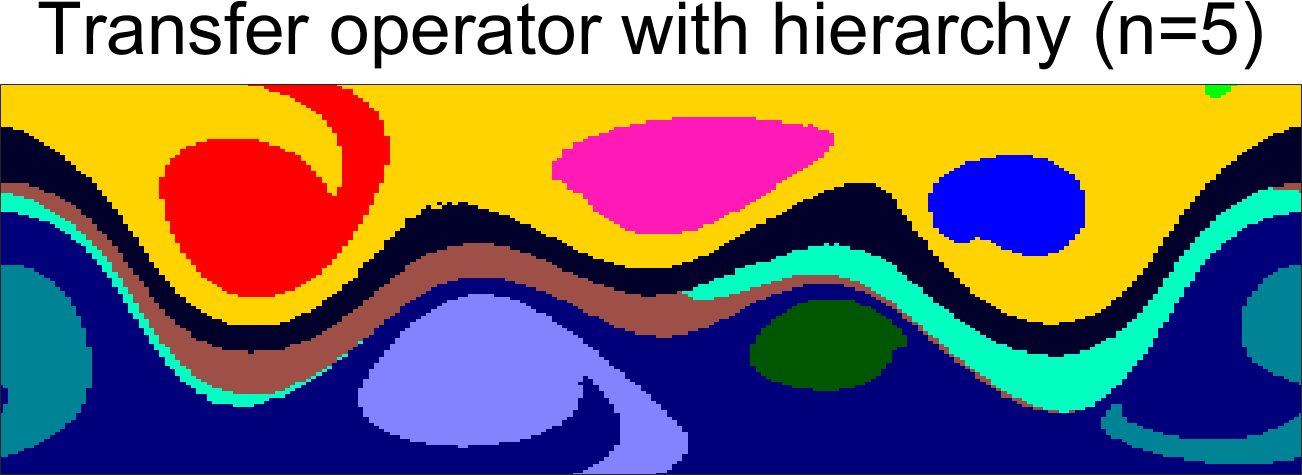}}\; \subfloat[\label{fig:BickleyFCM}]{\includegraphics[width=0.42\textwidth]{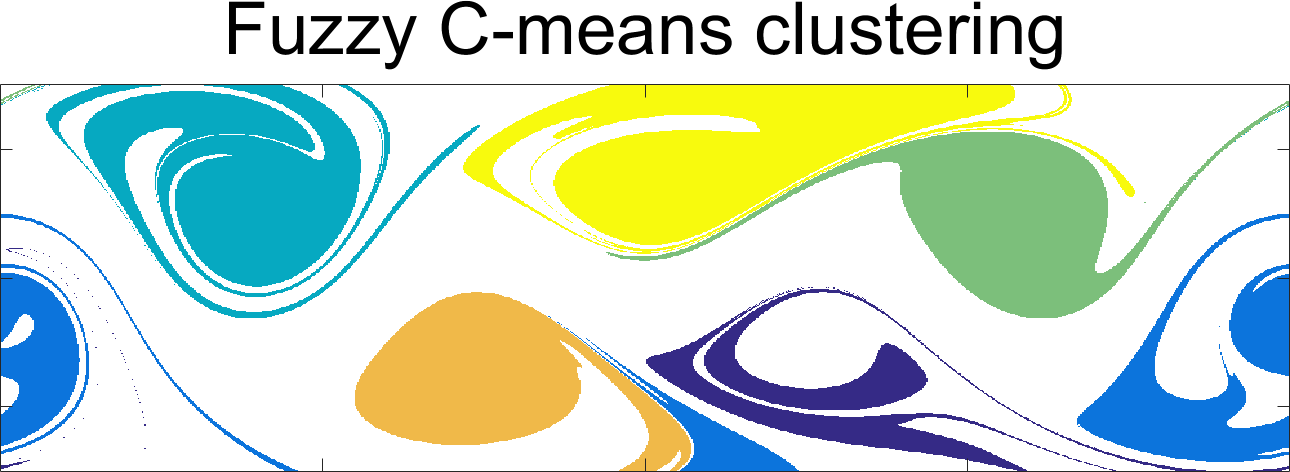}}\; \subfloat[\label{fig:BickleySpectral}]{\includegraphics[width=0.42\textwidth]{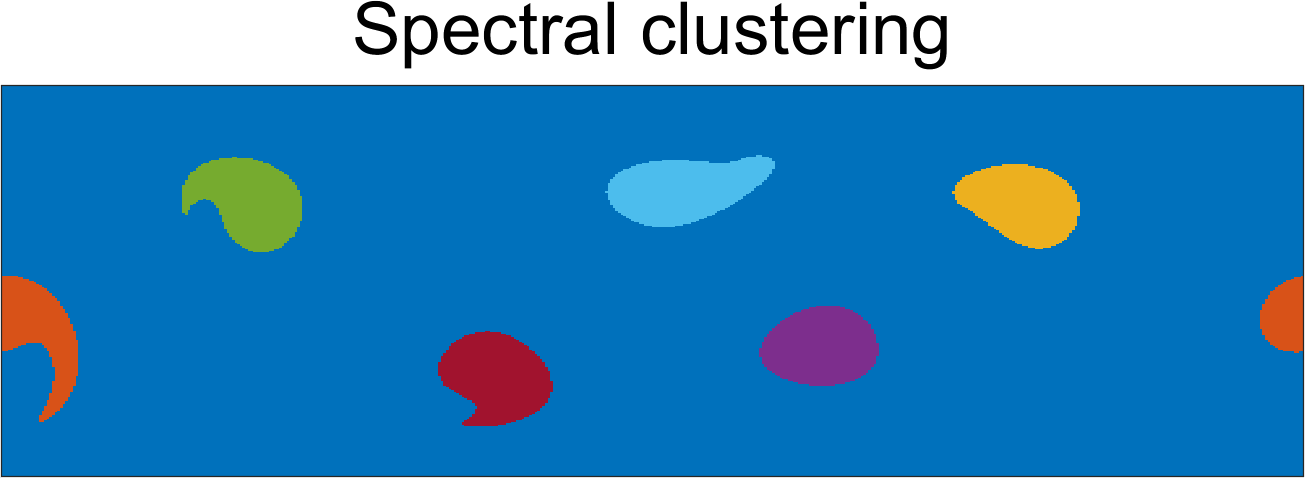}}\; \subfloat[\label{fig:BickleyGeodesic}]{\includegraphics[width=0.42\textwidth]{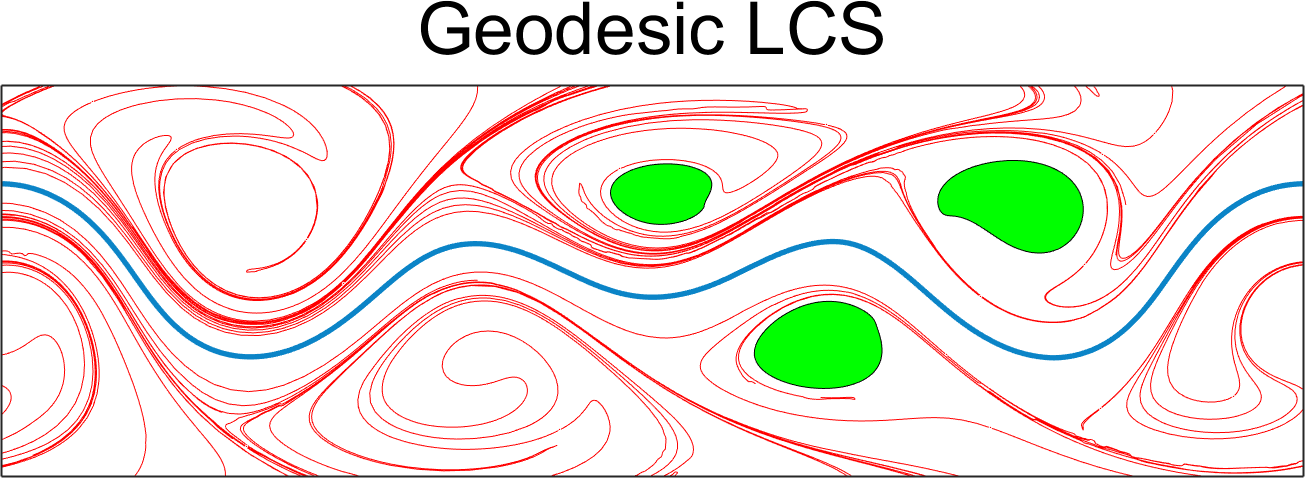}}\; \subfloat[\label{fig:BickleyLAVD}]{\includegraphics[width=0.42\textwidth]{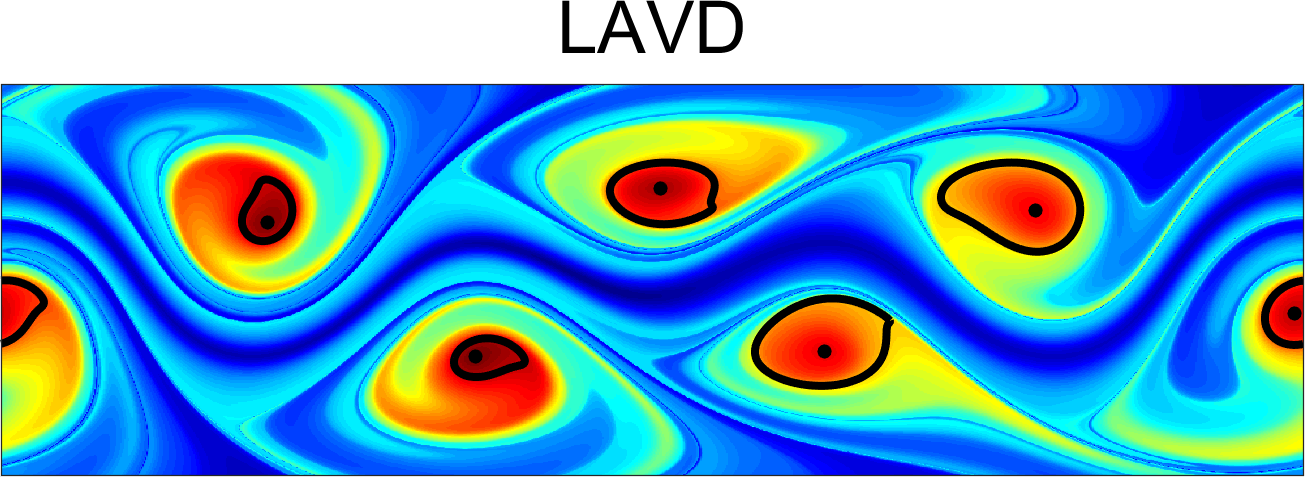}}\; \caption{Comparison of Lagrangian methods on the quasiperiodic Bickley jet
example (forward-time calculation only).}
\label{fig:Bickleycombined} 
\end{figure}
\end{turnpage}

We generate $5\times 10^5$ trajectories from a grid of initial conditions
in the domain $[0,20]\times[-3,3]$. For the FTLE, mesochronic analysis,
shape coherence and geodesic LCS methods, this means using a grid of $500\times250$ grid points with $4$ auxiliary points at each grid point for
finite-differencing that approximates the gradient of the flow map. FSLE
similarly requires $4$ auxiliary points in addition to the main grid
points to measure the minimal separation time $\tau$ between
the auxiliary points and the main grid. In contrast, the arclength
function, trajectory complexity, fuzzy C-means clustering, spectral
clustering and LAVD methods are computed on a $1000\times500$ grid to ensure that the same number of points are used in the comparison.
We compute the transfer operator and its hierarchical version using
a partition of $250\times125$ boxes, with $16$ particles per box.
We show the results for all methods in \Cref{fig:Bickleycombined}. 

The majority of diagnostic scalar fields in \Cref{fig:Bickleycombined} indicate
the presence of six vortices. Out of those offering more specific
definitions for coherent structure boundaries, however, the mesochronic analysis, the shape coherence, the transfer operator and the
geodesic method miss some or all of the vortices. Below, we discuss
these exceptions in more detail.

\Cref{fig:BickleyMesochronic} shows the mesochronic partitioning
of the domain into three different regions: mesohyperbolic without
rotation (blue), mesoelliptic (green) and mesohyperbolic with rotation
(red). Following the criterion proposed by Mezi{\'{c}} \cite{Mezic14},
we seek coherent vortex regions as nested sequences of alternating
mesoelliptic and mesohyperbolic annuli with smooth boundaries (i.e.,
no saddle-type critical points of the mesochronic plot should be embedded
in the boundary of at least three annuli of different colors). Examining
\cref{fig:Bickleymesocontour}, we observe saddle-type critical
points for the mesochronic field in \emph{all} the vortex regions,
resulting in a lack of smooth annular region boundaries. Hence, when
precisely implemented, the mesochronic analysis put forward in Refs.~\onlinecite{Mezic10,Mezic14}
does not indicate any coherent vortex in this example, even though
the topology of mesochronic contours gives a good general indication of
the vortical regions identified by objective methods. The mesochronic plot also fails to identify the hyperbolic
and parabolic LCSs identified by other diagnostics, such as the FTLE
field.

\Cref{fig:BickleyShapeCoherent} shows candidate regions (red) where
shape coherent sets may exist at the initial time $t_{0}=0$. In these
regions, the splitting angle between the dominant eigenvectors of
the forward-time and the backward-time Cauchy--Green strain tensor is smaller than $5.7^{\circ}$.
As mentioned earlier, these candidate regions are supposed to encompass
vortex boundaries that have significant shape coherence over the time
interval $[t_{0},t_{1}]$ of interest. In \Cref{fig:BickleyShapeCoherent},
however, all candidate regions are of spiral shapes, and hence cannot
contain closed curves encircling the candidate regions. Consequently, the
shape coherence method captures none of the coherent vortices for
the Bickley jet, given that even the weakened version of the underlying
criterion provides domains that cannot contain closed boundaries for these vortices.

\Cref{fig:BickleyTransferOperator} shows the two coherent sets
identified by the transfer operator method in this example. These
two sets are precisely the upper and lower parts of the flow domain
separated by the core of the jet. The jet core is identified very
sharply, but the method misses the coherent vortices identified by
most other methods. Higher singular vectors of the transfer operator
do indicate the presence of all these vortices, even if the actual
boundaries of these vortices will depend on what thresholding one
uses to extract structures from the eigenfunctions. It is a priori
unclear, however, how many singular vectors one needs to consider
to obtain an indication of all vortices in the problem (but see below
for more detail on how to make the exploration of singular vectors
systematic).

\Cref{fig:BickleyHierarchy} provides a successive partitioning
of the coherent sets obtained from the hierarchical transfer operator
method into further coherent sets. At the fifth level of hierarchy
$(n=5)$, the method captures the three most coherent vortices in
the problem. These vortices will be further partitioned under subsequent
steps in the hierarchical construction, unless one has a sense of
the ground truth and hence knows when to stop. The increased hierarchy
also dilutes the sharpness of the jet core identified by the transfer
operator method. A steadily growing number of patches appear that
are hard to justify physically in a perfectly homogeneous shear jet.

\begin{figure}[!t]
\subfloat[\label{fig:BickleyTransferOperatortf}]{\includegraphics[width=0.45\textwidth]{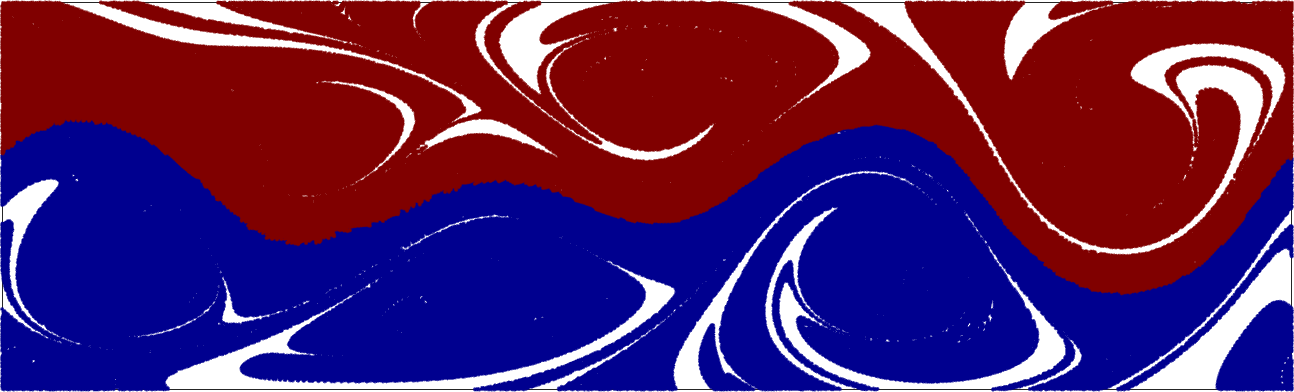}}
\; \subfloat[\label{fig:BickleyHierarchytf}]{\includegraphics[width=0.45\textwidth]{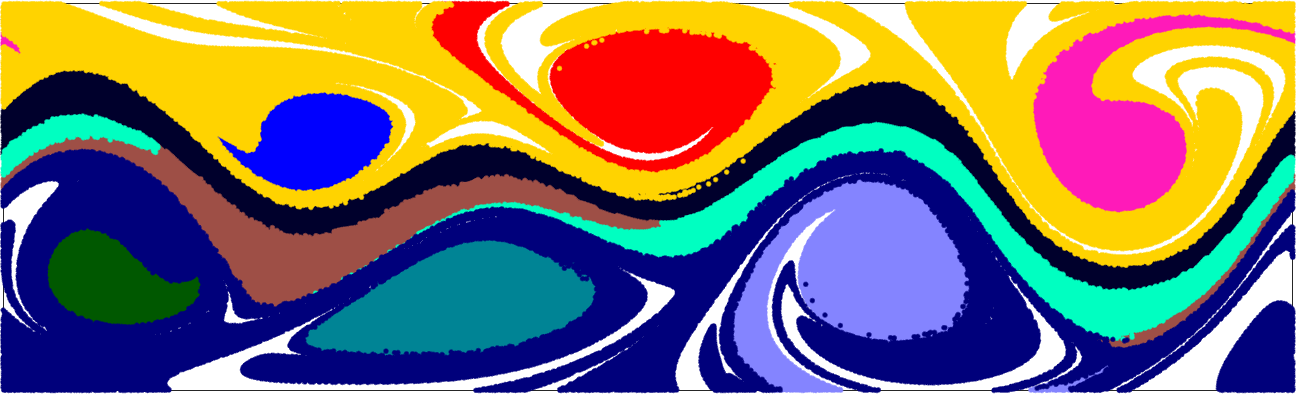}}
\; \subfloat[\label{fig:BickleyFCMtf}]{\includegraphics[width=0.45\textwidth]{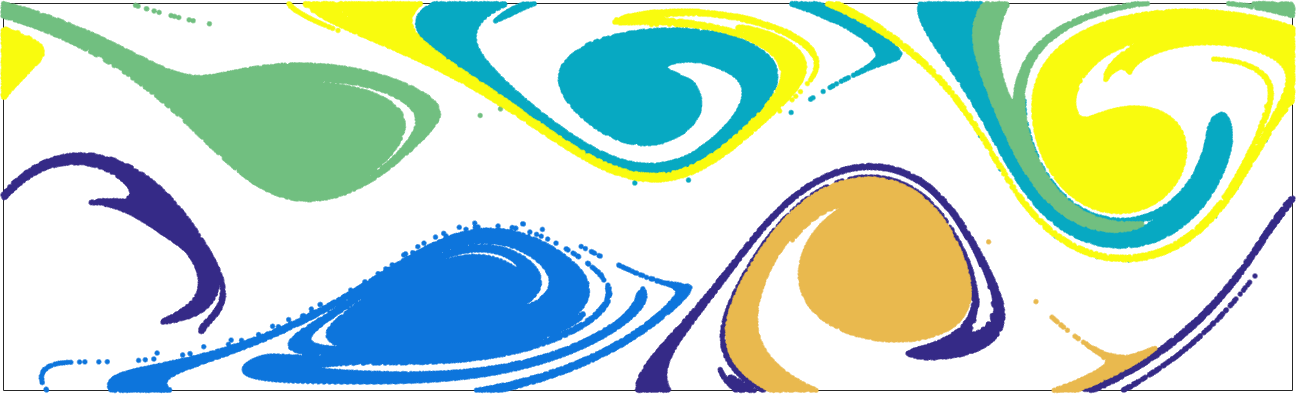}}
\; \subfloat[\label{fig:Bickleyclusteringtf}]{\includegraphics[width=0.45\textwidth]{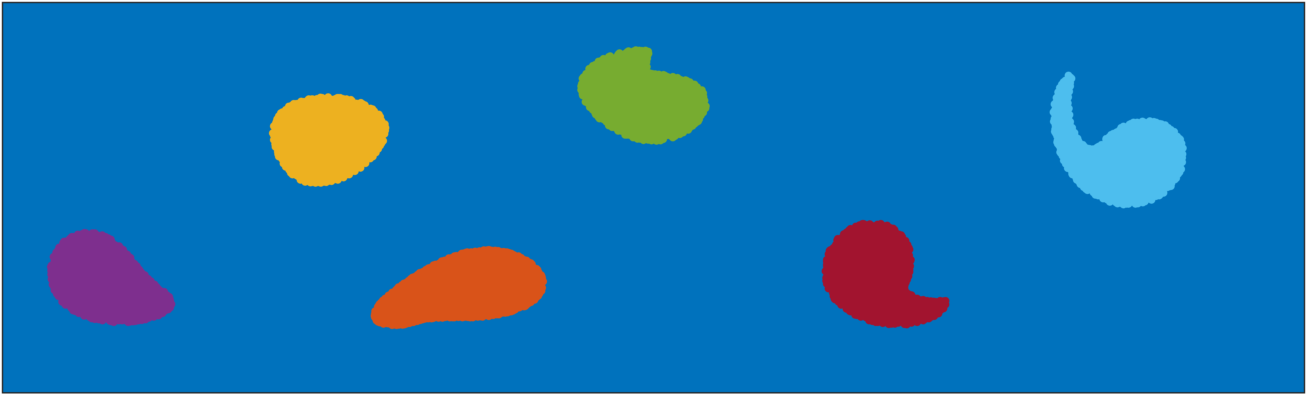}}
\; \; \subfloat[\label{fig:Bickleygeodesictf}]{\includegraphics[width=0.45\textwidth]{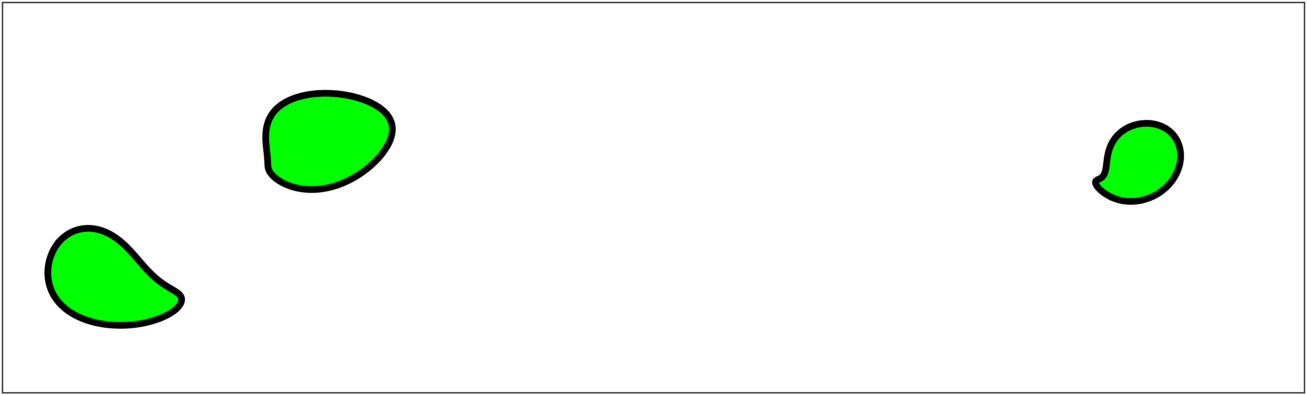}}
\; \subfloat[\label{fig:BickleyLAVDtf}]{\includegraphics[width=0.45\textwidth]{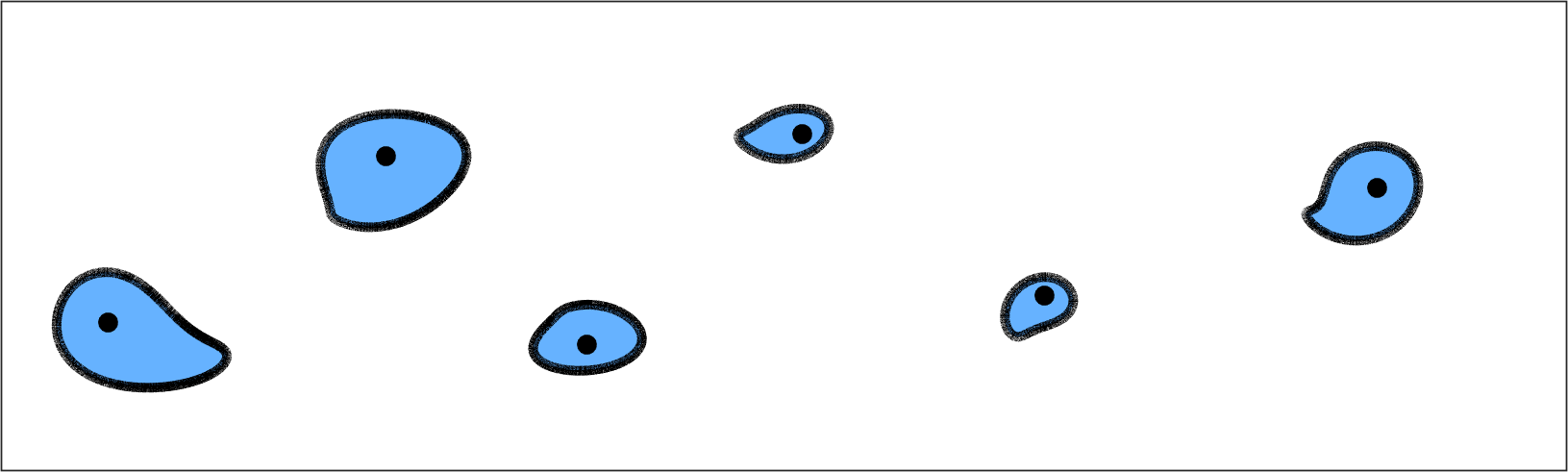}}
\caption{Advected images of Lagrangian coherent structures at the final time
$t_{1}=11$ day for six different methods: (a) Probabilistic transfer
operator (Multimedia
view) (b) Hierarchical transfer operator (Multimedia
view) (c) Fuzzy clustering (Multimedia
view) (d)
Spectral clustering (Multimedia
view) (e) Geodesic (Multimedia
view) and (f) LAVD (Multimedia
view). See also \Cref{fig:leftrightvecs}
(right) and \Cref{fig:allthresh}b for the transfer operator. Plots (a) and (b) have lower resolution because the total number of trajectories used in all computations were selected equal for a fair comparison.}
\end{figure}

\begin{figure}[!tbp]
\begin{minipage}[t]{0.45\textwidth}%
 \includegraphics[width=1\textwidth]{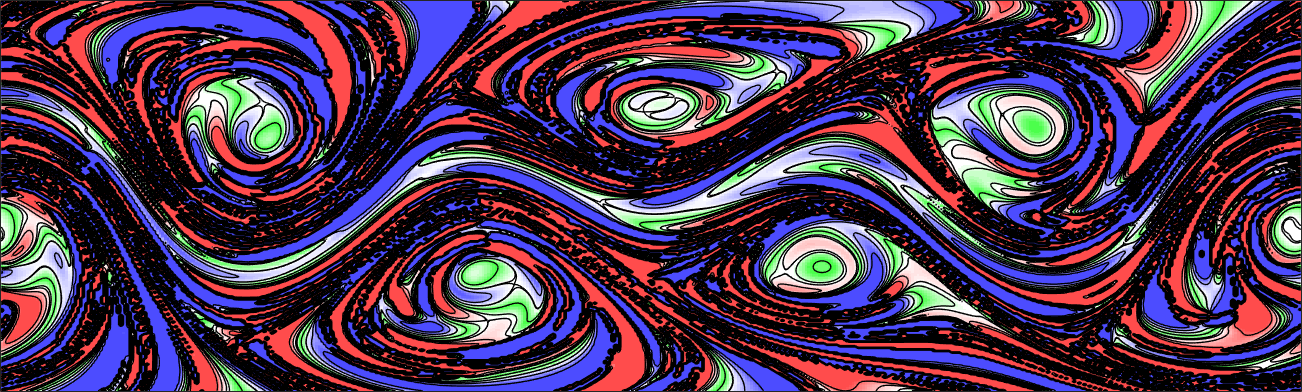} \caption{Mesochronic plot and its contours at the initial time $t_{0}=0$ for
the quasiperiodic Bickley jet.}
\label{fig:Bickleymesocontour} %
\end{minipage}\; %
\begin{minipage}[t]{0.45\textwidth}%
 \includegraphics[width=1\textwidth]{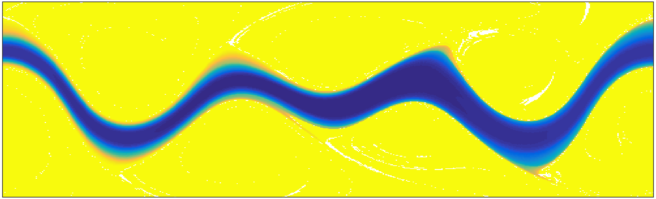} \caption{The seventh generalized eigenvector of the graph Laplacian $L$ obtained
from the spectral clustering analysis for the quasiperiodic Bickley
jet.}
\label{fig:BickleySpectralEig7} %
\end{minipage}
\end{figure}

\Cref{fig:BickleyFCM} shows the results from fuzzy clustering ($K=6$,
$m=1.25$). The method gives a good general sense for all coherent
vortices, but indicates no well-defined coherent vortex core with
a closed boundary. Instead, convoluted boundaries are detected for
all vortical regions, suggesting a lack of regular, convex domains
that stay tightly packed under advection. The sharp jet core detected
by the transfer operator method is also absent in these results. The
detected structures remain convoluted under advection in \Cref{fig:BickleyFCMtf}
(Multimedia view), except for their subsets contained in coherent
vortices signaled by other methods.

\Cref{fig:BickleySpectral} shows that the spectral clustering method
consistently detects all vortices involved, improving on the estimates
on their sizes given by other method. All these Lagrangian vortices
do remain coherent, as confirmed by their advection in \Cref{fig:Bickleyclusteringtf}
(Multimedia view). At the same time, the method gives no indication
of the coherent meandering jet in the dominant eigenvectors $u_{1},\ldots,u_{6}$
of the graph Laplacian $L$. The seventh eigenvector $u_{7}$ does
reveal the meandering jet in the flow (see \Cref{fig:BickleySpectralEig7}),
but there is no a priori indication from the spectrum that it should.
The reason is that the jet particles separate from each other due
to shear, which creates notably weaker within-class-similarity for
the jet than for the vortices.

\Cref{fig:BickleyGeodesic} shows the result for the geodesic LCS
analysis, where elliptic LCSs (material vortices), a parabolic LCS
(material jet core) and repelling hyperbolic LCSs (stable manifolds)
are shown in green, blue and red, respectively. In this example, the
geodesic method identifies only three out of six vortical regions
as coherent. Indeed, as seen in \Cref{fig:Bickleygeodesictf} (Multimedia
view), only three material vortex cores with no filamentation can
be found under advection to the final time $t_{1}=11$ day. (As seen
in \Cref{fig:Bickleycombined}, these three vortices also happen
to be the ones most clearly identified by the hierarchical transfer
operator method.) That said, \Cref{fig:Bickleyclusteringtf} (Multimedia
view) and \Cref{fig:BickleyLAVDtf} (Multimedia view) show that
the actual number of arguably coherent material vortices is six, which
indicates that the variational principle behind the geodesic method
is too restrictive for some of the vortices of the Bickley jet flow.

\Cref{fig:BickleyLAVD} shows that the LAVD method captures all
vortices accurately and the detected structures only show tangential
filamentation under advection (cf. \Cref{fig:BickleyLAVDtf} (Multimedia
view)), as they should by construction. At the same time, the LAVD
method is unable to detect the intended main feature of this model
flow, the meandering jet in the middle. More generally, the LAVD method
is not designed to detect hyperbolic or parabolic LCSs.

As for jet identification, we observe that most methods offer some
indication of the central jet, except for the shape coherence, fuzzy
clustering, spectral clustering and LAVD methods. The majority of
methods, however, do not offer a systematic approach to extracting
the jet core or jet boundaries. The only exceptions are the FTLE, geodesic and the transfer operator methods that give a sharp boundary for the jet core (see \Cref{fig:BickleyGeodesic,fig:BickleyTransferOperator,fig:BickleyHierarchy}).

\begin{figure}
\includegraphics[width=1\textwidth]{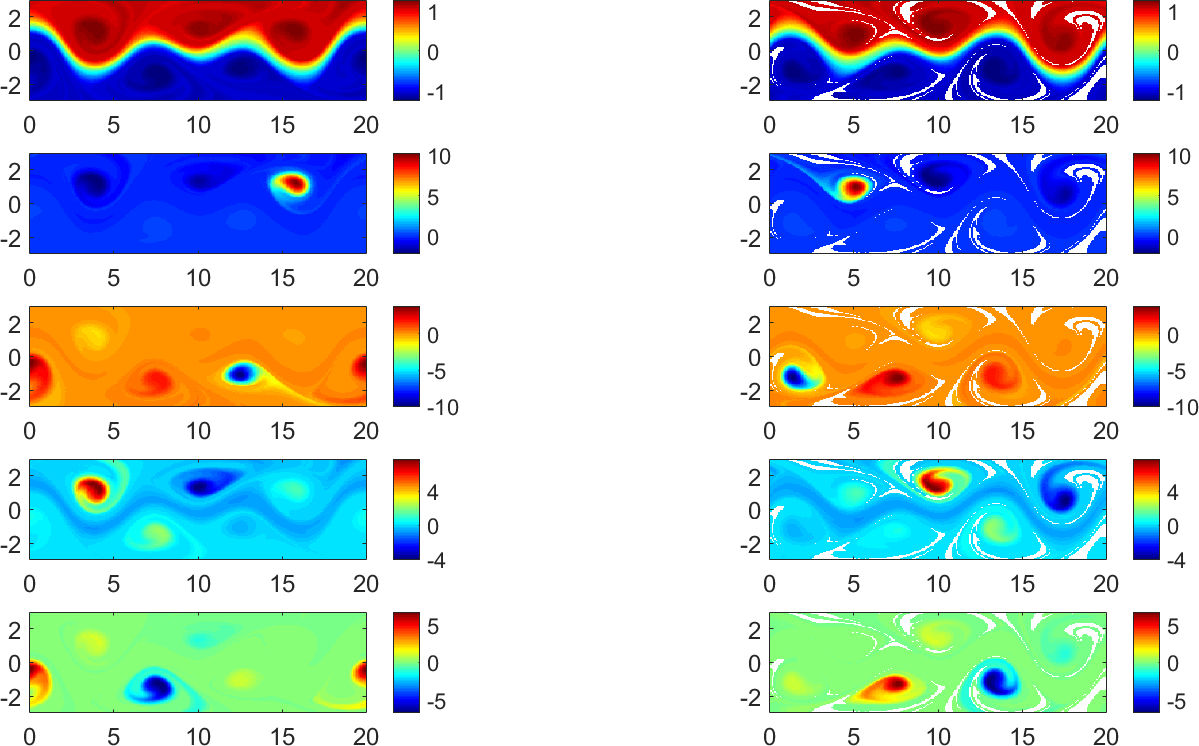} \caption{The first five nontrivial singular vectors of the transfer operator
for the Bickley jet example. Left column: Vectors $u_{2},u_{3},u_{4},u_{5},u_{6}$
(top to bottom). Right column: Vectors $v_{2},v_{3},v_{4},v_{5},v_{6}$
(top to bottom). Various finite-time coherent sets are highlighted
at the initial time (left column) and final time (right column).}
\label{fig:leftrightvecs} 
\end{figure}

On this example, we also illustrate how a consideration of the higher
singular vectors of the transfer operator yields additional insight
into the structure of the two main coherent sets revealed by its second
singular vector in \Cref{fig:BickleyTransferOperator}. The initial
domain $X=[0,20]\times[-2.5,2.5]$ (with left and right edges identified)
is gridded into 125,000 identical squares (250 grid boxes in the $x$-direction
and $125$ grid boxes in the $y$-direction). We use 16 uniformly
distributed sample points per grid box and compute Lagrangian trajectories,
recording the terminal points after time $t_{1}=24$ days. The image
domain $Y=T(X)$ is gridded into squares of the same size, and is
covered by 132,131 grid boxes. The grid-to-grid transition matrix
$P$ (see Ref.~\onlinecite{Froyland10} for details) is therefore
a row-stochastic $132,131\times125,000$ rectangular matrix. The leading
singular vectors $u_{k}$ (resp.\ $v_{k}$), $k=2,\ldots,6$ of the
transfer operator are shown in the left (resp.\ right) columns of
\Cref{fig:leftrightvecs}. The top row of \Cref{fig:leftrightvecs}
shows a clear separation of the upper and lower parts of the flow
domain.

\begin{figure}
\subfloat[]{\includegraphics[width=0.45\textwidth]{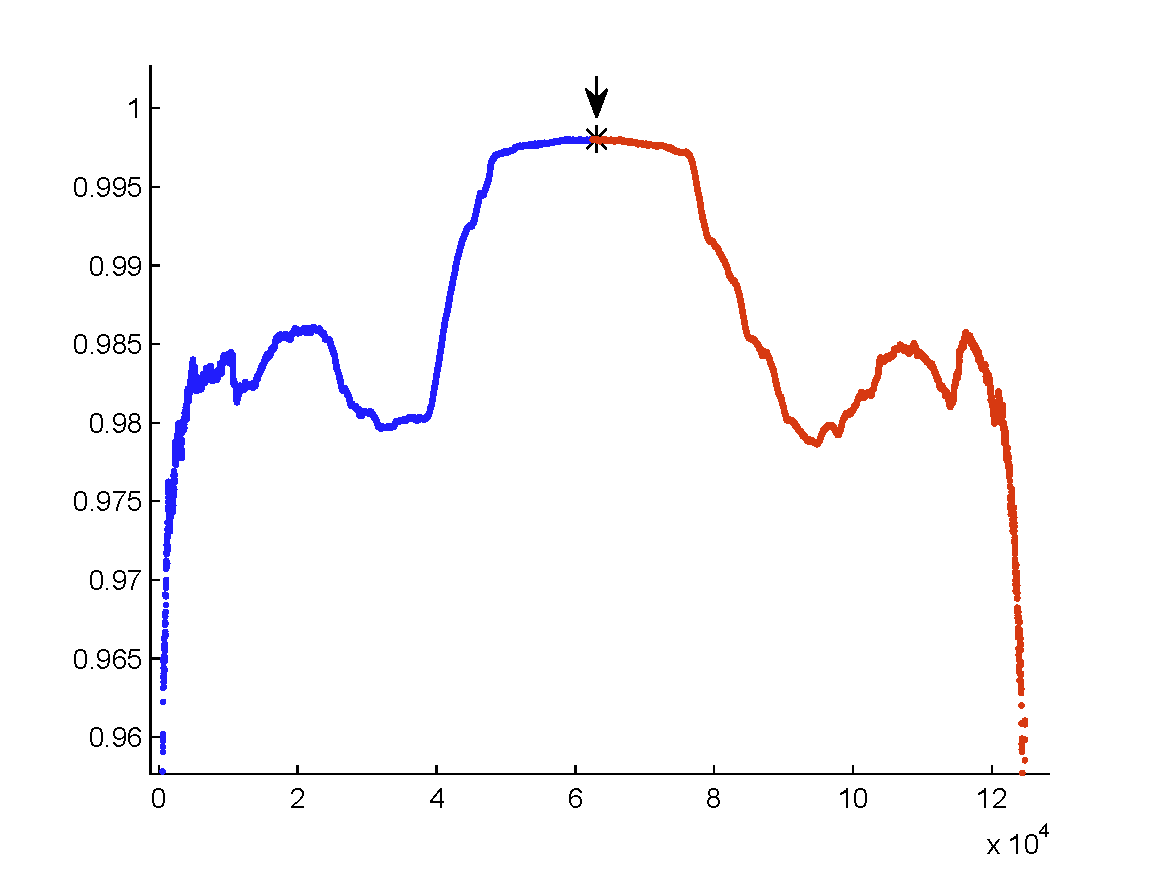}}
\subfloat[]{\includegraphics[width=0.45\textwidth]{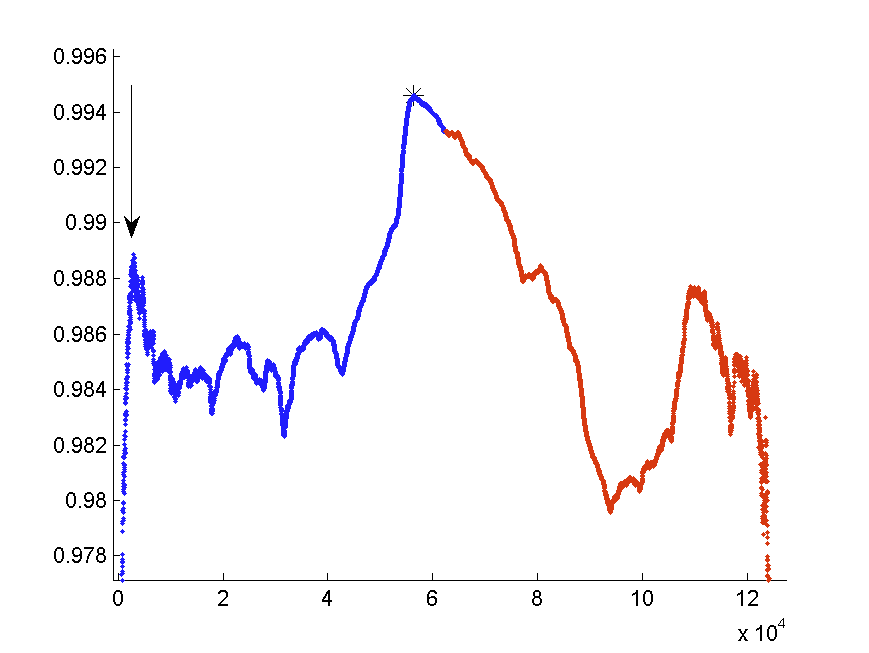}}
\caption{(a) Plot of $\rho(A_{c},\tilde{A}_{c})$ vs.\ $c$ based on $u_{2},v_{2}$.
The global maximum of $\rho(A_{c},\tilde{A}_{c})$ is indicated with
a vertical arrow and black asterisk. This corresponds to the upper/lower
separation shown in pale yellow/pale orange in \Cref{fig:allthresh}.
(b) Plot of $\rho(A_{c},\tilde{A}_{c})$ vs.\ $c$ based on $u_{3},v_{3}$.
The first local maximum of $\rho(A_{c},\tilde{A}_{c})$, starting
from the largest value of $u_{3}$ and descending, is indicated with
a vertical arrow. This corresponds to the red set in the lower right
of the upper panel of \Cref{fig:allthresh}, and its image in
the lower left of the lower panel of \Cref{fig:allthresh}.}
\label{fig:thresh} 
\end{figure}

\begin{figure}
\subfloat[]{\includegraphics[width=0.45\textwidth]{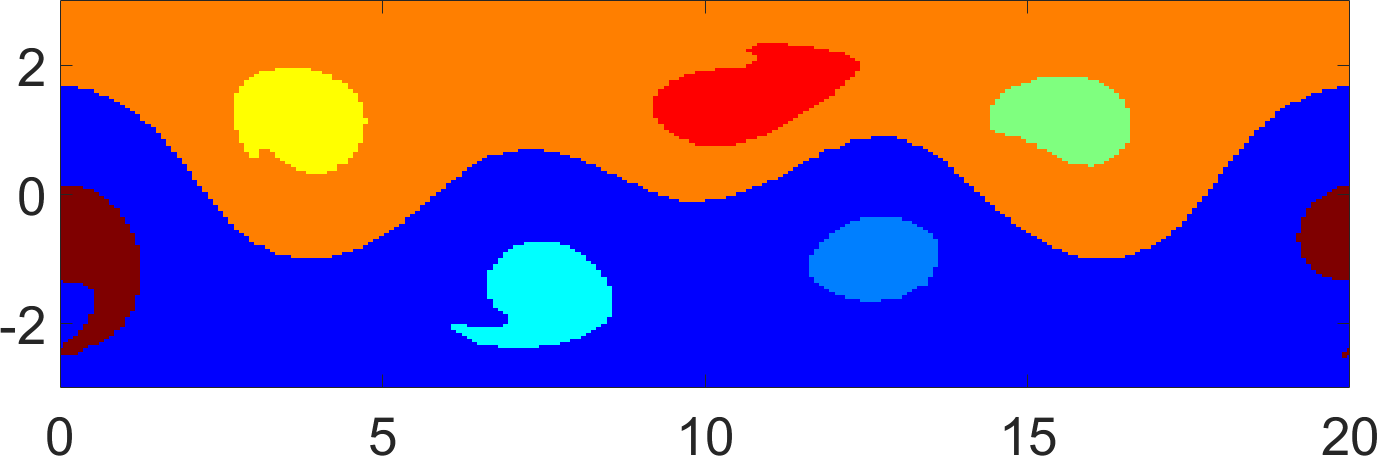}}\;
\subfloat[]{\includegraphics[width=0.45\textwidth]{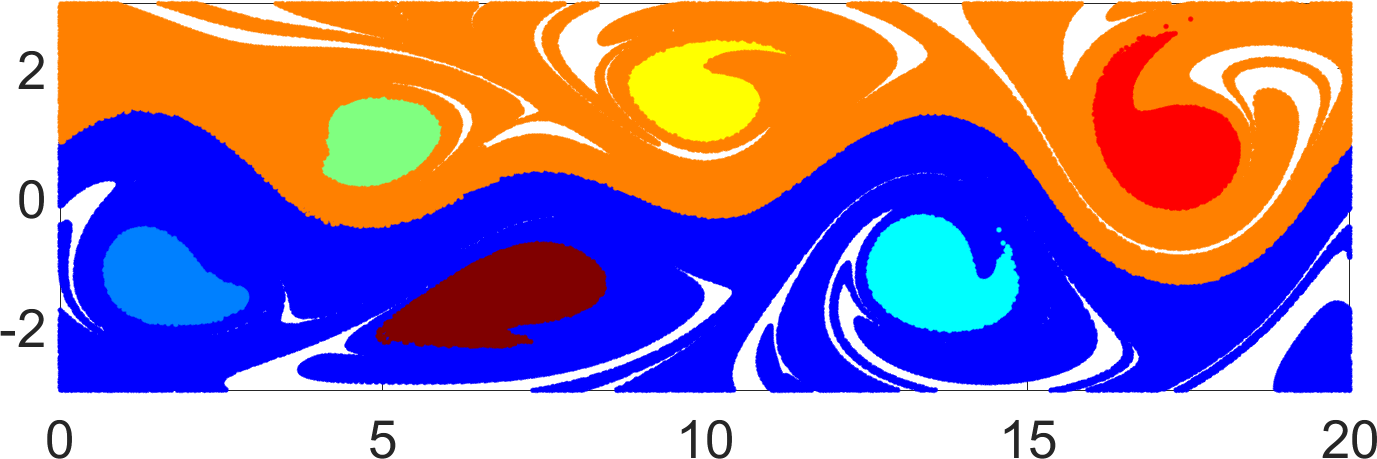}}
\caption{(a) The finite-time coherent sets extracted from the singular vectors
$u_{k},v_{k},k=2,\ldots,6$ for the Bickley jet example at the initial
time. (b) Advected image of the coherent sets at the final time. }
\label{fig:allthresh} 
\end{figure}

We threshold the vectors $u_{2},v_{2}$ according to the algorithm
proposed in Ref.~\onlinecite{Froyland10} by letting $c\in\{1,\ldots,12,500\}$
represent a sorted box index. We then plot the coherence ratio $\rho(A_{c},\tilde{A}_{c})$
vs.\ $c$ in \Cref{fig:thresh} (left), where $A_{c}$, $\tilde{A}_{c}$
are super/sub-level sets of $u_{2}$, $v_{2}$ (see Algorithm 1 in
Ref.~\onlinecite{Froyland10} for details). In \Cref{fig:thresh},
the blue curve indicates grid sets sorted in descending value of $u_{2}$
from the maximum of $u_{2}$, while the red curve indicates grid sets
sorted in ascending value of $u_{2}$ from the minimum of $u_{2}$;
the two curves meet where the mass of the partition sets are both
equal to 1/2. The maximum value of $\rho(A_{c},\tilde{A}_{c})$ is
indicated by the vertical arrow and the black asterisk. The resulting
spatial partition is the pale yellow/pale orange separation shown
in \Cref{fig:allthresh}.

The vectors $u_{k},v_{k},k=3,\ldots,5$ in \Cref{fig:leftrightvecs}
(second and lower rows) highlight other smaller features. In order
to extract these smaller features, there are two main approaches.
First, one could restrict the domain to a smaller domain, a little
larger than twice the size of the highlighted feature (see, e.g.,
the experiments in the atmosphere \cite{Froyland10} and the ocean
\cite{FHRSS12,Froyland15_3}). One then recomputes $u_{2}$ and $v_{2}$
and because other coherent features have been eliminated from the
domain, these dominant nontrivial vectors capture the required feature.
Note that this procedure is different to Ref.~\onlinecite{Ma13}.

Second, one could retain the original domain and use the vectors $u_{k},v_{k},k=2,\ldots,5$
directly. Various techniques have been devised to extract information
from multiple vectors (see, e.g., the references in 3.1 of Ref.~\onlinecite{Froyland09}).
One could, for instance, fuzzy cluster the embedded vectors $u_{k}$,\cite{Froyland05}.
Here we take a vector by vector approach. In the present example,
there are clear features highlighted through the extreme negative
or positive values of $u_{k},v_{k},k=3,\ldots,5$. In general, given
a particular sufficiently coherent spatial feature, one should always
be able to find a vector which highlights that feature through an
extreme negative or positive value (for example, see Fig. 4 of Ref.~\onlinecite{FSV14}
for computations on the global ocean). A simple approach is to look
for the first local maxima of $\rho(A_{c},\tilde{A}_{c})$ in the
thresholding figures computed from $u_{k},v_{k},k=3,\ldots,5$, starting
at either the negative or positive end of the vector that corresponds
to the spatial feature one wishes to extract.

For example, the second row of \Cref{fig:leftrightvecs} highlights
a small red feature, which corresponds to a extreme positive values
of $u_{3},v_{3}$. Thus, we threshold starting from the maxima of
$u_{3},v_{3}$ and descend, looking for the first local maximum of
$\rho(A_{c},\tilde{A}_{c})$. Figure \ref{fig:thresh} (right) shows
the full plot of $\rho(A_{c},\tilde{A}_{c})$ vs.\ $c$, with the
first local maximum indicated with a vertical black arrow. The corresponding
spatial feature is shown in the lightest red in \Cref{fig:allthresh}.
This approach is repeated for all remaining highlighted features in
\Cref{fig:leftrightvecs} . The extracted finite-time coherent
sets are displayed in \Cref{fig:allthresh}.

\subsection{Two-dimensional turbulence}

As our second example, we consider a flow without any temporal recurrence.
We solve the forced Navier\textendash Stokes equation 
\begin{equation}
\partial_{t}v+v\cdot\nabla v=-\nabla p+\nu\Delta v+f,\quad\nabla\cdot v=0,
\end{equation}
for a two-dimensional velocity field $v(x,t)$ with $x=(x_{1},x_{2})\in U=[0,2\pi]\times[0,2\pi]$.
We use a pseudo-spectral code with viscosity $\nu=10^{-5}$ on a $512\times512$
grid, as described in Ref.~\onlinecite{Farazmand13}. A random-in-phase
velocity field evolves in the absence of forcing ($f=0$) until the
flow is fully developed. At that point, a random-in-phase forcing
is applied. For the purposes of the following Lagrangian analysis,
we identify this instance with the initial time $t=0$. The finite
time interval of interest is then $t\in[0,50]$.

\begin{turnpage}
\begin{figure}
\subfloat[\label{fig:2DturbFTLE}]{\includegraphics[width=0.26\textwidth]{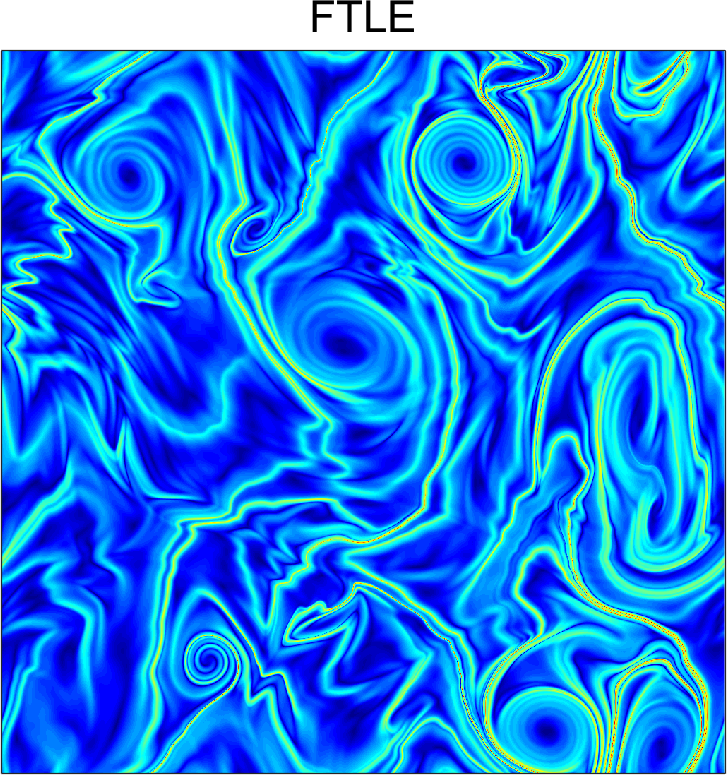}}\; \subfloat[\label{fig:2DturbFSLE}]{\includegraphics[width=0.26\textwidth]{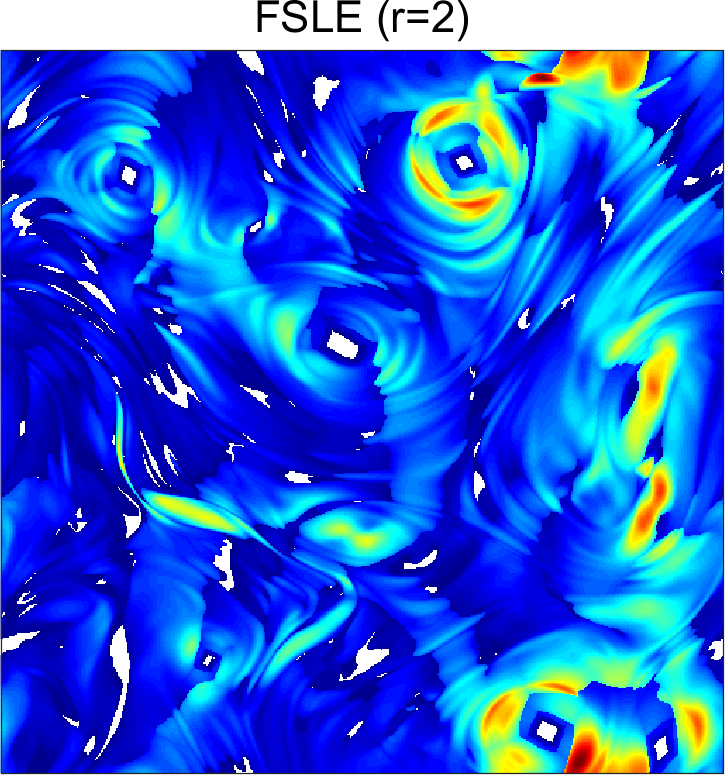}}\;  \subfloat[\label{fig:2DturbMesochronic}]{\includegraphics[width=0.26\textwidth]{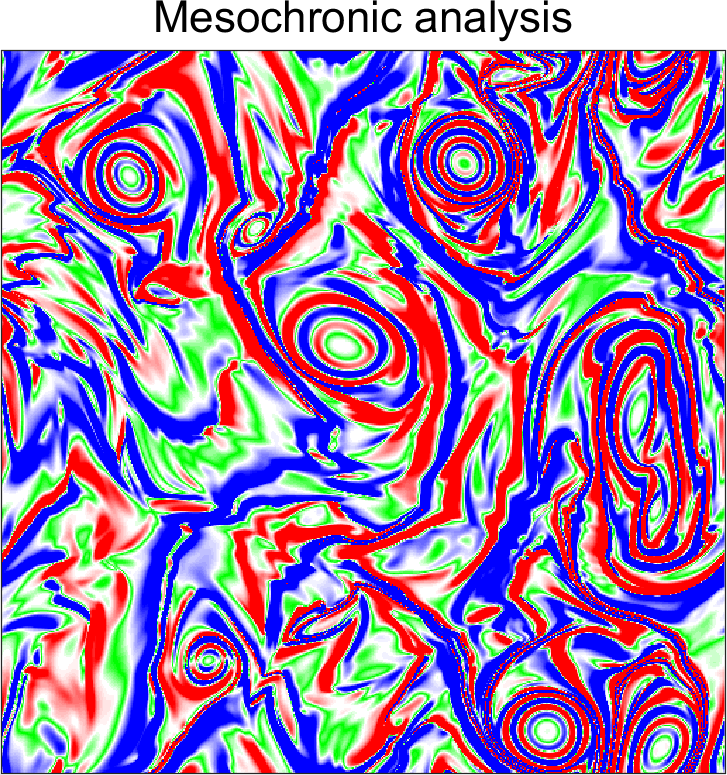}}\;
\subfloat[\label{fig:2DturbMfunction}]{\includegraphics[width=0.26\textwidth]{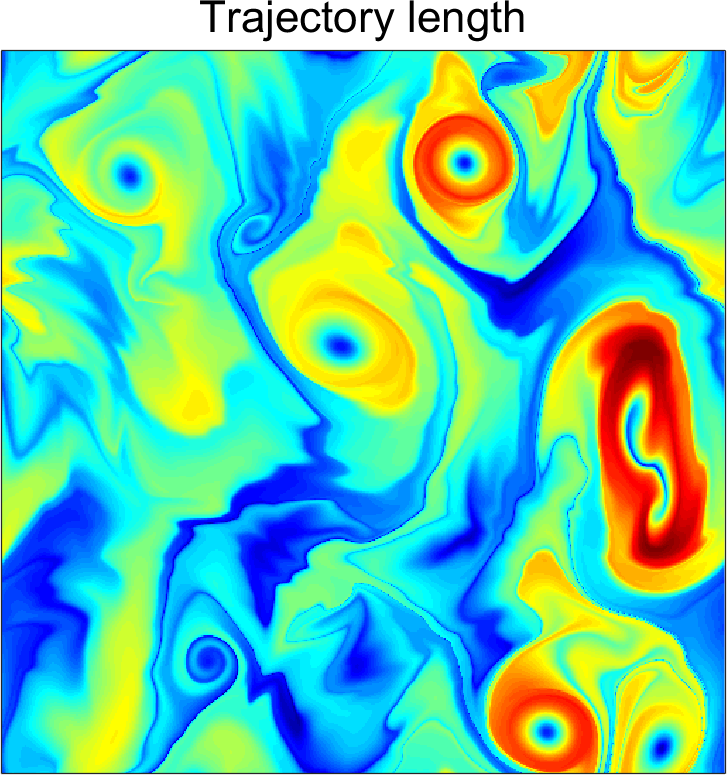}}\\
\subfloat[\label{fig:2DturbCM}]{\includegraphics[width=0.26\textwidth]{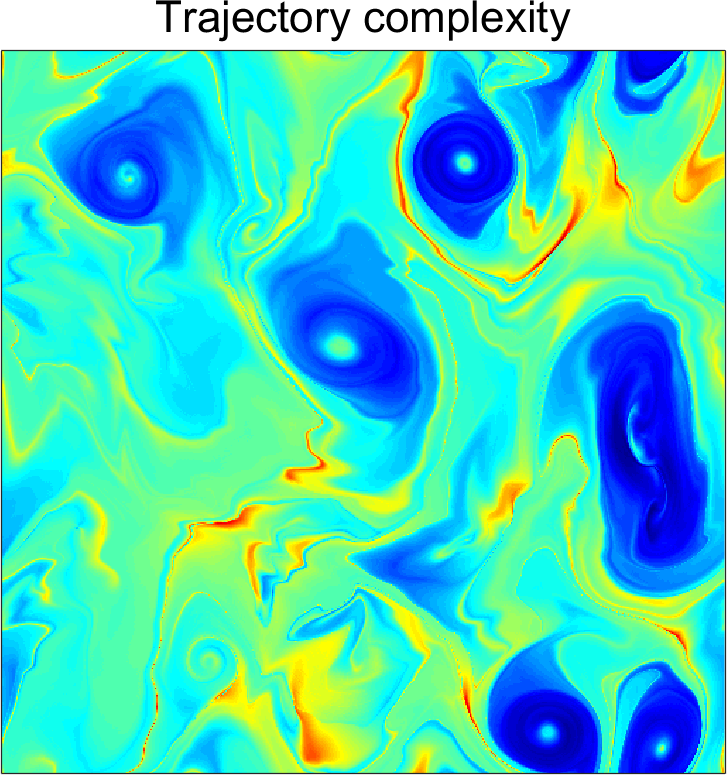}}\;
\subfloat[\label{fig:2DturbShapeCoherence}]{\includegraphics[width=0.26\textwidth]{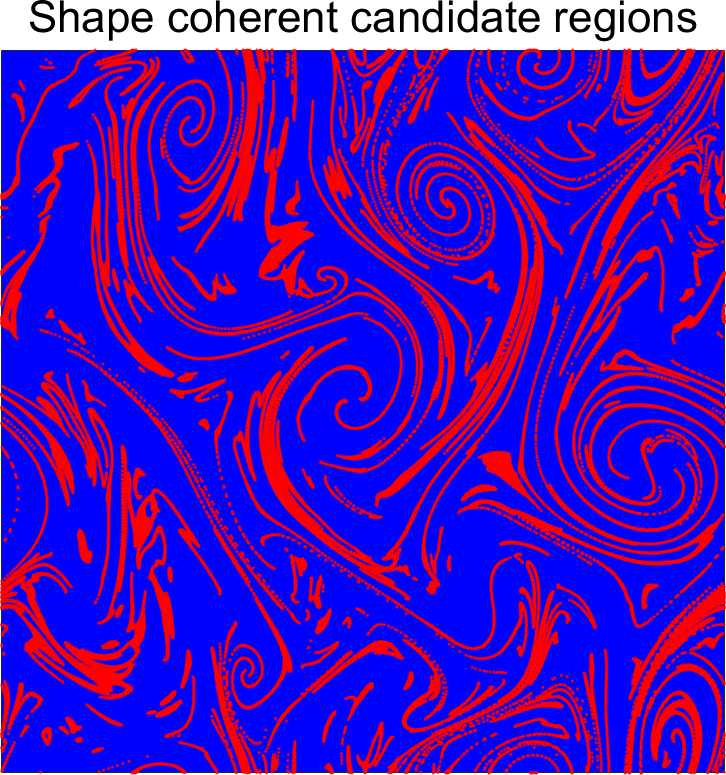}}\; \subfloat[\label{fig:2DturbTransferOperator}]{\includegraphics[width=0.26\textwidth]{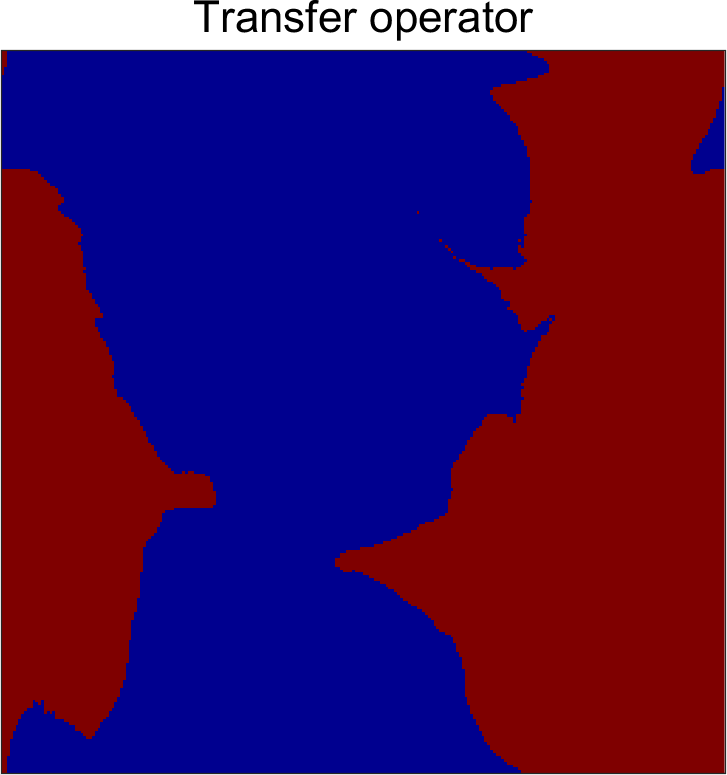}}\; \subfloat[\label{fig:2DturbHierarchy}]{\includegraphics[width=0.26\textwidth]{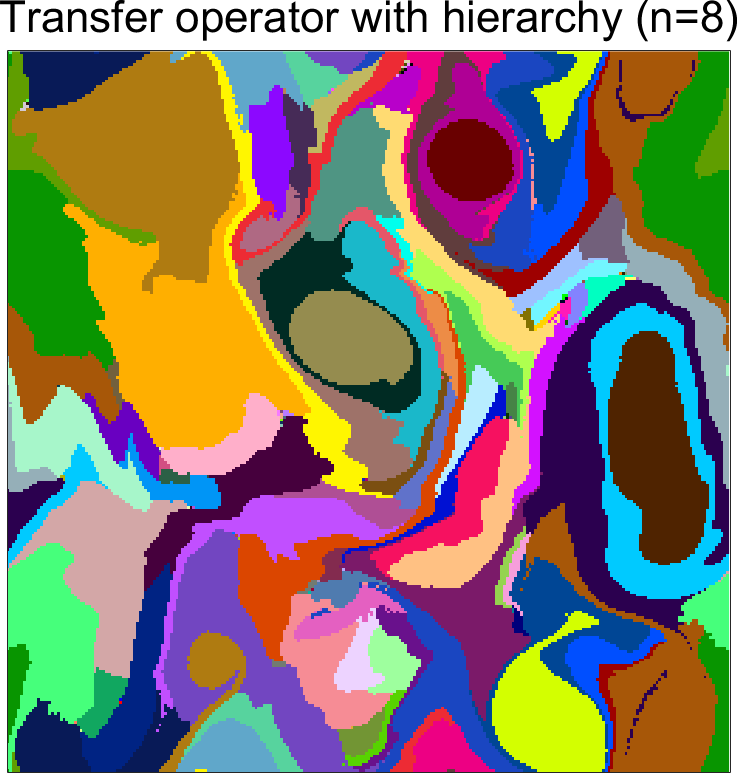}}\; \subfloat[\label{fig:2DturbFCM}]{\includegraphics[width=0.26\textwidth]{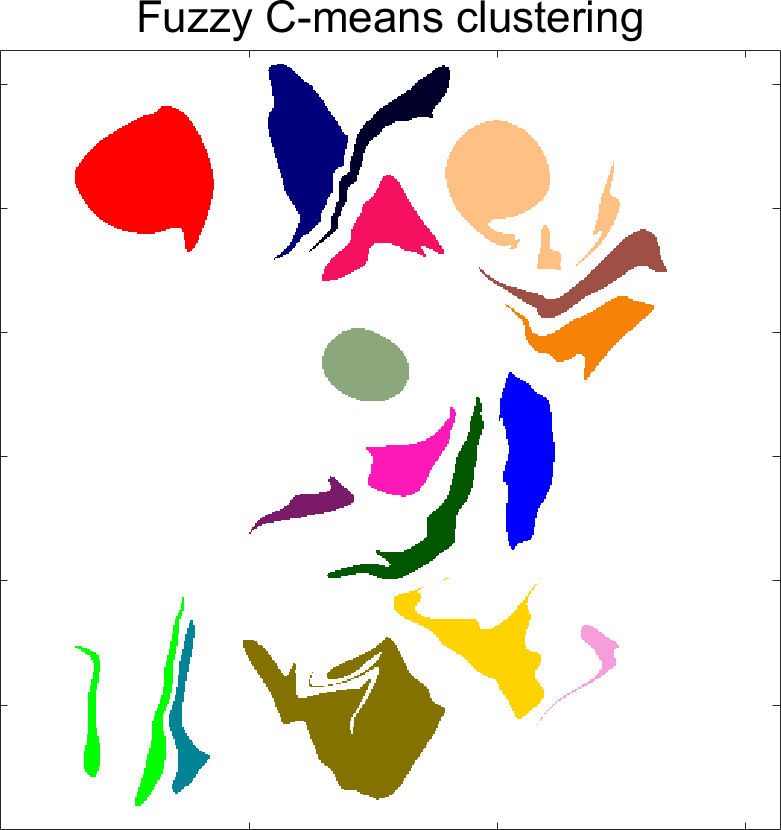}}\; \subfloat[\label{fig:2DturbSpectral}]{\includegraphics[width=0.26\textwidth]{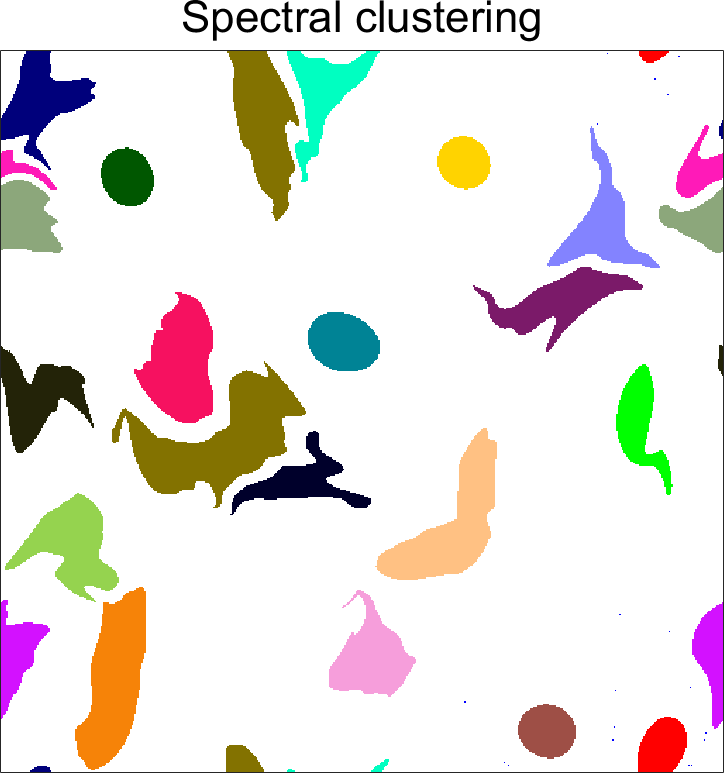}}\; \subfloat[\label{fig:2DturbGeodesic}]{\includegraphics[width=0.26\textwidth]{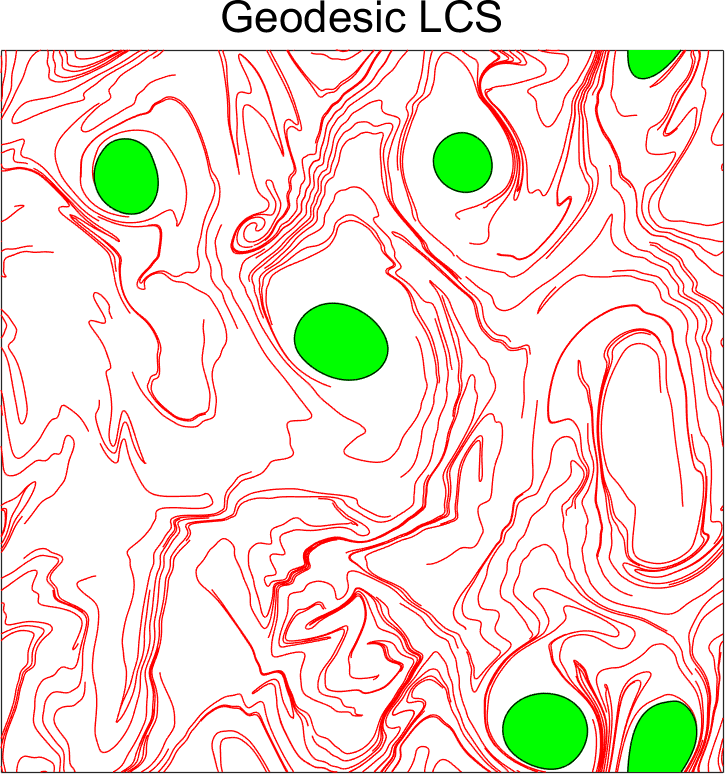}}\; \subfloat[\label{fig:2DturbLAVD}]{\includegraphics[width=0.26\textwidth]{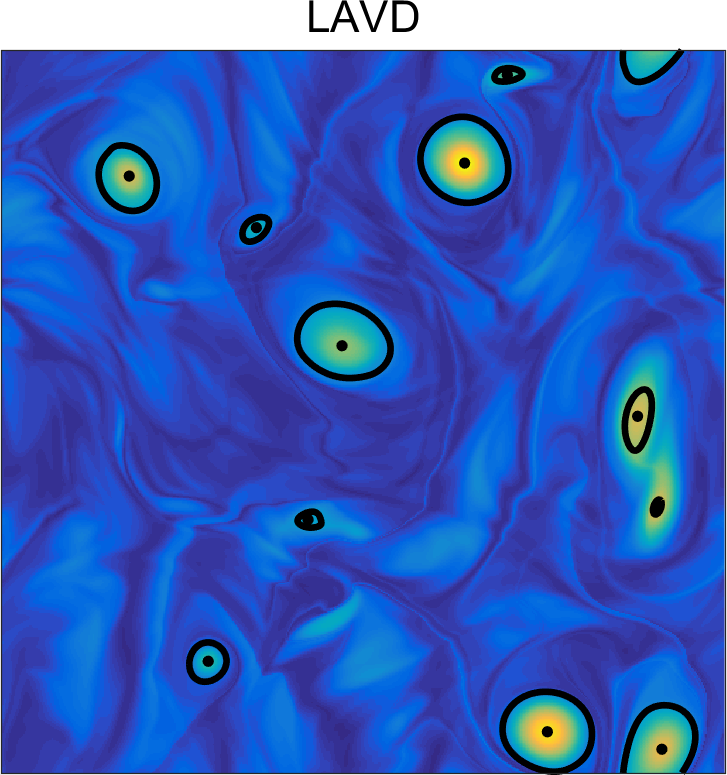}}\; \caption{Comparison of Lagrangian methods on the two-dimensional turbulence
simulation example.}
\label{fig:2Dturbcompare} 
\end{figure}
\end{turnpage}

\Cref{fig:2Dturbcompare} shows the result from various Lagrangian
methods applied to the resulting finite-time dynamical system $\dot{x}=v(x,t)$.
We use the auxiliary gird approach with the distance $\rho=10^{-3}$
to construct the FTLE, FSLE, mesochronic and shape coherence diagnostic
fields. The same auxiliary distance is used to compute the Cauchy\textendash Green
strain tensor as well as the vorticity for the geodesic and LAVD methods,
respectively. 

Most plots in \Cref{fig:2Dturbcompare} indicate several
vortex-type structures, except for the shape coherence and transfer
operator methods. While the boundaries of the large-scale coherent
sets identified by the latter method indeed do not grow significantly
under the finite-time flow, these sets are unrelated to the vortices
that are generally agreed to be the coherent structures of two-dimensional
turbulence. These vortices only appear in some of the higher singular
vectors of the transfer operator, similarly to \Cref{fig:leftrightvecs}
and \Cref{fig:allthresh}. Just as in the case of the Bickley jet,
however, there is no clear indication from the spectrum of singular
values for the number of singular vectors to be considered to recover
all vortices.

\begin{figure}
\includegraphics[width=0.3\textwidth]{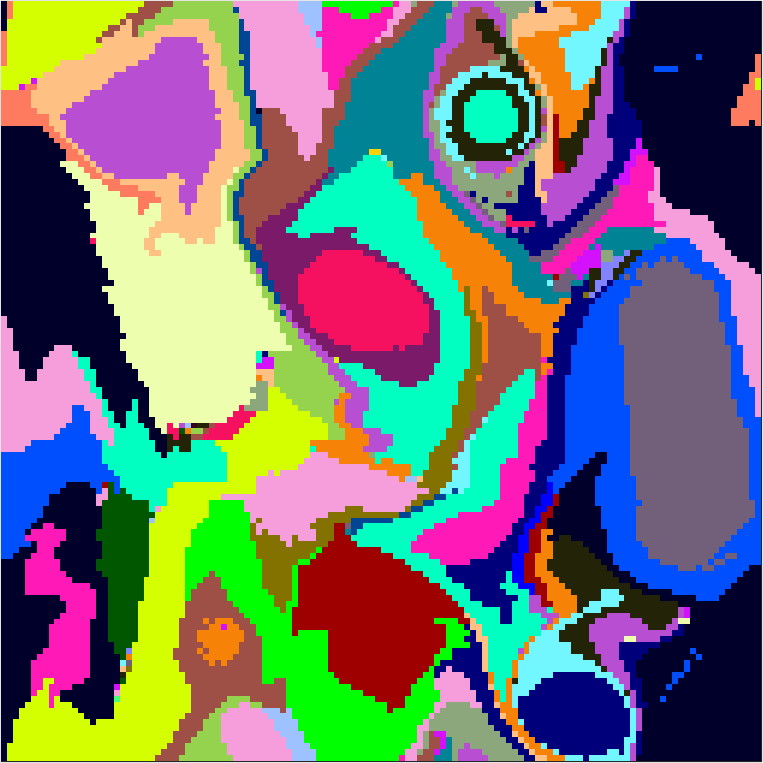}\; \includegraphics[width=0.3\textwidth]{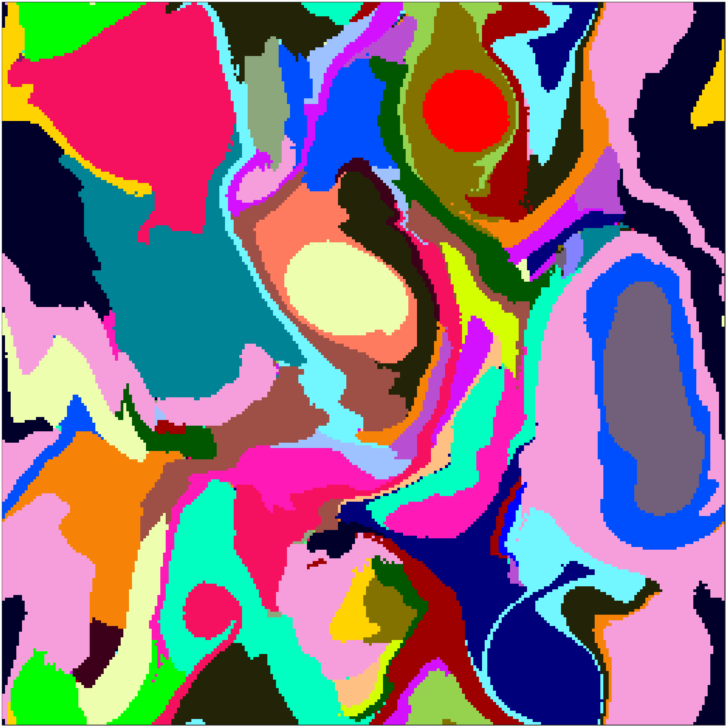}\;
\includegraphics[width=0.3\textwidth]{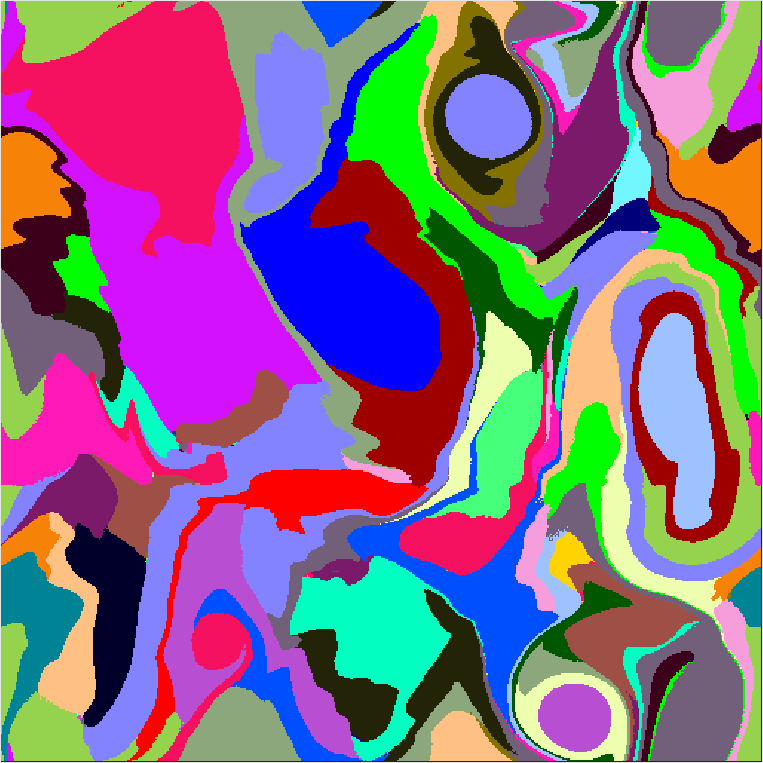} \caption{Hierarchical transfer operator method with $128$ (left), $256$ (middle)
and $512$ (right) boxes. For all cases: Each box contains $16$ points;
$8$ levels of hierarchy where used; and $\mu$-tolerance is set to
$5\times10^{-2}$.}
\label{fig:2Dturbhier} 
\end{figure}

The hierarchical application of the transfer operator method \cite{Ma13}
also signals vortex-like structures but these no longer stand out
of the many additional patches it labels as coherent sets. Most of
these patches appear to be examples of coincidental, rather than physical,
coherence with respect to the coherence metric imposed by the method.
An additional issue with the hierarchical transfer operator method
\cite{Ma13} is its convergence on this example. The method sets a
threshold on the relative improvement of the coherence with respect
to the reference probability measure $\mu$, which needs to be computed
and satisfied over consecutive refinements of coherent pairs. However, at each iteration, $\mu$ depends
on the initial numerical diffusion imposed by the box covering. As
a consequence, identifying similar coherent sets under various box
covering resolutions requires different threshold values. \Cref{fig:2Dturbhier}
shows the hierarchical coherent sets obtained with a fixed termination
threshold for three different box covering resolutions. \Cref{fig:2Dturbhier}
indicates no overall convergence, except in some minor details. 

\begin{figure}
\subfloat[\label{fig:2DturbTransferOperatortf}]{\includegraphics[width=0.3\textwidth]{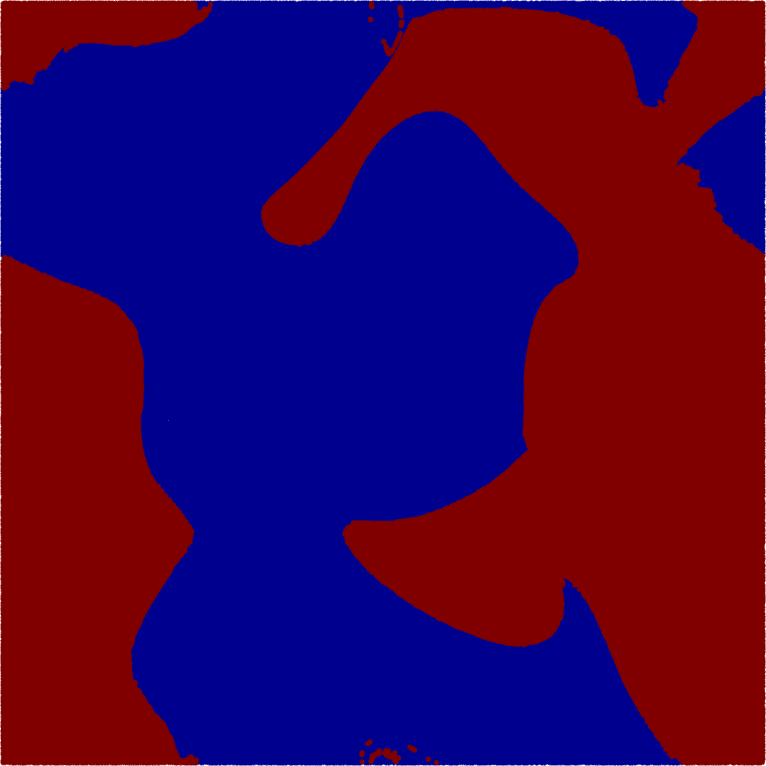}}\;
\subfloat[\label{fig:2DturbHierarchytf}]{\includegraphics[width=0.3\textwidth]{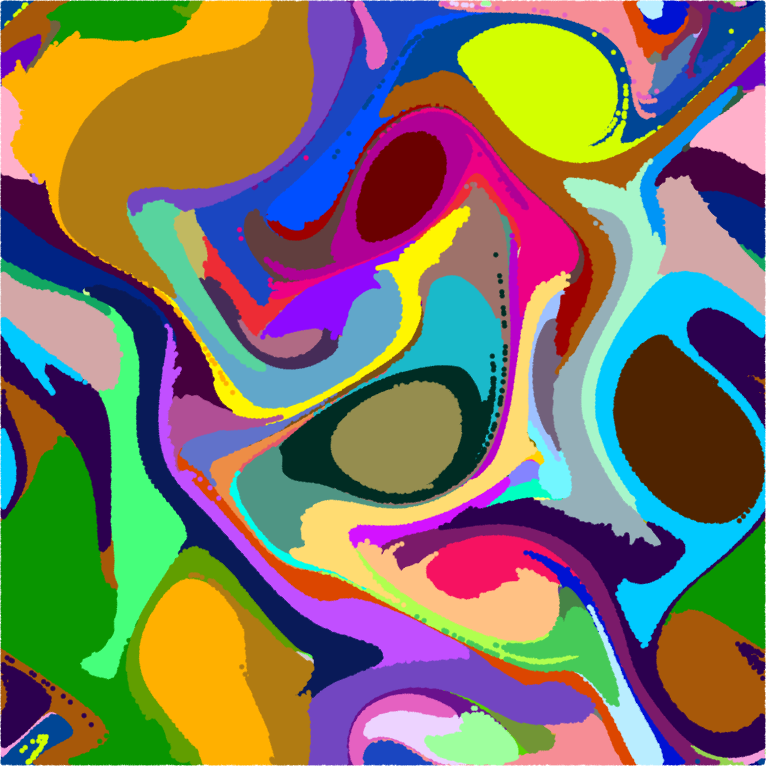}}\;
\subfloat[\label{fig:2DturbFCMtf}]{\includegraphics[width=0.3\textwidth]{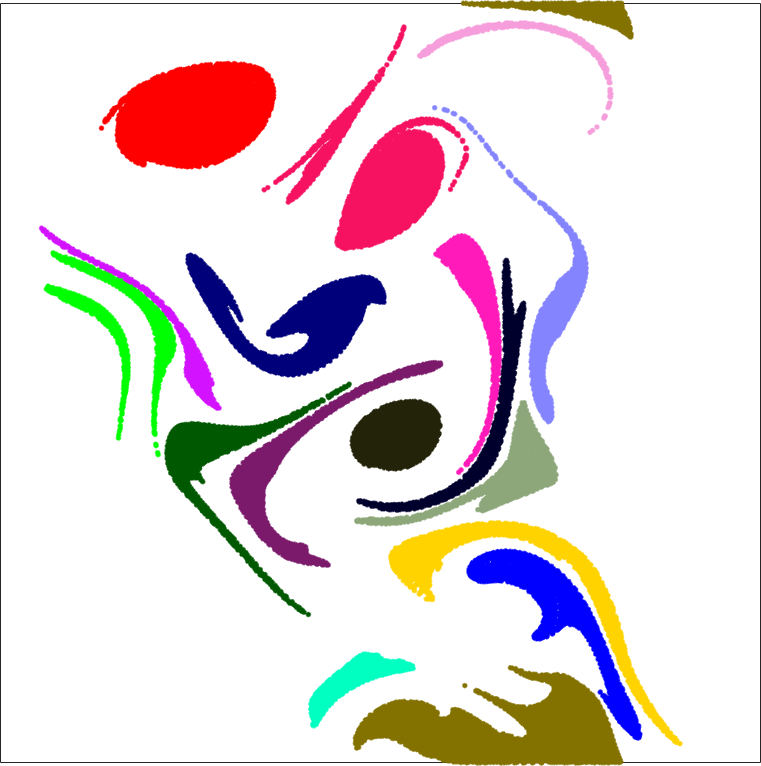}}\;
\subfloat[\label{fig:2Dturbclusteringtf}]{\includegraphics[width=0.3\textwidth]{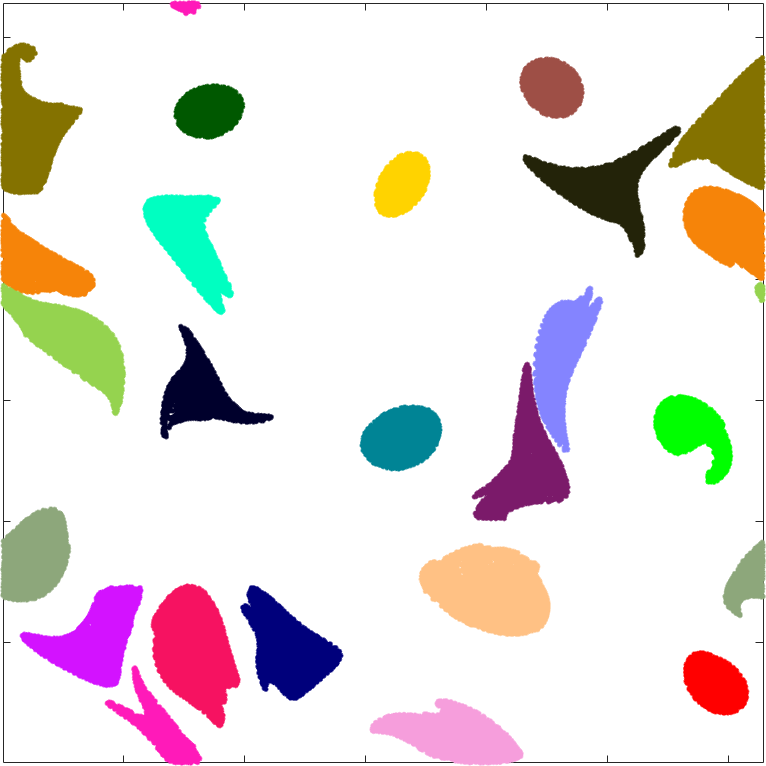}}\;
\subfloat[\label{fig:2Dturbgeodesictf}]{\includegraphics[width=0.3\textwidth]{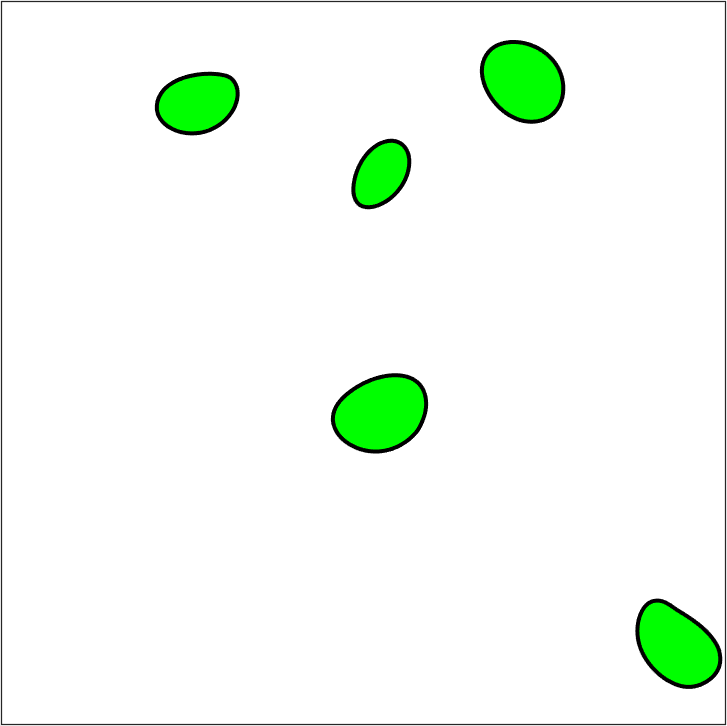}}\;
\subfloat[\label{fig:2DturbLAVDtf}]{\includegraphics[width=0.3\textwidth]{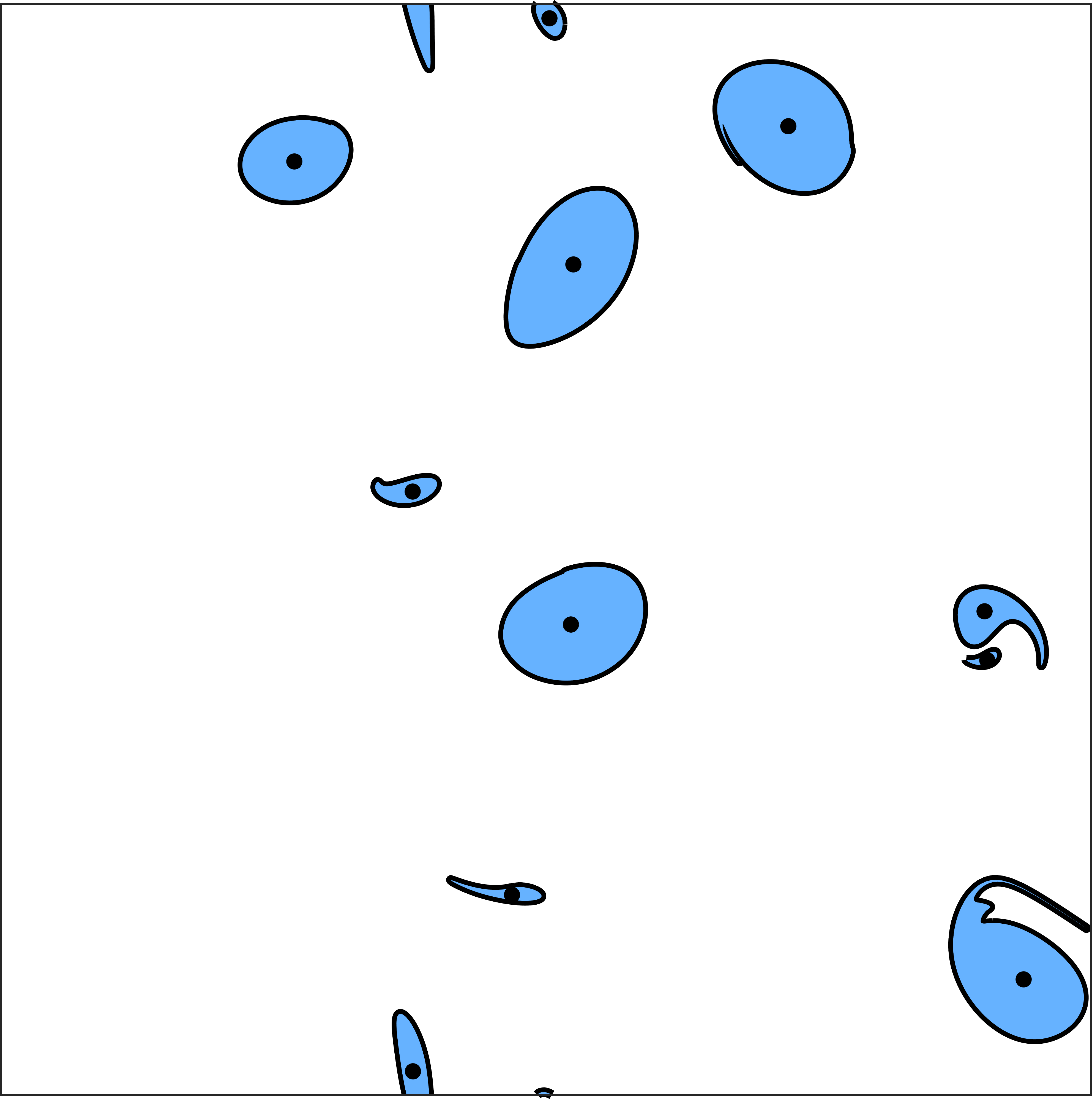}}
\caption{Advected images of Lagrangian coherent structures at the final time
$t_{1}=50$ for six different methods: (a) Probabilistic transfer operator (Multimedia
view) (b) Hierarchical transfer operator (Multimedia
view) (c) Fuzzy clustering (Multimedia
view) (d) Spectral clustering (Multimedia
view) (e) Geodesic (Multimedia
view) and (f) LAVD (Multimedia
view). Plots (a) and (b) have
lower resolution because the total number of trajectories used in all computations were selected equal for a fair comparison.}
\end{figure}

\Cref{fig:2DturbShapeCoherence} shows candidate regions (red) where
shape coherent sets may exist at the initial time $t_{0}=0$. In these
regions, the angle between stable and unstable foliations is smaller
than $5.7^{\circ}$, and hence all vortex boundaries should be fully
contained in these regions. Inspection of \Cref{fig:2DturbShapeCoherence},
however, reveals that these candidate regions are spirals, and hence
no closed vortex boundaries satisfying the shape coherence requirement
exist. This is unsurprising as the underlying coherence principle
is only arguable for flows whose behavior is the same in forward and
backward time, which is not the case for the present example.

\Cref{fig:2DturbSpectral} shows the coherent sets detected by the
spectral clustering method at the initial time. These coherent sets
include the vortices captured by the Geodesic and LAVD methods, as
well as some additional structures. \Cref{fig:2Dturbclusteringtf}
(Multimedia view) illustrates that the advected image of these additional
coherent sets indeed show limited dispersion at the final time $t_{1}=50$.
As in the case of hierarchical transfer operator method, some of these
moderately dispersive sets are of irregular, physically unexpected
shape. A systematic comparison with the results of the FTLE analysis
(see \Cref{fig:2DturbFTLE}) shows that all these irregularly shaped
regions are valleys of low FTLE values among FTLE ridges. Therefore,
beyond coherent vortices, spectral clustering also identifies domains
that are trapped between finite-time stable manifolds of saddle-type
(hyperbolic) trajectories. This feature may make spectral clustering
the method of choice in applications with a well-defined time scale
of interest (e.g., fixed-time forecasting problems). At the same time,
there is no a priori constraint in a turbulent flow that keeps stable
manifolds of different hyperbolic trajectories close to each other.
For this reason, several of the irregularly shaped sets identified
from spectral clustering may change substantially under changes in
the extraction interval.

\Cref{fig:2DturbFCM} shows that fuzzy clustering (with $m=1.5$
and $K=20$) also identifies both regularly and irregularly shaped
coherent sets. Three of these clearly indicate coherent vortices,
containing the coherent vortices indicated by other methods in these
locations. Since these larger vortices predicted by fuzzy clustering
only show tangential filamentation (cf. \Cref{fig:2DturbFCMtf}
(Multimedia view) ), this method gives the sharpest, least conservative
assessment of coherence for these vortices relative to the results
returned by other methods. That said, the method also completely misses
the remaining two, highly coherent larger vortices. Furthermore, the
irregularly shaped domains identified by fuzzy clustering lose their
coherence by the end time of the extraction interval, showing stretching
and filamentation in \Cref{fig:2DturbFCMtf} (Multimedia view).
The total number of extracted sets (the number $K$ of clusters) is
an input parameter for the method, so the number of inaccurate coherence
predictions are influenced by choices made by the user.

\Cref{fig:2DturbGeodesic} shows the geodesic Lagrangian vortex
boundaries (green) as well as the repelling hyperbolic LCSs (red)
at the initial time $t_{0}=0$. Coherent Lagrangian
vortex boundaries (black) are defined as the outermost members of nested elliptic LCS families. In \Cref{fig:2Dturbgeodesictf}
(Multimedia view), we confirm the sustained coherence of the geodesic
vortex boundaries by advecting them to the final time $t_{1}=50$.
At the same time, other methods (e.g., the LAVD method discussed below)
reveal additional vortices that should also be considered coherent
based on their advection properties, as they only exhibit limited
tangential filamentation. The geodesic method, is therefore, too conservative
to detect these smaller vortices.

\Cref{fig:2DturbLAVD} shows the Lagrangian vortex boundaries extracted
using the LAVD method at the initial time $t_{0}=0$. In this computation,
we have set the minimum arc-length, $l_{min}=0.3$ and convexity deficiency
bound $d_{max}=0.005$. In \Cref{fig:2DturbLAVDtf} (Multimedia
view), we confirm the Lagrangian rotational coherence of these vortex
boundaries by advecting them to the final time $t_{1}=50$. As guaranteed
by the derivation of the LAVD method, the vortex boundaries display
only tangential filamentation. With this relaxed definition of coherence,
the LAVD approach identifies additional smaller vortices missed by
the geodesic LCS method (cf. \Cref{fig:2DturbGeodesic}). At the
same time, the LAVD method only targets vortices, missing other patches
of trajectories that remain closely packed over the same time interval
(see, e.g., the discussion on spectral clustering above).

Beyond showing the results of various methods, we also use this example
to investigate whether contours of diagnostic tools such as the trajectory length function or mesochronic field can be used for the purpose of vortex
boundary detection. Specifically, we extract the contours of these
two diagnostic methods for two select vortex regions at initial time
$t_{0}=0$, and advect them to the final time $t_{1}=50$. In addition,
we make a comparison with the geodesic vortex boundaries obtained
for the same regions.

\begin{figure}[!htpb]
\includegraphics[width=0.31\textwidth]{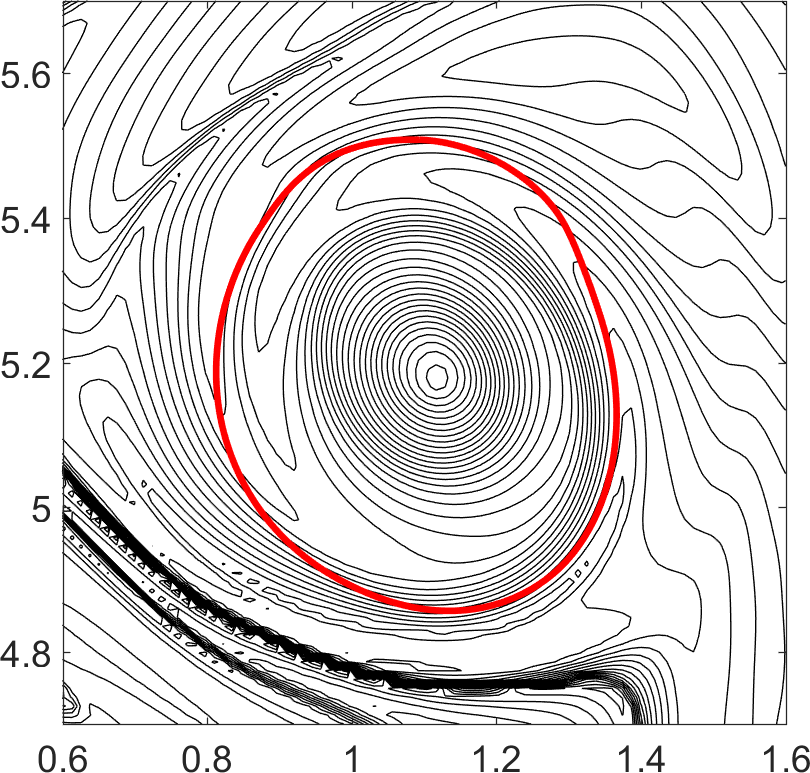} \raisebox{0.13\textwidth}{\includegraphics[height=1cm]{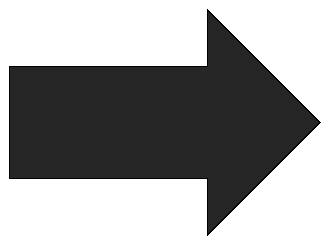}}
\includegraphics[width=0.31\textwidth]{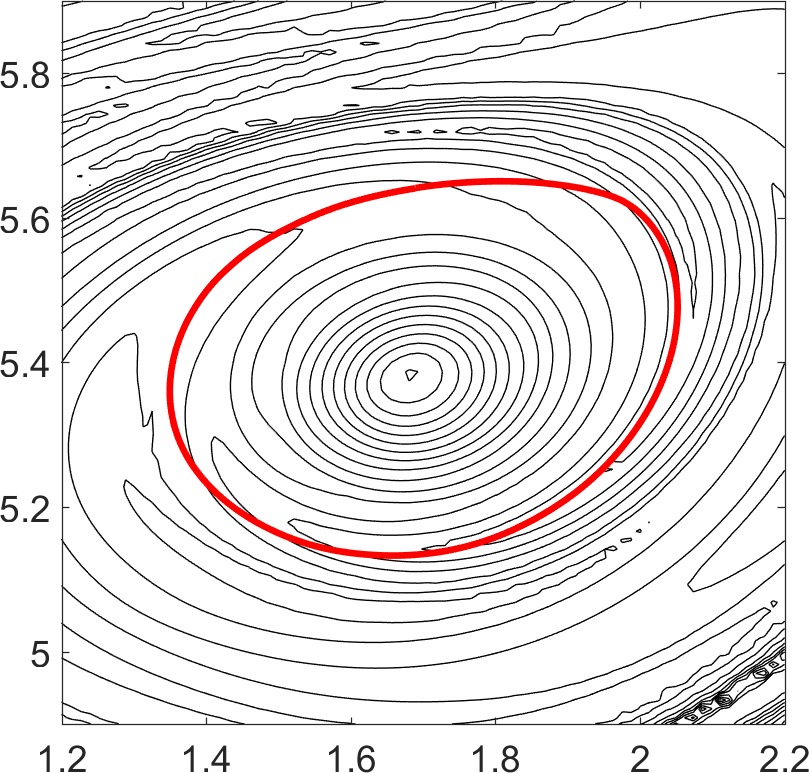}\\
 \vspace{0.02\textwidth}
 \includegraphics[width=0.31\textwidth]{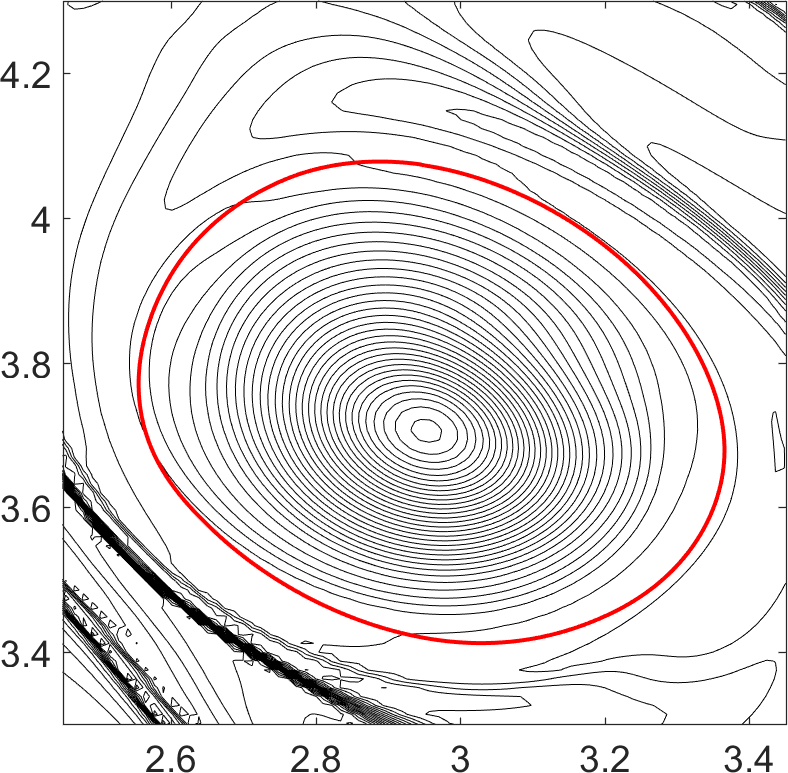} \raisebox{0.13\textwidth}{\includegraphics[height=1cm]{arrow}}
\includegraphics[width=0.31\textwidth]{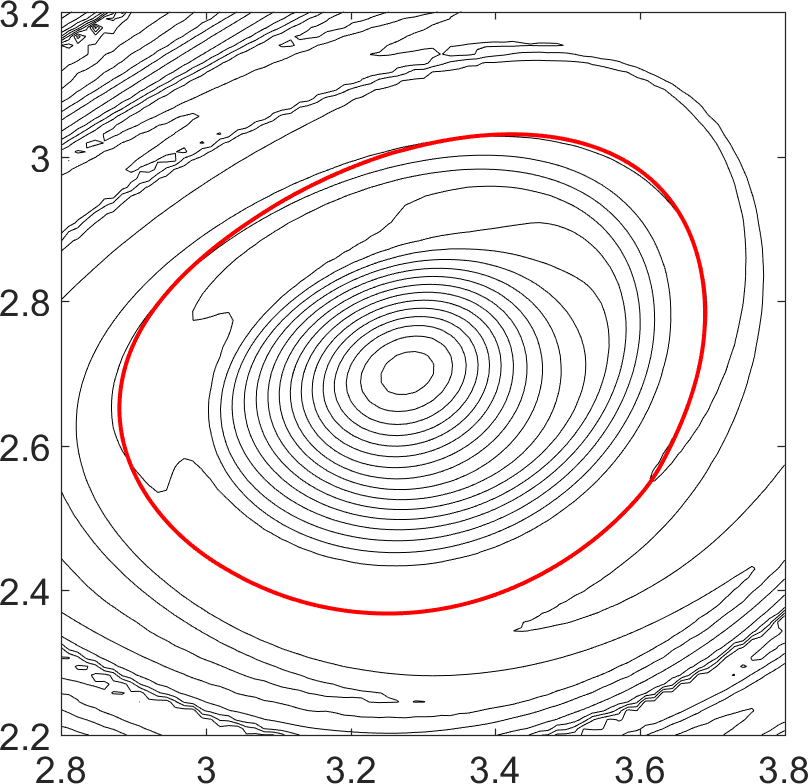} \caption{M-function contours (black curves) and the geodesic vortex boundary
(red curves) at the initial time $t=0$ (left) and at the final time
$t=50$ (right). }
\label{fig:2DturbMfunc}

\includegraphics[height=0.31\textwidth]{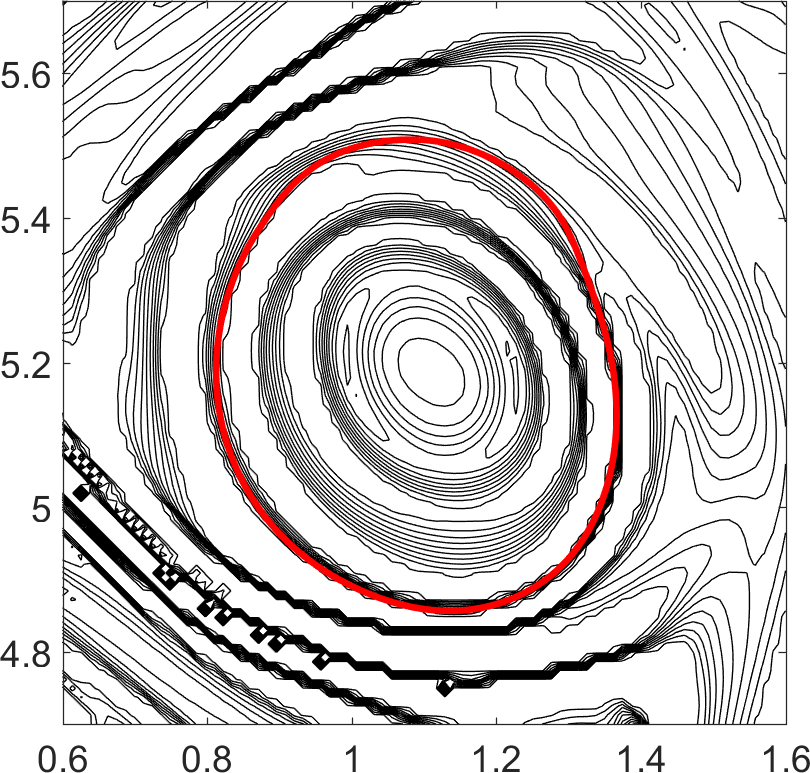} \raisebox{0.13\textwidth}{\includegraphics[height=1cm]{arrow}}
\includegraphics[height=0.31\textwidth]{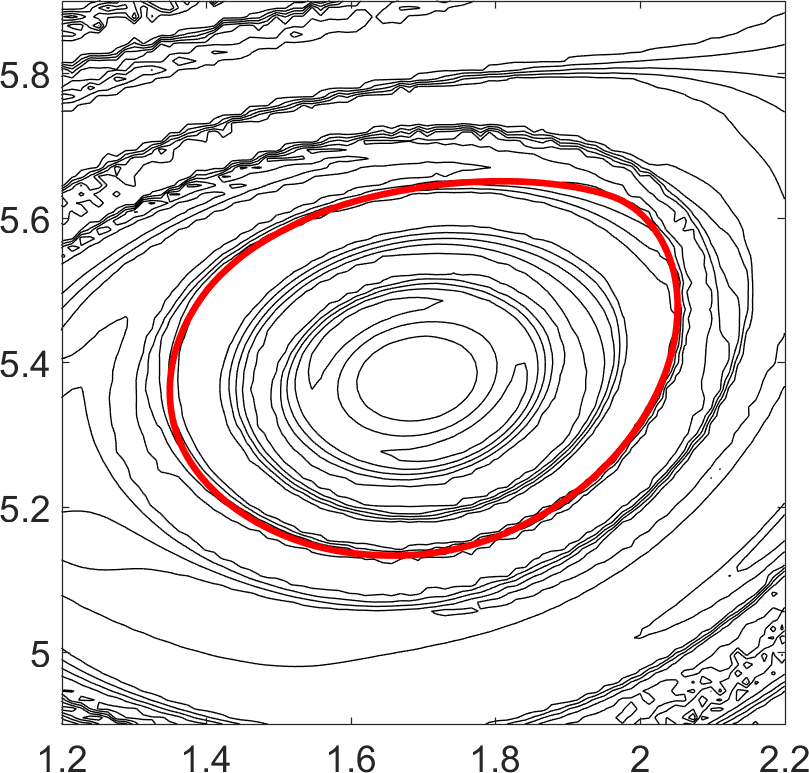}\\
 \vspace{0.02\textwidth}
 \includegraphics[height=0.31\textwidth]{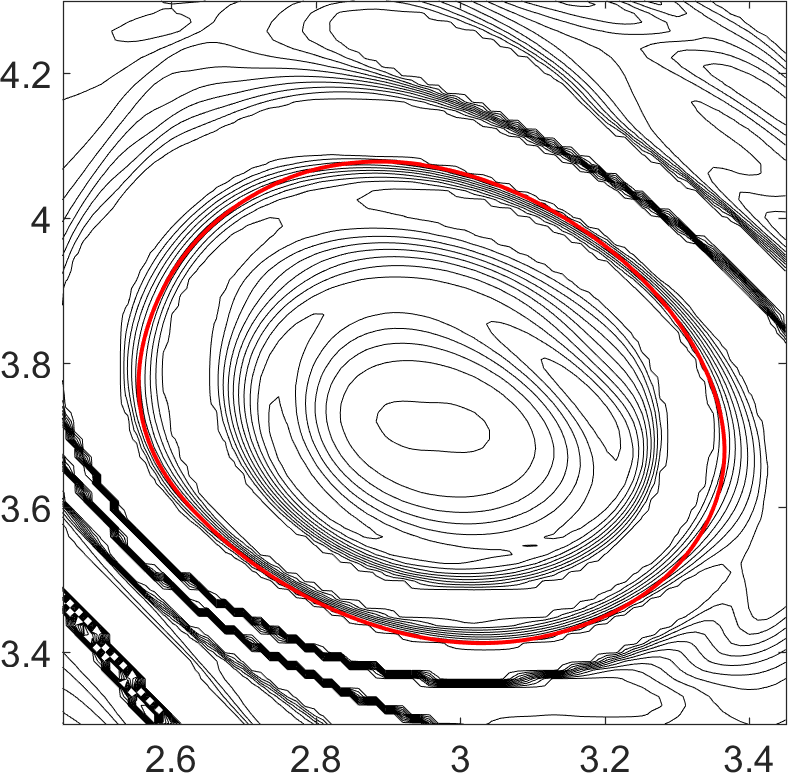} \raisebox{0.13\textwidth}{\includegraphics[height=1cm]{arrow}}
\includegraphics[height=0.31\textwidth]{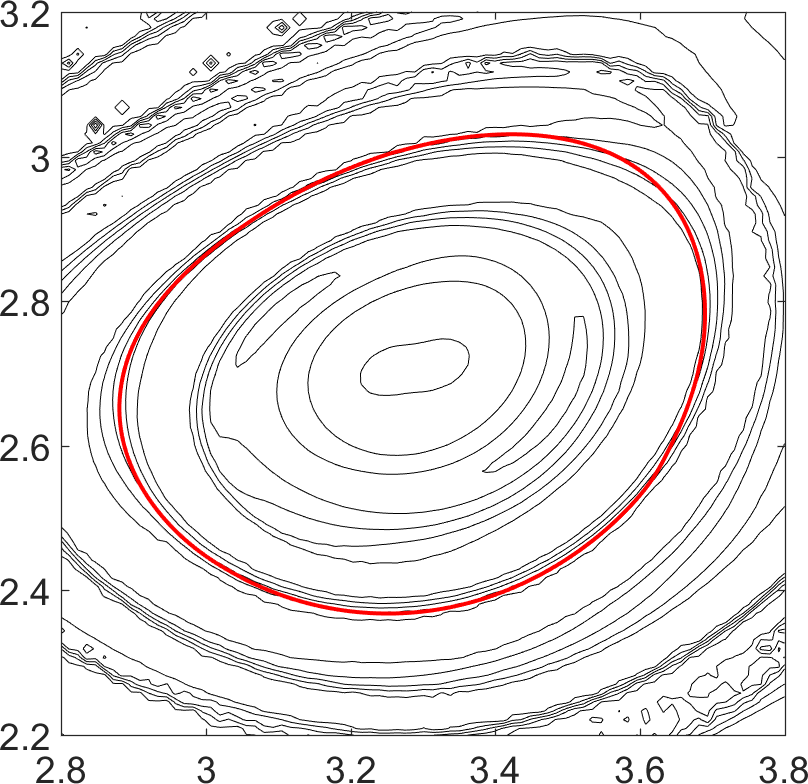} \caption{Contours of the mesochronic scalar (black curves) and the geodesic
vortex boundary (red curves) at the initial time $t=0$ (left) and
at the final time $t=50$ (right).}
\label{fig:2Dturbmeso} 
\end{figure}

\Cref{fig:2DturbMfunc} shows the advection of the level-curves
of the trajectory length function $M_{0}^{50}$ around two select vortices.
The level-curves closer to the vortex core remain coherent for both
vortices. A comparison with the geodesic vortex boundary, however,
shows that the contours of $M_{0}^{50}$ underestimate the size of
the upper vortex substantially. A precise implementation of the mesochronic
vortex criterion of Ref.~\onlinecite{Mezic14} shows again a lack
of vortex-type structures in the selected regions due to presence
of the saddle-type critical points. In contrast, a visual inspection
of the same regions in \Cref{fig:2Dturbcompare}, without implementing
the specific vortex criterion of Ref.~\onlinecite{Mezic14}, does
suggest coherent vortices in all vortical regions identified by the
geodesic and the LAVD method. The actual boundaries of the vortices, however, cannot be inferred based on such an inspection.

\subsection{Wind field from Jupiter's atmosphere}
\begin{figure}
\subfloat[\label{fig:JupiterFTLE}]{\includegraphics[width=0.31\textwidth]{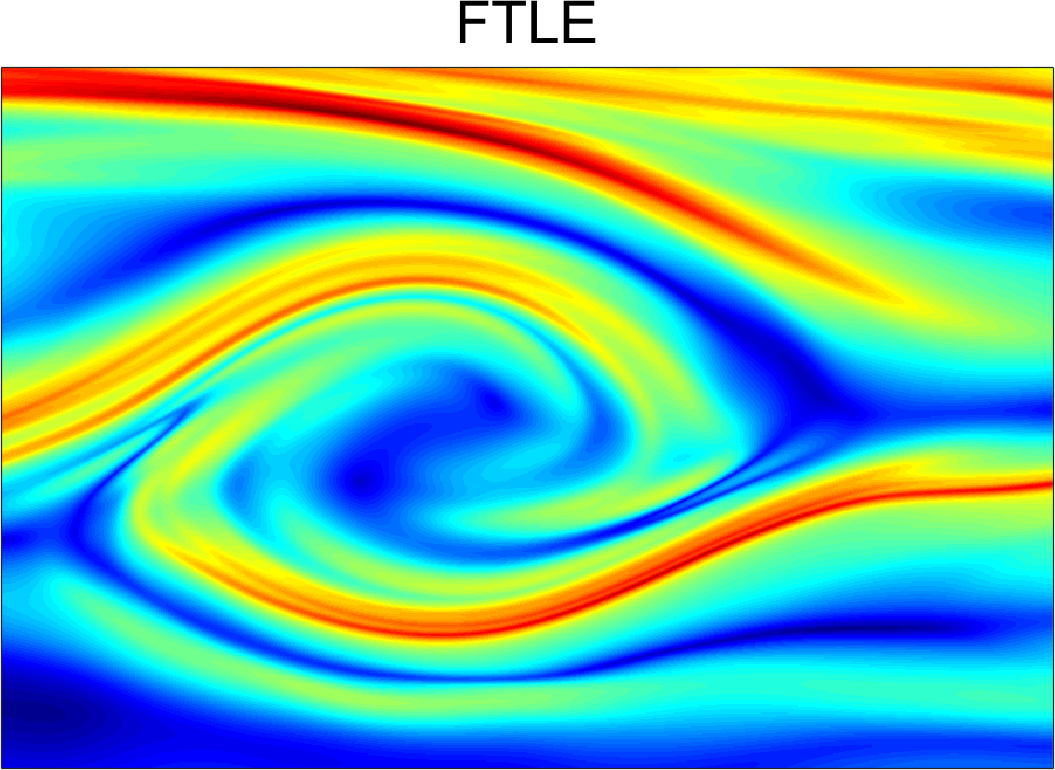}}\; \subfloat[\label{fig:JupiterFSLE}]{\includegraphics[width=0.31\textwidth]{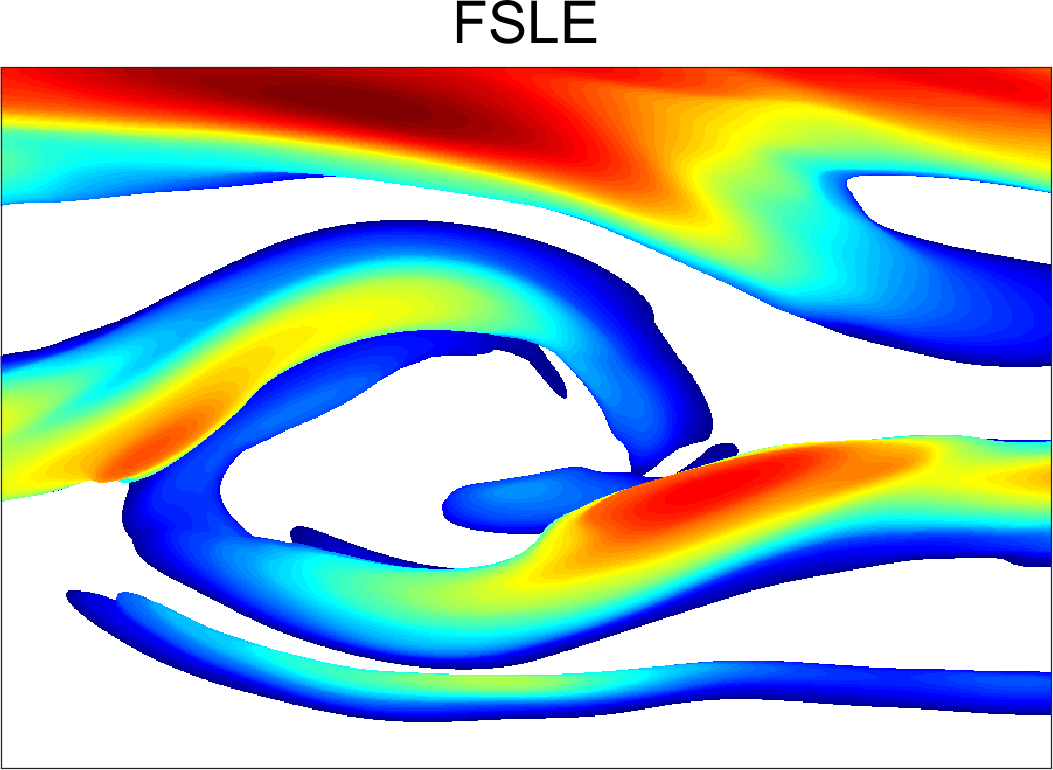}}\; 
\subfloat[\label{fig:JupiterMesochronic}]{\includegraphics[width=0.31\textwidth]{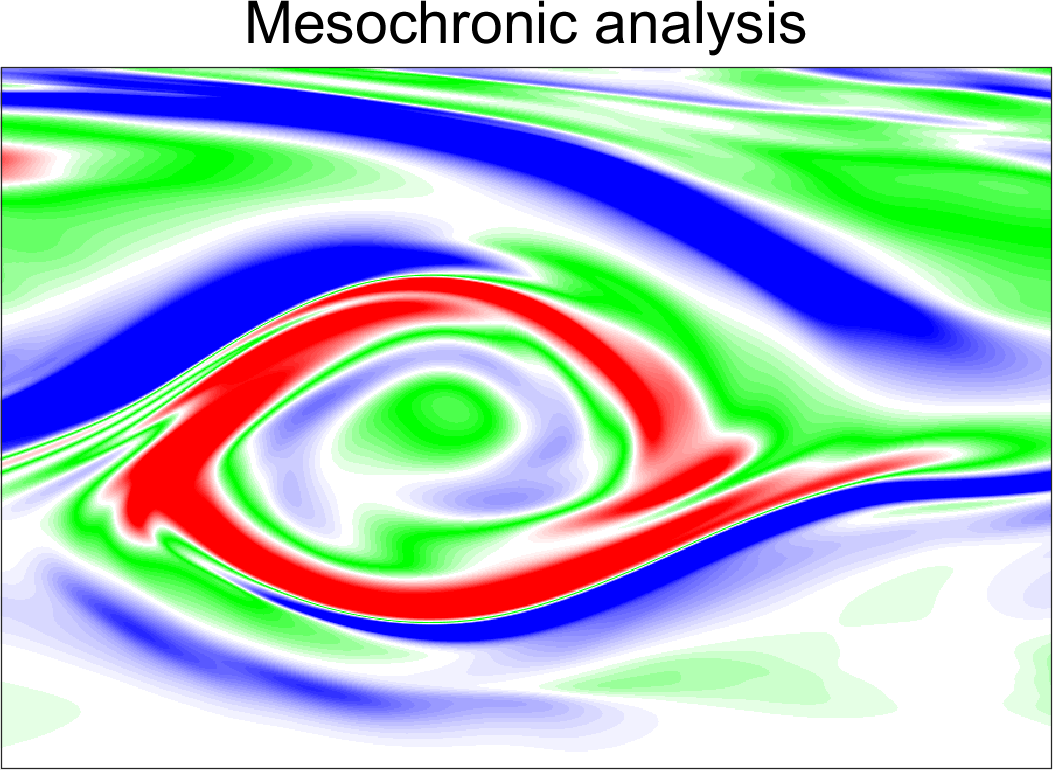}}\\
\subfloat[\label{fig:JupiterMfunction}]{\includegraphics[width=0.31\textwidth]{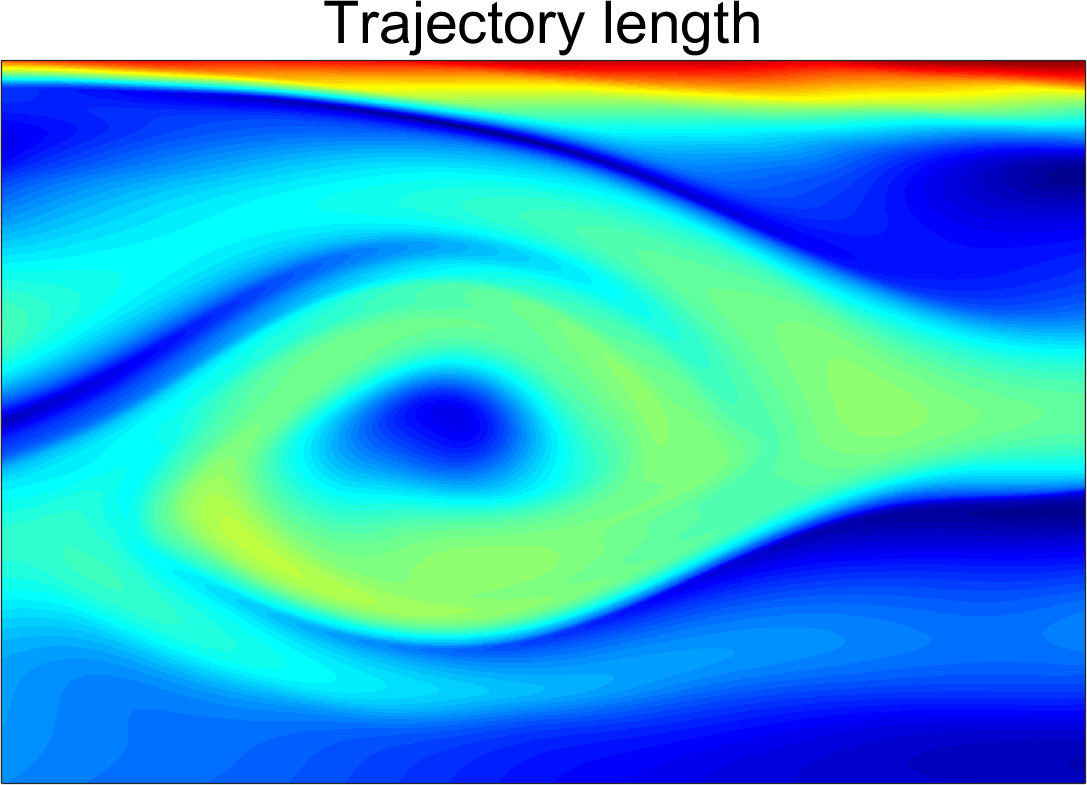}}\;
\subfloat[\label{fig:JupiterCM}]{\includegraphics[width=0.31\textwidth]{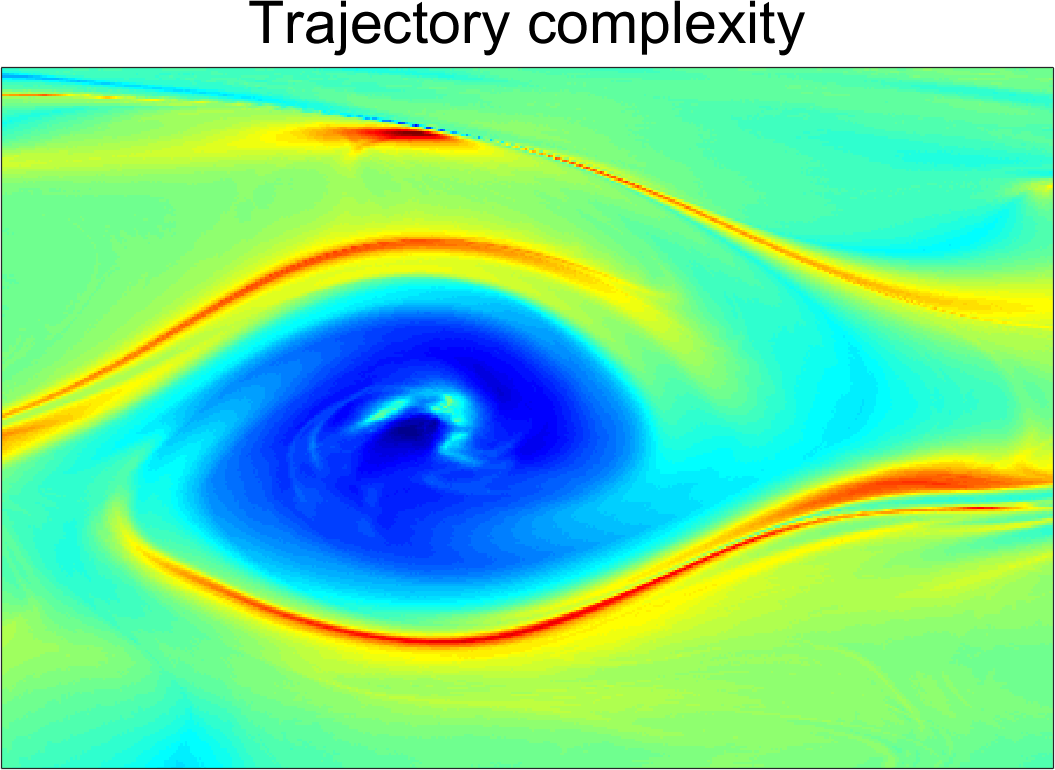}}\; \subfloat[\label{fig:JupiterShapeCoherence}]{\includegraphics[width=0.31\textwidth]{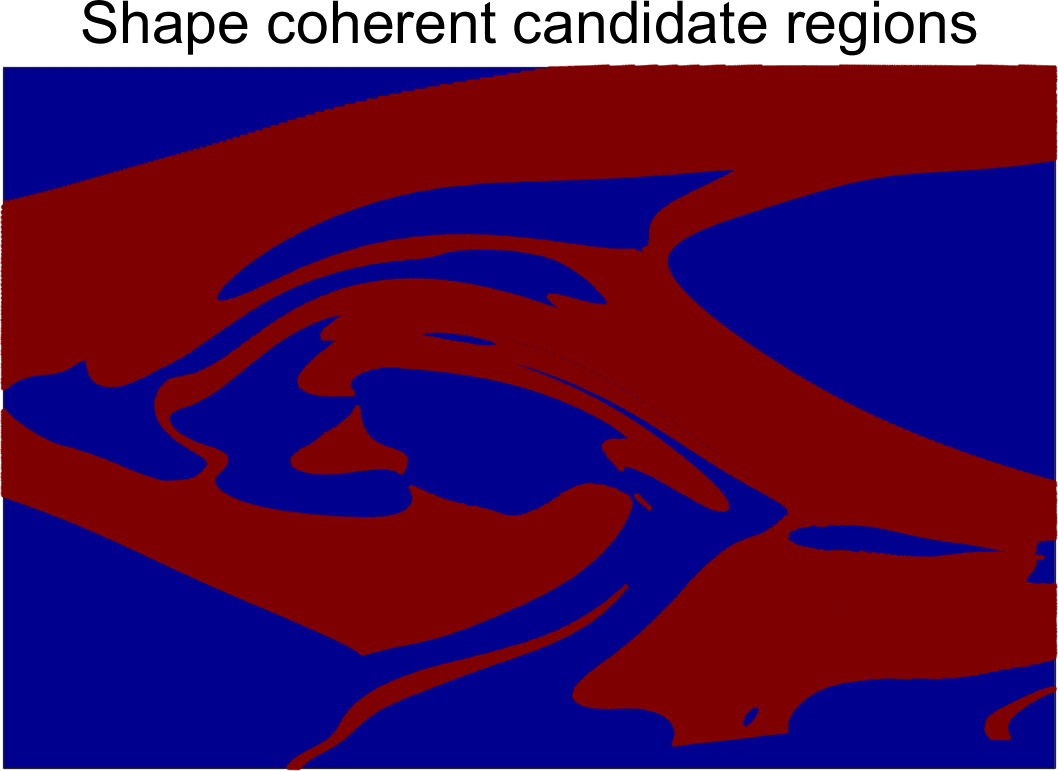}}\\ 
\subfloat[\label{fig:JupiterTransferOperator}]{\includegraphics[width=0.31\textwidth]{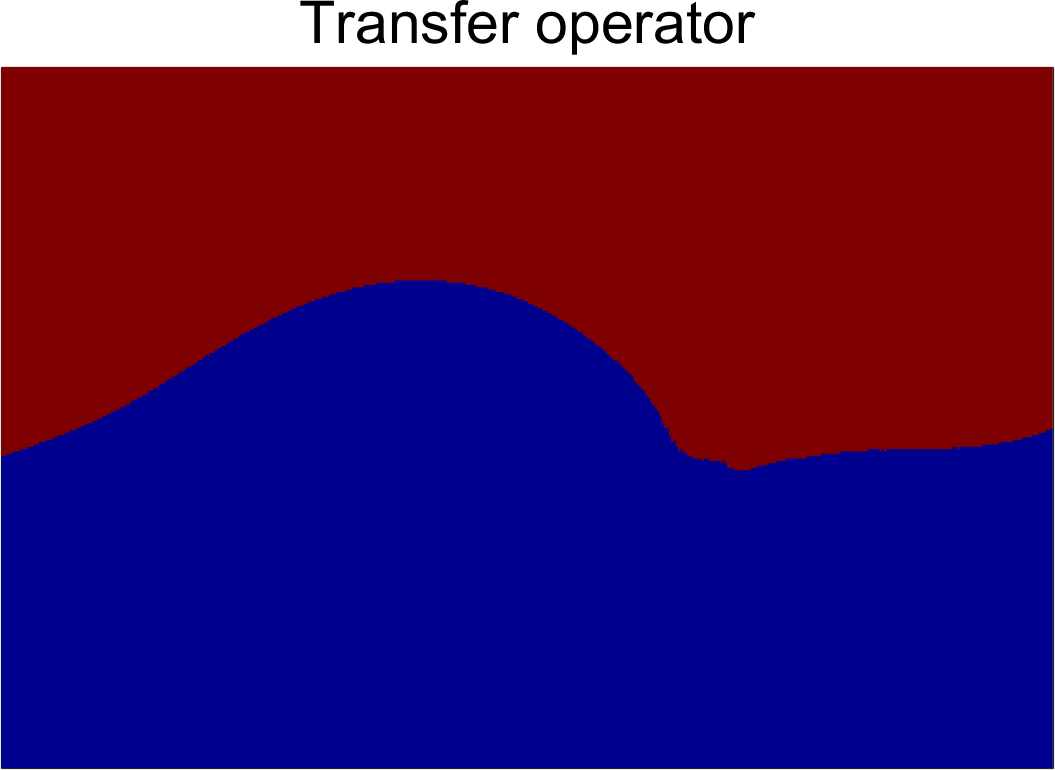}}\; \subfloat[\label{fig:JupiterHierarchy}]{\includegraphics[width=0.31\textwidth]{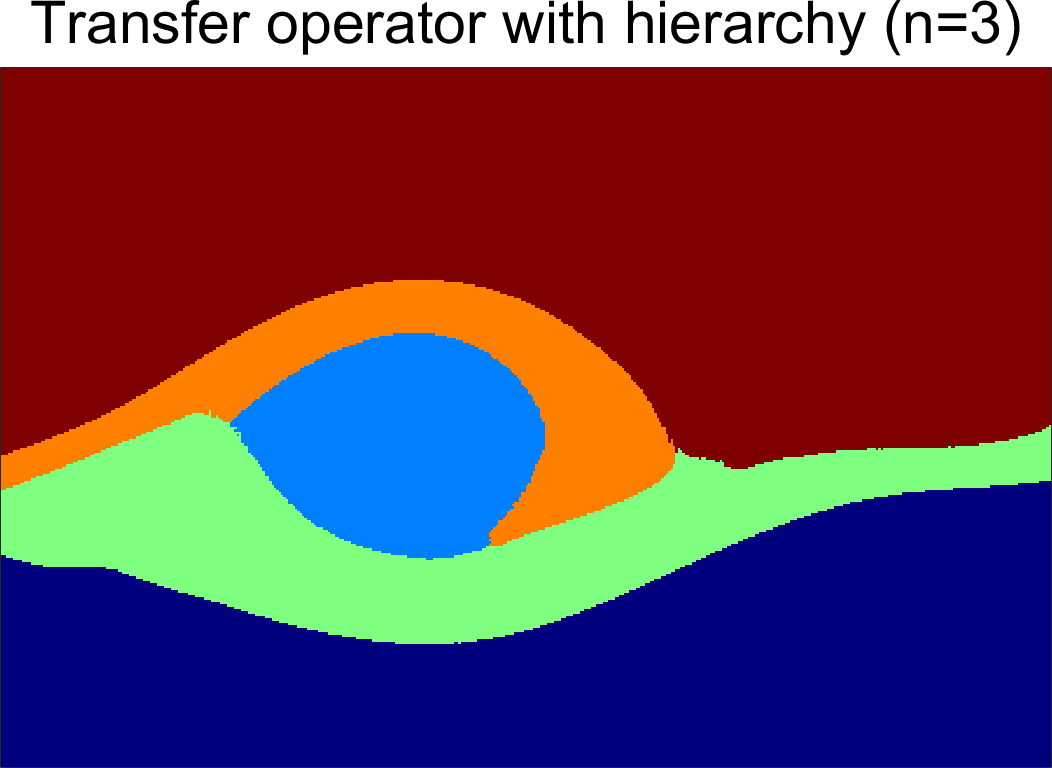}}\; \subfloat[\label{fig:JupiterFCM}]{\includegraphics[width=0.31\textwidth]{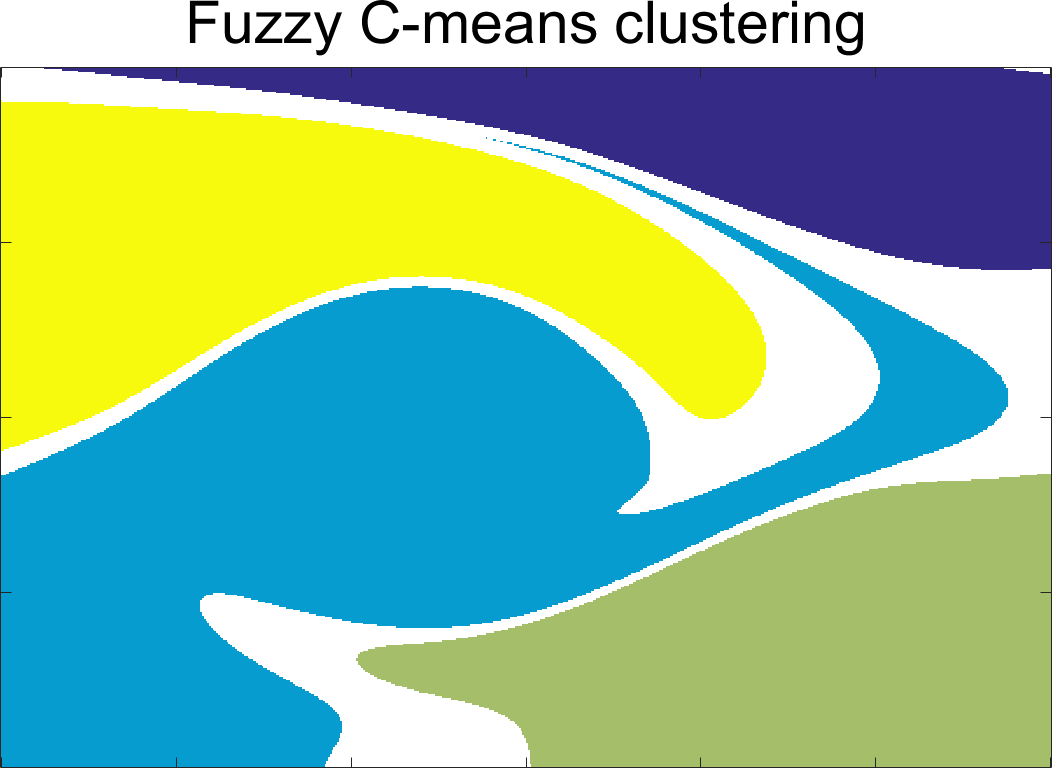}}\\
\subfloat[\label{fig:JupiterSpectral}]{\includegraphics[width=0.31\textwidth]{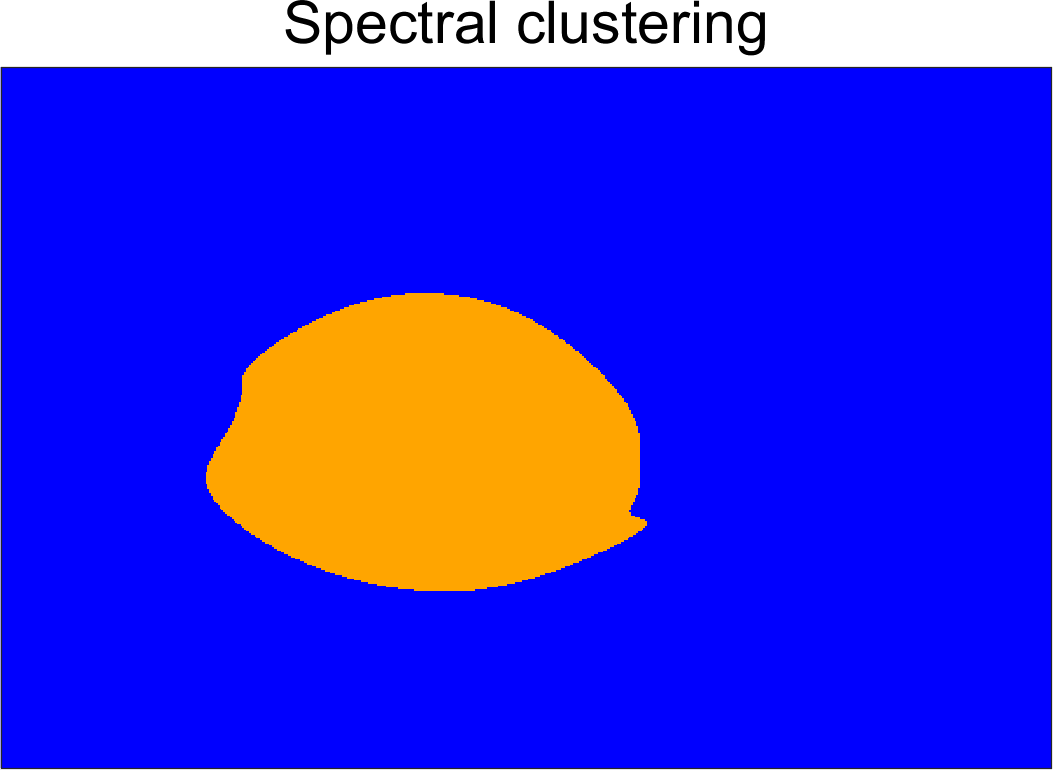}}\; \subfloat[\label{fig:JupiterGeodesic}]{\includegraphics[width=0.31\textwidth]{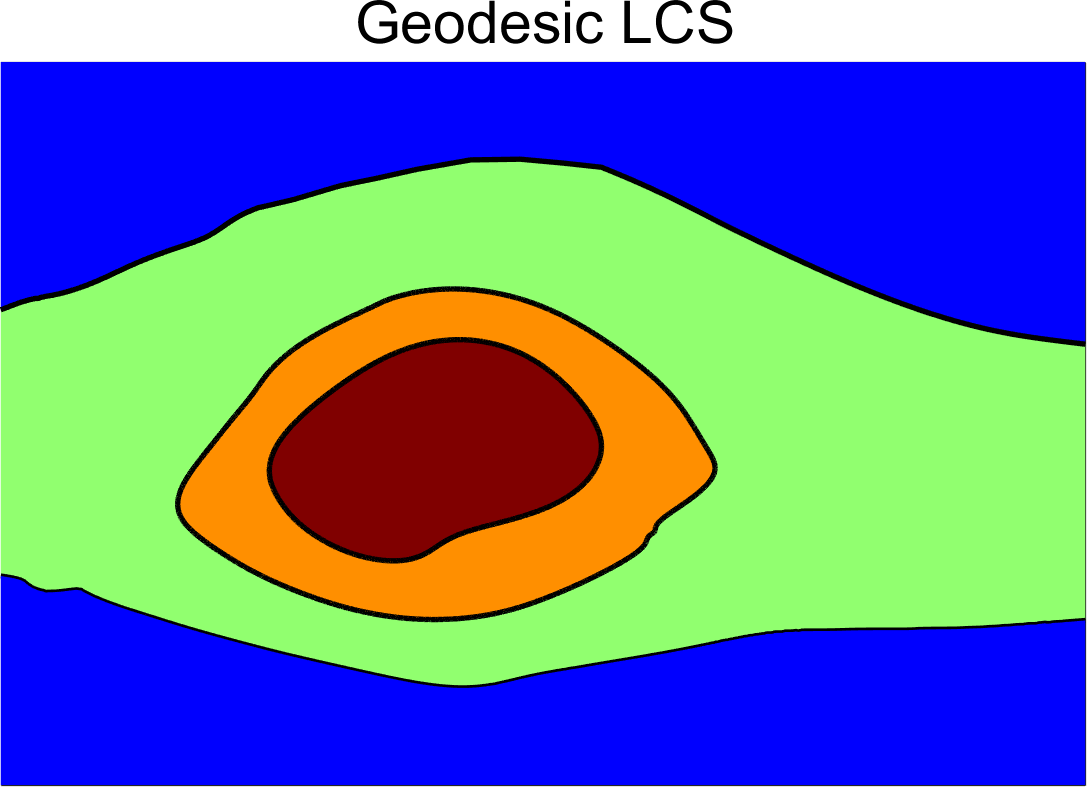}}\; \subfloat[\label{fig:JupiterLAVD}]{\includegraphics[width=0.31\textwidth]{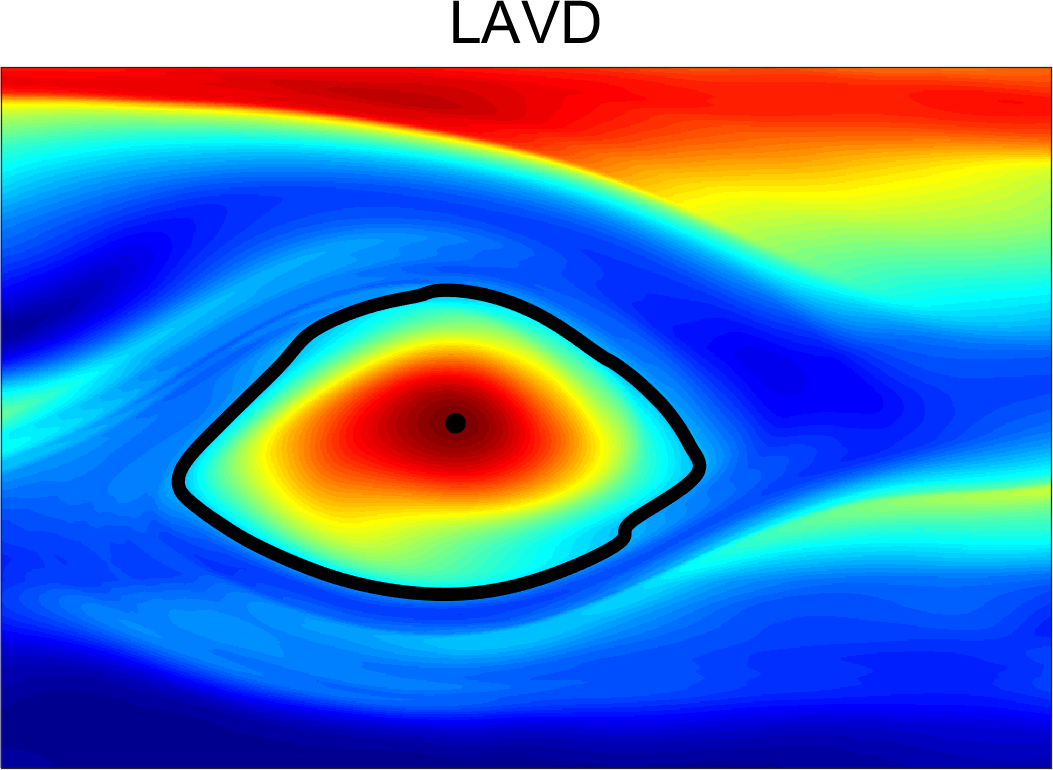}}
\caption{The output of all methods at initial time $t_{0}$ for Jupiter's wind-velocity
field of Ref.~\onlinecite{Hadjighasem16}.}
\label{fig:Jupiterpicbook} 
\end{figure}

In our third example, we compare the twelve Lagrangian structure
detection methods on an unsteady velocity field extracted from video
footage of Jupiter's atmosphere. The video footage was acquired by
the Cassini spacecraft, covering $24$ Jovian days, ranging from October
31 to November 9 in year 2000. To reconstruct the velocity field,
we used the Advection Corrected Correlation Image Velocimetry (ACCIV)
method \cite{Asay09} to obtain high-density, time-resolved velocity
vectors (cf. Ref.~\onlinecite{Hadjighasem16} for details). This
is a characteristically finite-time problem: no further video footage
and hence no further time-resolved velocity data are available outside
the time interval analyzed here. Furthermore, the data was acquired
in a frame orbiting around Jupiter, and hence the frame-invariance
of the results is a crucial requirements.

In this example, we use a total number of $1800\times1200$ particles
for all the methods. The spatial domain $U$ in question ranges from
$-61.6^{\circ}$ W to $-31.6^{\circ}$ W in longitude and from $-8.9^{\circ}$
S to $-28.9^{\circ}$ S in latitude. We perform the computation of
gradient-based approaches, such as FTLE, FSLE, mesochronic, shape
coherence and geodesic LCS analysis, using an auxiliary grid to ensure high-precision and numerical stability in the finite differencing.
Specifically, an embedded grid of resolution $900\times600$ is used
to construct the corresponding scalar fields. In contrast, we use
a uniform grid of $1800\times1200$ for the gradient-free methods.
As for the transfer-operator-based approaches, we use a grid of $450\times300$
boxes, with $16$ uniformly uniformly sampled points per grid box.
Here, we use a variable-order Adams\textendash Bashforth\textendash Moulton
solver (ODE113 in MATLAB), with relative and absolute tolerances of
$10^{-6}$, for trajectory advection. We obtain the velocity field
at any given point by interpolating the velocity data set using bilinear
interpolation.

As seen in \Cref{fig:Jupiterpicbook}, several methods that offer
specific structure boundary definitions signal a localized, vortex-type
coherent structure corresponding to the Great Red Spot (GRS) of Jupiter.
Exceptions to this rule are the transfer operator, shape coherence,
fuzzy clustering and the mesochronic method. The 5th singular vector
(not shown here) of the transfer operator does give an indication
of the GRS, similarly to \Cref{fig:Bickleycombined,fig:2Dturbcompare}.
As in our pervious examples, however, an inspection of the singular
value spectrum of the transfer operator does not a priori suggest
a distinguished role for the 5th singular vector. 

As in our previous example, the hierarchical transfer operator method
also signals a localized vortex-like structure (see \Cref{fig:JupiterHierarchy}).
The precise implementation of the mesochronic vortex criterion of
Ref.~\onlinecite{Mezic14} provides again no coherent vortex boundary
due to the lack of a nested sequence of smooth closed contours. An
intuitive visual inspection of the mesochronic plots still suggests
a vortical structure to the extent that other heuristic diagnostics
do (the FTLE, FSLE, M-function and trajectory complexity methods).

Spectral clustering, geodesic LCS detection and the LAVD method give
very close results for the boundary of the GRS in this example. This
suggests that the core of the Great Red Spot is a fairly well defined
material vortex with negligible material filamentation.

As for jet identification, most diagnostic methods in \Cref{fig:Jupiterpicbook}
give some indication of two jets passing north and south of the GRS.
However, since these methods give no clear recipe for jet identification,
we could not go beyond a general visual assessment of the results.
In contrast, the transfer operator, the hierarchical transfer operator and the geodesic LCS methods suggest clear jet cores or jet boundaries. The ones signalled by the geodesic LCS method also coincide with zonal jet cores observed visually in Jupiter's atmosphere \cite{Hadjighasem16}.

\begin{figure}
\subfloat[\label{fig:JupiterTransferOperatortf}]{\includegraphics[width=0.48\textwidth]{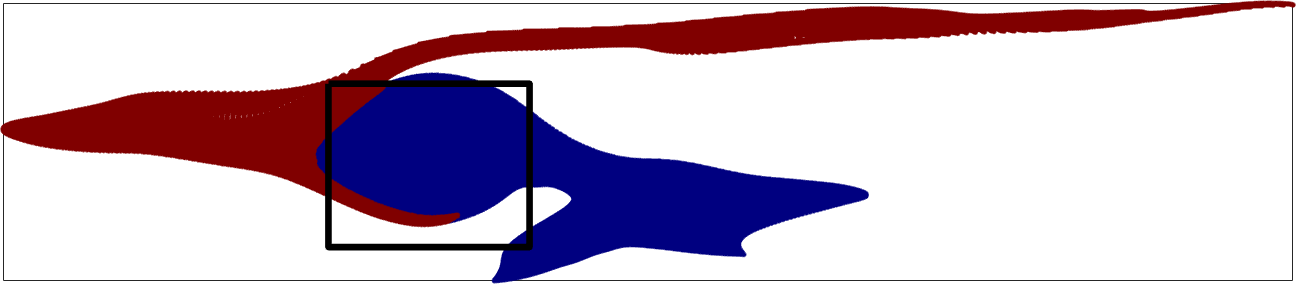}}\; \subfloat[\label{fig:JupiterHierarchytf}]{\includegraphics[width=0.48\textwidth]{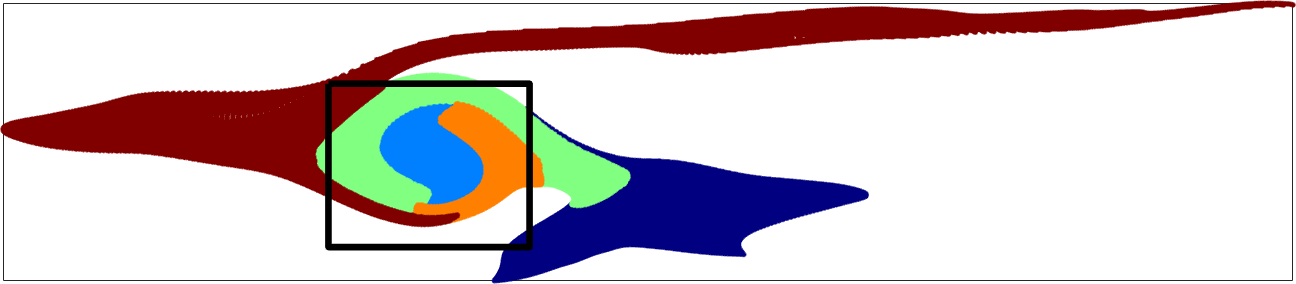}}\\
\subfloat[\label{fig:JupiterFCMtf}]{\includegraphics[width=0.48\textwidth]{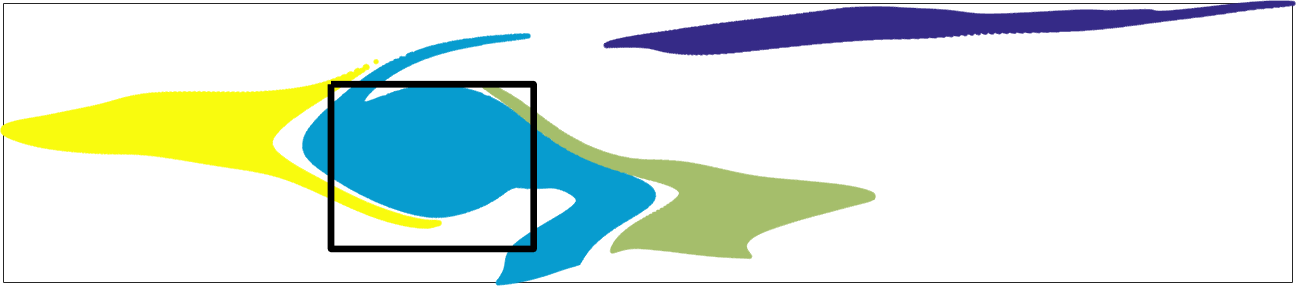}}\\
\subfloat[\label{fig:Jupiterclusteringtf}]{\includegraphics[width=0.25\textwidth]{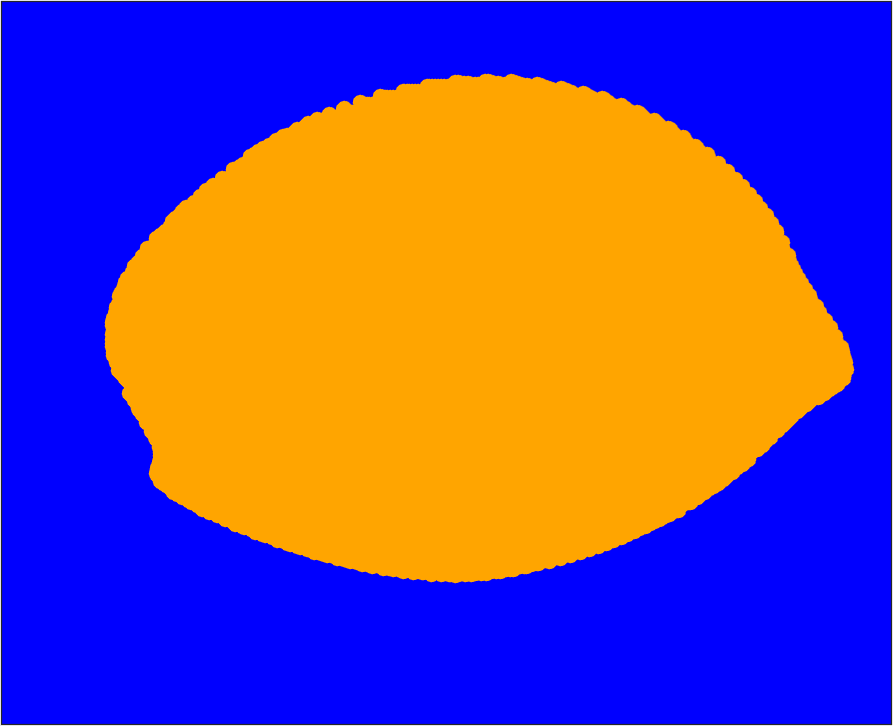}}\; \subfloat[\label{fig:Jupitergeodesictf}]{\includegraphics[width=0.25\textwidth]{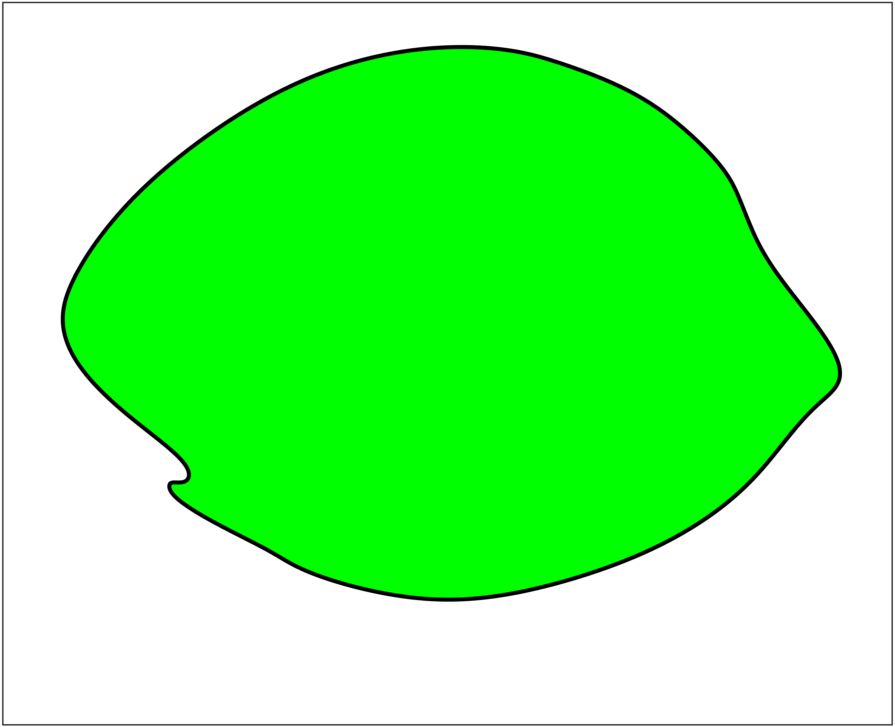}}\; \subfloat[\label{fig:JupiterLAVDtf}]{\includegraphics[width=0.25\textwidth]{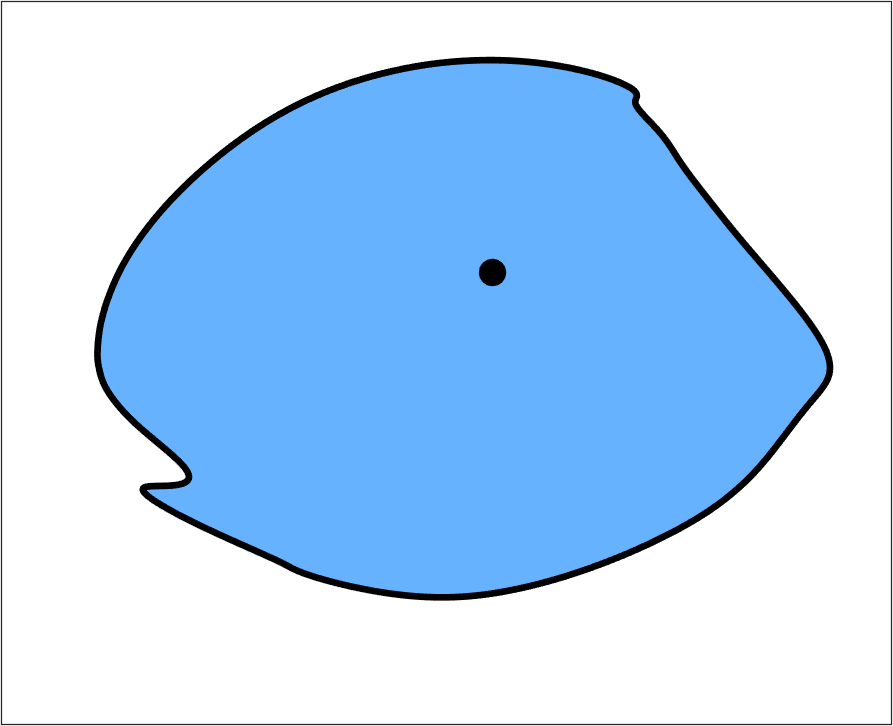}}
\caption{Advected images of Lagrangian coherent structures at the final time $t_{1}=24$ for six different methods: (a) Probabilistic transfer operator (Multimedia
view) (b) Hierarchical transfer operator (Multimedia
view) (c) Fuzzy clustering (Multimedia
view) (d) Spectral clustering (Multimedia
view) (e) Geodesic (Multimedia
view) and (f) LAVD (Multimedia
view). Plots (a), (b) and (c) are constructed using the advected image of the original rectangular domain of \Cref{fig:JupiterTransferOperator,fig:JupiterHierarchy,fig:JupiterFCM}, respectively. In these plots, the box framing the plots (d), (e), and (f) is shown in black for reference.}
\label{fig:Jupiteradvection} 
\end{figure}

As in the earlier two examples, fuzzy clustering (with $m=1.25$ and
$K=4$) gives convoluted structure boundary candidates, some of which
stretch out significantly under advection in \Cref{fig:JupiterFCMtf}.
The observed stretch in these boundaries is unsurprising, given that
they are transverse to known transport barriers (shear jets) in the
atmosphere of Jupiter. Together with our earlier Bickley jet example,
the present example indicates a difficulty for fuzzy clustering to
identify vortical and jet-type structures when both are present.

In \Cref{fig:JupiterTransferOperatortf,fig:JupiterHierarchytf,fig:JupiterFCMtf} (Multimedia view), we advect to the final time $t_{1}=24$ the initial rectangular domain and the partitioning of this domain into sets identified as coherent by the probabilistic transfer operator, hierarchical transfer operator, and fuzzy clustering methods. 
A zoom-in of the advected image of the single subset identified as coherent by the spectral clustering, geodesic LCS, and LAVD methods is shown in \Cref{fig:Jupiterclusteringtf,fig:Jupitergeodesictf,fig:JupiterLAVDtf} (Multimedia view). 
As noted above, coherent regions predicted by the transfer operator and  the transfer operator with hierarchy have common boundaries that remain short by construction. At the same time, the advected domain boundaries (which are beyond the control of these methods)  stretch substantially over the time interval of advection. In contrast, all boundary components of the  regions identified by the fuzzy clustering as coherent either are long at the initial time or stretch under advection, defying expectations for coherence.  Finally, all structure boundaries predicted by the spectral clustering, geodesic LCS and LAVD methods stay coherent.

\section{Assessment}\label{sec:Assessment}

Based on our three benchmark examples and available evidence in the
literature, we now summarize the inferred strengths and weaknesses
of the twelve Lagrangian methods compared here:

\subsubsection{FTLE method}
\begin{description}
\item [{Strengths}] simple and objective algorithm; FTLE ridges capture
hyperbolic LCSs under additional mathematical conditions (cf. Ref.~\onlinecite{Haller11}); FTLE trenches tend to approximate jet cores (but see Ref.~\onlinecite{Farazmand14} for exceptions).
\item [{Weaknesses}] no reliable detection of elliptic LCSs; ridges connect
locations of high stretching with those of high shear, and hence also
produce false positives for hyperbolic LCS.
\end{description}

\subsubsection{FSLE method}
\begin{description}
\item [{Strengths}] simple and objective algorithm; requires no a priori
time scale of integration; can be focused on a length scale of interest; requires no differentiation of the flow map with respect to initial conditions. 
\item [{Weaknesses}] correspondence to actual hyperbolic LCS is limited
(cf. Ref.~\onlinecite{Karrasch13}); no reliable detection of elliptic
and parabolic LCSs; highlights structures arising over different time
intervals; has unremovable jump discontinuities (cf. Ref.~\onlinecite{Karrasch13}). 
\end{description}

\subsubsection{Mesochronic analysis}
\begin{description}
\item [{Strengths}] simple algorithm; visual inspection often reveals features
generally consistent with material vortices detected by objective methods.
\item [{Weaknesses}] nonobjective; unclear mathematical meaning for non-periodic trajectories (cf. \Cref{Mesochronic_method}); no reliable
detection of hyperbolic and parabolic LCS; elliptic and hyperbolic
classification of trajectories inconsistent with classic notions of
stability (cf. \Cref{Mesochronic_method}); precise implementation of additional vortex criterion of
Ref.~\onlinecite{Mezic14} eliminates most visually inferred material vortex candidates. 
\end{description}

\subsubsection{Trajectory length method}
\begin{description}
\item [{Strengths}] simplest of all to implement; visual inspection often
reveals features consistent with output of other methods; requires
no differentiation of flow map with respect to initial conditions. 
\item [{Weaknesses}] nonobjective; a number of known counterexamples in
simple flows show inconsistencies with the method (cf. Ref.~\onlinecite{Ruiz-Herrera15,Ruiz-Herrera16});
unclear definition of a coherent structure.
\end{description}

\subsubsection{Trajectory complexity method}
\begin{description}
\item [{Strengths}] simple and objective algorithm; underlying principle
is physically intuitive; topology is consistent for all vortical features;
requires no differentiation of the flow map with respect to initial
conditions.
\item [{Weaknesses}] delivers no clear structure boundaries; lacks clear
mathematical connection to coherence.
\end{description}

\subsubsection{Shape coherence method}
\begin{description}
\item [{Strengths}] objective; intuitive for steady and time-periodic flows.
\item [{Weaknesses}] assumes that stretching history is the same in forward
and backward time; as a consequence, misses coherent structures in time-dependent flow data; no clear recipe for extracting closed structure boundaries.
\end{description}

\subsubsection{Transfer operator/dynamic Laplacian method}
\begin{description}
\item [{Strengths}] objective method with an appealing mathematical foundation; supported by rigorous estimates for material coherence; 
applies in any dimensions; gives sharp structure boundaries when a
given flow region can be partitioned into precisely two coherent sets
(e.g., two sides of a jet core-type barrier); higher-order eigenfunctions
reveal further coherent structures; can be applied to diffusive problems
as well (probabilistic transfer operator only); requires no differentiation of the flow map with respect
to initial conditions; requires a small number of user inputs.
\item [{Weaknesses}] computationally expensive (this does not apply to
dynamic Laplacian method); does not generally detect hyperbolic LCSs;
first nontrivial singular vector will always partition the domain
into just two coherent sets; an a priori unclear number of further
singular vectors need to be deployed and thresholded to recover coherent
features revealed by some other methods. 
\end{description}

\subsubsection{Hierarchical transfer operator method}
\begin{description}
\item [{Strengths}] objective method with an appealing mathematical foundation; 
not limited to flows with two coherent sets; incremental implementation
possible until required granularity is reached; no inspection of an
a priori undetermined number of higher eigenfunctions is required;
requires no differentiation of flow map with respect to initial conditions; requires a small number of user inputs.
\item [{Weaknesses}] lack of overall convergence under increasing hierarchy;
the number of identified coherent subsets increases endlessly even
in regions that are clearly homogeneous; unphysical output tends to
arise over a certain level of granularity. 
\end{description}

\subsubsection{Fuzzy clustering of trajectories}
\begin{description}
\item [{Strengths}] simple implementation, appealing theoretical foundation;
objective method; requires no differentiation of the flow map with
respect to initial conditions.
\item [{Weaknesses}] most detected structures have convoluted shapes that
differ from known coherent structure boundaries; some of the detected
structures lose their coherence further via stretching under advection;
inability to detect hyperbolic LCS and difficulty in detecting elliptic
and parabolic LCSs simultaneously; robustness of structures need to
be checked over different parameters; number of coherent structures
to be located is an input parameter.
\end{description}

\subsubsection{Spectral clustering of trajectories}
\begin{description}
\item [{Strengths}] objective method with an appealing mathematical foundation; simple implementation; number of coherent structures is output; requires no differentiation
of the flow map with respect to initial conditions; requires a small number of user inputs; consistently finds open, low-dispersion regions beyond elliptic LCS.
\item [{Weaknesses}] requires a well-defined spectral gap; computationally
expensive for a large number of trajectories; inability to detect
hyperbolic and parabolic LCSs; also produces low-dispersion structures whose robustness is unlikely under variations in the extraction time; the size of the spectral gap varies with the choice of the sparsification radius.
\end{description}

\subsubsection{Geodesic LCS method}
\begin{description}
\item [{Strengths:}] automated and objective detection of hyperbolic,
elliptic and parabolic LCS; supported by exact variational principles;
perfect lack of filamentation is guaranteed for elliptic LCS under
advection.
\item [{Weaknesses:}] computationally involved; detects only the most coherent
elliptic LCSs, misses those with non-uniformly stretching boundaries;
unlike all other methods reviewed, it does not extend to three dimensions; automated implementation in [32] requires a large number of numerical parameters and requires a parameter-sensitive identification of Cauchy--Green singularities (but see Ref.~\onlinecite{Serra16} for a recent implementation eliminating all these issues)
\end{description}

\subsubsection{LAVD method}
\begin{description}
\item [{Strengths:}] automated, simple and objective algorithm; low computational
cost; requires no differentiation of the flow map with respect to
initial conditions; precise mathematical relationship to material
rotation; requires a small number of user inputs.
\item [{Weaknesses:}] inability to detect hyperbolic and parabolic LCS;
relies on derivatives of the velocity field; requires a minimal spatial
scale and a maximal convexity deficiency parameter; assumes large
enough computational domain for spatial mean vorticity to be representative.
\end{description}

\section{Conclusions} \label{sec:conclude} 
In addition to the specific evaluations we have given for twelve coherent structure methods in the previous section,
we now discuss some general aspects of Lagrangian coherence detection.

We have found that the performance of randomly chosen scalar fields
compares favorably with that of the heuristic Lagrangian diagnostic
tools surveyed here. This is no coincidence: the significance of LCSs
is precisely that they leave observable footprints in \emph{any} generic
scalar field advected by the flow. Such footprints can clearly be
observed in various physical processes in the ocean, ranging from
larval transport \cite{Harrison13} and algal blooms \cite{Olascoaga08}
to massive transport of salinity and temperature via coherent structures
\cite{Beal11}. These imprints, however, reveal the consequence, rather
than the root cause, of observed material coherence in unsteady flows
\cite{Beron-Vera15,Haller15}. Accordingly, the emergence of features
in a heuristic diagnostic field for specific examples does not constitute
a validation of the intuitive arguments used in constructing that
diagnostic.

As is generally accepted, the material deformation of a fluid (or
any continuum) cannot depend on the frame of the observer \cite{Deville12}.
This implies that questions inherently linked to material deformation
(such as material coherence, material transport, material mixing or
lack thereof) should be expressible in terms of objective physical
quantities. No matter how straightforward this requirement might sound,
several of the Lagrangian diagnostic tools developed over the past few
years fail to satisfy it (see, e.g., the trajectory arclength and
the mesochronic approaches discussed in the present comparison). We
believe that, just as all newly proposed constitutive laws in continuum
mechanics, newly proposed Lagrangian (i.e., material) coherence principles
and computational methods should be required to pass the requirement
of objectivity (see Ref.~\onlinecite{Peacock15} for more arguments
supporting this requirement).

As a second requirement, we believe that a new Lagrangian coherence detection method should have a specific quantitative statement on what a coherent structure is, and how it can be extracted systematically. This would help in moving beyond the current trend of visually inspecting colorful pictures, a practice that is inherently subjective and forgiving towards false positives and false negatives.

As a third requirement, we believe that a Lagrangian coherence detection
method should deliver on capturing at least the majority of structures
in truly aperiodic finite-time data sets, such as the three benchmark
flows treated in this paper. This implies moving beyond the current
practice of illustrating a proposed approach on the simplest two-dimensional,
bounded and time-periodic flows (typically a time-periodic double
gyre model). Trajectories in such flows can be run forever in forward
and backward times, displaying the characteristically recurrent, and
highly idealized, patterns of time-periodic flows. Several diagnostics
proposed for aperiodic flow data, in fact, crucially depend on assuming
such recurrence to justify implicit assumptions in their derivations
(see, e.g., the shape coherence method and the mesochronic analysis
reviewed here).

As a fourth requirement, the actual material coherence of structures
delivered by any method at the initial time $t_{0}$ should be confirmed
by simple material advection. The advected image of a coherent set
should then satisfy the exact coherence principle laid down at the
derivation of the underlying method. It is this last step that may
hold even a well-argued, mathematical method to task by exposing the
weakness of its underlying coherence principle.

If any of the above four requirements fails, it appears to make little
sense to propose a method for exploring a priori unknown material
coherent structures in complex unsteady flows. This is especially
true when a Lagrangian method is intended for now-casting, short-term
forecasting, flow control, or real-time decision making in sensitive
situations.

\begin{acknowledgements}
We are grateful to Daniel Karrasch, Igor Mezi{\'{c}}, and Irina Rypina, for useful conversations and suggestions.
\end{acknowledgements}

\bibliography{ThesisLib_new}

 \end{document}